\documentclass[12pt]{article} \usepackage[xdvi]{graphicx}
\usepackage{amssymb}
\makeatletter
\@addtoreset{equation}{section}
\makeatother

\begin{document}

\begin{titlepage}
\begin{flushright} 
HIP-2008-09/TH \\
\end{flushright}

\vspace{3cm}

\begin{center}{\Large\bf
Gauge and Yukawa couplings in 6D supersymmetric SU(6) models
}
\end{center}

\vspace{1ex}

\begin{center}
{\large 
Nobuhiro Uekusa}
\end{center}
\begin{center}
{\it Department of Physics, 
University of Helsinki  \\
and Helsinki Institute of Physics, \\
P.O. Box 64, FIN-00014 Helsinki, Finland} \\
\textit{E-mail}: nobuhiro.uekusa@helsinki.fi
\end{center}


\vspace{1ex}

\begin{abstract}
We study six-dimensional (6D) SU(6) supersymmetric models
where the doublet-triplet splitting, 
quark-lepton mass relations and
gaugino-mediated supersymmetry breaking are taken into account.
We find that effective 4D gauge coupling constants have 
highly nontrivial
behavior between two compactification scales.
It is shown that realistic patterns of Yukawa
coupling constants are obtained for valid values of parameters
and that hierarchical numbers are generated via
suppression by extra-dimensional effects.

\end{abstract}
\end{titlepage}

\newpage
\pagenumbering{arabic}

\section{Introduction}

Approaching grand unification and supersymmetry in higher dimensions
has been an intriguing possibility.
One of the problems to be solved in grand unification 
is the doublet-triplet splitting.
In the four-dimensional (4D) minimal SU(5) grand unified model,
an adjoint Higgs field is responsible 
for breaking the unified gauge group to 
the standard model gauge group, whereas
a fundamental Higgs field leads to
breaking electroweak symmetry.
This fundamental Higgs field includes a color-triplet Higgs
field as well as the weak-doublet Higgs field
under the standard model gauge group.
At tree level, the triplet and doublet Higgs fields 
acquire their
masses depending on the parameters in the potential of
the original adjoint and fundamental Higgs fields.
It is natural that these masses are of
the same order.
Because the doublet Higgs field whose vacuum expectation value 
is developed at the electroweak scale should be lighter than 
the triplet Higgs field,
a tuning would be needed 
unless the hierarchy is generated by any mechanism.
In addition, even if the masses are tuned at tree level,
radiative corrections can break it.
In theory with extra dimensions,
higher-dimensional gauge invariance can consist of
a unified gauge group while the standard model
gauge group only survives on 4D.
The unified gauge group is reduced to the standard model
gauge group via boundary conditions in the direction
of the extra dimensions.
If the fundamental SU(5) Higgs field  
in the role of the electroweak symmetry breaking
propagates in higher dimensions,
the mass splitting of the triplet and doublet Higgs fields
can be obtained as a result of the boundary 
conditions~\cite{Kawamura:2000ev}.
Such higher-dimensional grand unified models have been
widely 
studied~\cite{Kawamura:2000ir}-\cite{Kobakhidze:2006zq}.

Employing boundary conditions provides
various interesting application.
If in constructing 4D grand unified models
one requires that the unified gauge group is a simple group
or a direct product of simple groups,
that it contains the standard model gauge group
as a subgroup,
that its rank is four and
that it has complex representation,
the candidates of gauge group could be  not only SU(5) but also
SU(3)$_\textrm{\scriptsize C}$$\times$SU(3)$_\textrm{\scriptsize W}$.
An unfavorable reason of adopting  
SU(3)$_\textrm{\scriptsize C}$$\times$SU(3)$_\textrm{\scriptsize W}$ 
would be that 
the inclusion of matter is not minimal.
If quarks are 
transformed  as $(3,3)$ under 
SU(3)$_\textrm{\scriptsize C}$$\times$SU(3)$_\textrm{\scriptsize W}$,
the electric charge matrix can be assigned as
$Q=\textrm{diag}({2\over 3},-{1\over 3},-{1\over 3})$.
Then integer electric charges 
are also made from $Q$ because
the adjoint representation includes components with the integer charge 
${2\over 3}-(-{1\over 3})=1$
and  the third symmetric representation also includes components with
the integer charge ${2\over 3}+{2\over 3}-{1\over 3}=1$.  
If leptons are assigned in these representations,
extra fields in addition to standard model fields would be involved.
Such additional fields should be decoupled at low energies
as they are heavy. 
In the case where 
the theory is based on higher-dimensionsional gauge invariance,
the mass splitting of  extra fields and matter fields
can be obtained as a result of boundary conditions
similar to the mass splitting for the Higgs field.
This type of decoupling is also used for avoiding
another disputable feature in the 4D minimal SU(5) unified model:
the fermion mass relations.
At a unification scale, down-type quarks and charged leptons 
have the identical Yukawa coupling
(matrix in flavor space, while up-type Yukawa matrix is
symmetric). Their mass eigenvalues are equal.
For one-loop mass correction arising from
fermion self-energy with helicity flip,
the ratio of the down-type quark masses to 
the charged lepton masses is described in powers of gauge 
coupling constants.
For the third generation,
it provides a successful prediction and
for the first two generations,
the prediction seems unfavorable.
In higher-dimensional gauge theory,
down-type quarks and charged leptons
for the first two generations 
may be taken to arise from distinct origin of  multiplets
as extra components are decoupled via boundary conditions.
Then the unfavorable fermion mass relation disappears.

Supersymmetry breaking 
transmitted via extra dimensions
can be a solution to no experimentally incompatible
flavor changing neutral current.
In gaugino mediation~\cite{Kaplan:1999ac}\cite{Chacko:1999mi},
supersymmetry is broken in a sector  
spatially separated from supersymmetric standard model sector
and gauginos acquire  masses at high energy.
For squarks and sleptons,
the positive masses squared are generated at low energy by
renormalizaiton group flow. 
The regularities required to avoid flavor changing neutral currents
are automatically obtained since the gauge interactions 
do not distinguish generations.
For the renormalization group equations,
a simple possiblity of the intitial condition is that
the gaugino masses have unified values
as an input at high energy.
If this is taken seriously, 
it would be natural that
gaugino-mediated supersymmetry breaking is
incorporated into grand unified models.

If the doublet-triplet splitting by boundary conditions, 
no fermion mass relations 
for the first two generations 
and gaugino-mediated supersymmetry breaking are taken into account,
the simplest setup would be to consider two extra dimensions.
Although the doublet-triplet splitting and no fermion mass relations 
can be
simultaneously treated for one extra dimension,
the source of supersymmetry breaking 
in gaugino mediation should not  
be directly coupled to the matter superfields propagating 
in the extra dimension.
For such unified models, 
to contain the weak-doublet Higgs fields in 
an adjoint representation
(a possiblity of gauge-Higgs unification) and 
to introduce right-handed neutrino motivate
that the original higher dimensional gauge group
is larger than the standard model gauge group.

We consider 6D SU(6) supersymmetric models on an orbifold, where 
the sizes of two extra dimensions are different.
To solve the doublet-triplet splitting, Higgs fields
propagate in 5D or 6D. 
We choose the smaller compactification radius
as the unification scale where the unified gauge group is broken by
boundary conditions.
The other compactification radius is limited 
so that field-theoretical description is valid.
Our first model has the weak-doublet Higgs fields 
in 5D chiral superfields,
which is a version with gaugino-mediated supersymmetry
breaking of Ref.\cite{Hall:2002qw}
where the doublet-triplet splitting,
no proton decay from operators of 
dimension four or five,
no mass relations for the first two generations and
gauge coupling unification were achieved.
Our other model has the Higgs doublets in the 6D gauge multipet,
which is a 6D version with gaugino mediated supersymmetry 
breaking of Ref.\cite{Burdman:2002se}
where Yukawa couplings were given without conflicting 
with higher-dimensional gauge invariance 
and the sizes of the Yukawa couplings arised from
wave-function profiles of the matter zero modes determined
by bulk mass parameters.
In both models, we examine 
gauge coupling constants beyond 
the energy scale of the smaller compactification
radius which is the unification scale.
We find very nontrivial high energy behavior of
gauge coupling constants.
Within the region where the larger compactification radius has
such a size that the gauge coupling constants are not blow-up nor zero, 
we find realisic patterns of Yukawa coupling constants.
Proton stability is achieved by R invariance.
Although Higgs fields propagate in 5D or 6D,
gaugino-mediated supersymmetry breaking is 
quite similar to the minimal supersymmetric standard model
(MSSM) case with only gauge superfields propagating in bulk.

The paper is organized as follows.
In Section 2 we present an explicit 6D component action
starting with the action already known in the 4D superfield formalism.
General properties of orbifold parity, mode expansion,
couplings on fixed lines and fixed points,
localization and dceoupling of fields on fixed lines are 
shown also with explicit equations.
In Section 3 our approach to examine 
high energy behavior of gauge coupling constants 
is given.
Gauge coupling corrections, Yukawa couplings,
proton stability and supersymmetry breaking
are examined for the model with Higgs fields in 5D multiplets
in Section 4
and for the model with Higgs fields in the 6D gauge multiplet
in Section 5.
In Section 6 conclusion is given.

\section{6D supersymmetric theory on an orbifold}

We work with 6D $N=2$ theory which corresponds to $N=4$ in the language
of 4D. The theory is vector-like and it is manifest
to be free from 6D anomaly. 
 
\subsection{Superfields and 6D action}

We consider a vector multiplet in 6D bulk.
The field contents are 
the vector superfield $V$
and the three chiral superfields $\Sigma_5$, $\Sigma_6$ and $\Phi$.
These superfields are written in the Wess-Zumino gauge for $V$
in terms of 4D superfields as
\begin{eqnarray}
 V(x,x^i)&\!\!=\!\!&-\theta \sigma^m \bar{\theta} A_m(x,x^i)
   +i\theta\theta\bar{\theta} \bar{\lambda} (x,x^i)
   -i\bar{\theta}\bar{\theta}\theta \lambda (x,x^i)
   +\textrm{${1\over 2}$}\theta\theta\bar{\theta}\bar{\theta} D(x,x^i) ,
 \label{vx}
\\
 \Sigma_5(y,x^i)/\sqrt{2} &\!\!=\!\!&
   \sigma_5(y,x^i)+\sqrt{2}\theta \lambda_5(y,x^i)
   +\theta\theta F_5(y,x^i) ,
  \label{sig5y}
\\
 \Sigma_6(y,x^i)/\sqrt{2} &\!\!=\!\!&
  \sigma_6(y,x^i)+\sqrt{2}\theta \lambda_6(y,x^i)
   +\theta\theta F_6(y,x^i) .
\\
 \Phi(y,x^i)/\sqrt{2} &\!\!=\!\!&
  \phi(y,x^i)+\sqrt{2}\theta \chi(y,x^i)
   +\theta\theta F(y,x^i) ,
  \label{phiy}
\end{eqnarray}
where $i=5,6$. 
The two-component Weyl fermions $\lambda,\lambda_5,\lambda_6,\chi$
are left-handed. 
For the three left-handed chiral superfields,
the 4D coordinates are indicated
in the $y$-basis 
$y=x+i\theta \sigma\bar{\theta}$.
The coordinates $x^5$ and $x^6$ 
are labeled as a kind of internal spin
in 4D.
In the left-hand sides of 
the component expressions for $\Sigma_5$, $\Sigma_6$ and $\Phi$,
(\ref{sig5y})-(\ref{phiy}),
the factor $(1/\sqrt{2})$ is included
so that kinetic terms such as 
$-(\partial^n\phi)^*(\partial_n\phi)$ will be canonically normalized.
For the lowest components of
$\Sigma_5$ and $\Sigma_6$,
the complex scalar fields $\sigma_5$ and $\sigma_6$ are written as
\begin{eqnarray}
  \sigma_5(y,x^i) = \pi_1(y,x^i)+i\textrm{${1\over \sqrt{2}}$}
 A_5(y,x^i) ,\quad~
  \sigma_6(y,x^i) = \pi_2(y,x^i)+i\textrm{${1\over \sqrt{2}}$}
  A_6(y,x^i) .
\end{eqnarray}
The 4D real scalars $A_5$ and $A_6$
are identified with the fifth- and 
sixth-components of the 6D gauge field,
respectively.
The superfields $V$, $\Sigma_5$, $\Sigma_6$ and $\Phi$
have dimensionality 
$\left[V\right]=\left[\textrm{mass}\right]^0$
and
$\left[\Sigma_5\right]=\left[\Sigma_6\right]=
\left[\Phi\right]=\left[\textrm{mass}\right]$
which are the same as in the 4D case.
The difference of dimensionality from the 4D case is included
as the 6D gauge coupling has dimensionality 
$\left[g\right]=\left[\textrm{mass}\right]^{-1}$.
In the Wess-Zumino gauge with these superfields, 
the 6D supersymmetric action is written 
as~\cite{ArkaniHamed:2001tb}\cite{Jiang:2002at}
\begin{eqnarray}
 S_6&\!\!\!=\!\!\!&\int d^6 x \bigg\lbrace 
  2\textrm{Re}\left[
   \textrm{Tr}\left({1\over 16kg^2}W^\alpha W_\alpha
   +{1\over 2kg^2} (\Phi \partial_5\Sigma_6
    -\Phi\partial_6\Sigma_5
    +\Phi\left[\Sigma_5,\Sigma_6\right])\right)\right]_F
\nonumber
\\
  &\!\!\!+\!\!\!&{1\over 8kg^2}
  \textrm{Tr}\bigg[\sum_{i=5}^6
   \left(2(-\partial_i+\Sigma_i^\dag)e^{2V}
    (\partial_i+\Sigma_i)e^{-2V}
    +(\partial_i e^{2V})( \partial_i e^{-2V})\right)
    +2\Phi^\dag e^{2V} \Phi e^{-2V}\bigg]_D\bigg\rbrace ,
\nonumber
\\
  \label{actions6}
\end{eqnarray}
where extracting $F$-term is represented 
as the square brackets with the subscript $F$ such 
as $\left[~\right]_F$ and
extracting $D$-term is represented as $\left[~\right]_D$
with the notation of a vector superfield 
$V=C+{1\over 2}\theta\theta\bar{\theta}\bar{\theta}(D+{1\over 2}\Box C)
+\cdots$.
The action is invariant under the transformation with $\Lambda(x,x^i)$ 
\begin{eqnarray}
 && e^{2V}\to 
    e^{\Lambda^\dag}e^{2V}e^{\Lambda} ,~~
 W_\alpha \to e^{\Lambda^\dag}W_\alpha e^{\Lambda} ,
\\
 && 
\Sigma_5 \to e^{-\Lambda}(\partial_5+\Sigma_5) e^{\Lambda} ,~~ 
\Sigma_6 \to e^{-\Lambda}(\partial_6+\Sigma_6) e^{\Lambda} ,~~
\Phi \to e^{-\Lambda}\Phi e^{\Lambda} .
 \label{sig5}
\end{eqnarray}
In the action (\ref{actions6}),
the numerical coefficient in front of $W^\alpha W_\alpha$
is determined by the normalization of the gauge field kinetic term
$-1/(4kg^2)\cdot\textrm{Tr}\,F_{nm}F^{nm}$.
The numerical coefficient of the K\"ahler term including $\Sigma_5$
is determined by the normalization of
$-1/(2kg^2)\cdot\textrm{Tr}\,\partial_5 A_n \partial_5 A^n$.
The K\"ahler term with $\Sigma_5$ also includes 
$-(1/2kg^2)\cdot\textrm{Tr}\,\partial^n A_5 \partial_n A_5$.
The K\"ahler term with $\Phi$ includes 
$1/(kg^2)\cdot\textrm{Tr}\,F^*F$.
The numerical coefficient of the superpotential is 
determined by the normalization of 
$-{1/(2kg^2)}\cdot\textrm{Tr}\,\partial_5 A_6\partial_5 A_6$
after the auxiliary field $F$ is eliminated.

Since the theory contains 4D $N=2$ supersymmetry ($\subset$ 4D $N=4$),
two of the chiral superfields may be paired
via complex conjugate in which 
representation matrices of gauge group can be
$T^a$ and $-T^a {}^T$.
In the above notation, this property is implicit
and all the representation matrices of gauge group 
are taken as $T^a$.
The superfields are written as
$V=V^a T^a$, $\Sigma_5=\Sigma_5^a T^a$,
$\Sigma_6=\Sigma_6^a T^a$ and $\Phi=\Phi^a T^a$,
where $T^a={1\over 2}\lambda^a$ and $\textrm{Tr}\,T^a T^b=k\delta^{ab}$.
The generators $\lambda^a$ with $k=2$ and the structure constants 
are shown in Appendix~\ref{lam}.

From the action (\ref{actions6}),
the component action is written as
\begin{eqnarray}
 S_6&\!\!\!=\!\!\!&
 \int d^6x {1\over kg^2}
 \textrm{Tr}\bigg[
    -i\lambda \sigma^n{\cal D}_n\bar{\lambda}
    +{1\over 2}D^2-{1\over 4}F_{nm}F^{nm}
\nonumber
\\
&\!\!\!+\!\!\!& \bigg(
   \phi(\nabla_5 F_6-\nabla_6 F_5)
    -\chi(\nabla_5 \lambda_6- \nabla_6 \lambda_5)
\nonumber
\\
   &&\qquad
   +F(\partial_5\sigma_6-\partial_6\sigma_5
    +\sqrt{2}\left[\sigma_5,\sigma_6\right])
    -\sqrt{2}\phi\left[\lambda_5,\lambda_6\right]
   +\textrm{H.c.}\bigg)
\nonumber
\\
  &\!\!\!+\!\!\!& {1\over 2}
   \sum_{j=5}^6
    \bigg\lbrace -\partial_j A^n \partial_j A_n
    -\sqrt{2}i(\partial_j A^n)({\cal D}_n \sigma_j)
  +\sqrt{2}i({\cal D}^n \sigma_j)^*(\partial_j A_n)
  -2({\cal D}^n \sigma_j)^* ({\cal D}_n \sigma_j)
\nonumber
\\
 &&\qquad
 +\bigg( -2i(\partial_j\lambda)\lambda_j
   -\sqrt{2}(\partial_j D)\sigma_j
  -2\sqrt{2}i\bar{\lambda}_j\bar{\lambda}\sigma_j
   +2\sqrt{2}i(\bar{\lambda}_j\sigma_j\bar{\lambda})
  +\textrm{H.c.}\bigg)
\nonumber
\\
  &&\qquad
   -2i\bar{\lambda}_j \bar{\sigma}^n {\cal D}_n\lambda_j
     +2F_j^* F_j+2\sigma_j^* D\sigma_j
  -2\sigma_j^* \sigma_j D\bigg\rbrace
\nonumber
\\
 &\!\!\!-\!\!\!&
   ({\cal D}^n\phi)^*({\cal D}_n\phi)
    -i\bar{\chi}\bar{\sigma}^n{\cal D}_n\chi
   +F^* F
    +\phi^*D\phi -\phi^*\phi D
\nonumber
\\
  &&\qquad
 +\bigg(
  -\sqrt{2}i\bar{\chi}\bar{\lambda}\phi
  +\sqrt{2}i(\bar{\chi}\phi\bar{\lambda})
   +\textrm{H.c.}
 \bigg)\bigg] .
 \label{componentl6}
\end{eqnarray}
where
$\left[\sigma^*_j,\sigma_j\right]
=if^{abc}\sigma_j^{*b}\sigma_j^c T^a$.
Here the covariant derivatives ${\cal D}_n$
acts on fields as 
\begin{eqnarray}
 {\cal D}_n\sigma_i=\partial_n\sigma_i
    +i\left[A_n,\sigma_i\right] ,
\end{eqnarray}
where ${\cal D}_n\lambda_i$,
${\cal D}_n\phi$ and ${\cal D}_n\chi$
are similar and the derivative $\nabla_5$ are 
given by
\begin{eqnarray}
 \nabla_5 F_6&\!\!\!=\!\!\!&
 \partial_5F_6 +\sqrt{2}\left[\sigma_5,F_6\right] ,\qquad~
 \nabla_6 F_5=\partial_6F_5 +\sqrt{2}\left[\sigma_6,F_5\right] ,
\\
 \nabla_5 \lambda_6
 &\!\!=\!\!&\partial_5\lambda_6 +
 \sqrt{2}\left[\sigma_5,\lambda_6\right] ,\qquad~
 \nabla_6 \lambda_5
 =\partial_6\lambda_5 +
 \sqrt{2}\left[\sigma_6,\lambda_5\right] .
\end{eqnarray}
From the action (\ref{componentl6}),
the equations of motion for auxiliary fields are 
\begin{eqnarray}
 &&
  D^a 
  +\textrm{${\sqrt{2}\over 2}$}\sum_{j=5}^6 
\left(
  \partial_j\sigma_j^a
  +\partial_j\sigma_j^a{}^*
 +\sqrt{2}if^{bac}\sigma_j^{*b}\sigma_j^c\right)
  +if^{bac}\phi^{*b}\phi^c =0 ,
   \label{deq}
\\
 &&
 \partial_5\sigma_6^a -\partial_6\sigma_5^a
  +\sqrt{2}if^{abc}\sigma_5^b\sigma_6^c 
  +F^{*a}=0 ,
\\
 && 
   \partial_6\phi^a
   -\sqrt{2}\phi^b \sigma_6^c if^{abc}
   +F_5^{*a}
   =0 ,
  \label{f5eq}
\\
 && 
   -\partial_5\phi^a
   +\sqrt{2}\phi^b \sigma_5^c if^{abc}
   +F_6^{*a}
   =0 .
  \label{f6eq}
\end{eqnarray}
The equations of motion would change if
couplings of fixed lines and fixed points are included.
The corresponding equations and some formula in 
using superfields are given in Appendix~\ref{ap:sf}. 
In obtaining $\partial_6 \phi^a$
in Eq.(\ref{f5eq}) and $\partial_5 \phi^a$
in Eq.(\ref{f6eq}), partial integrals have been used .
In Eq.(\ref{deq}), 
$\partial_j\sigma_j^a+\partial_j\sigma_j^a{}^*$
leads to derivative terms over extra components
such as $(\partial_5 \pi_1)^2$
whereas it prevents  terms such as
$(\partial_5 A_5)^2$.
This correctly produces
the kinetic term of the 6D gauge field.

After equations of motion for auxiliary fields are 
employed, the action is written as
\begin{eqnarray}
  S_6 = \int d^6 x {1\over kg^2}\textrm{Tr}
   \left({\cal O}_B + {\cal O}_F\right)  .
 \label{actionbf}
\end{eqnarray}
We obtain the bosonic part as
\begin{eqnarray}
 {\cal O}_B
 &\!\!\!=\!\!\!&
   -{1\over 4}F_{NM}F^{NM}
   -({\cal D}^N\phi)^*({\cal D}_N\phi)
   -({\cal D}^N\varphi)^*({\cal D}_N\varphi)
\nonumber
\\
  &&-{1\over 2}\left(\left[\phi^*,\phi\right]\right)^2
  -\left[\varphi,\phi\right]  
   \left[\phi^*,\varphi\right]
  -\left[\varphi^*,\phi\right]
   \left[\phi^*,\varphi^*\right] .
 \label{bosop}
\end{eqnarray}
Here the field strength of 6D gauge field $A_M$ and
the covariant derivative of $\phi$ are given by
\begin{eqnarray}
 F_{NM}&\!\!\!\!=\!\!\!\!&\partial_N A_M
   -\partial_M A_N
   +iA_N A_M-iA_M A_N , 
\\
 {\cal D}_N \phi&\!\!\!\!=\!\!\!\!& 
 \partial_N \phi +iA_N \phi -i \phi A_N .
\end{eqnarray}
with the indices $N=0,1,2,3,5,6$ and 
the complex scalar field $\varphi$ is made out of 
$\pi_1$ and $\pi_2$ as
\begin{eqnarray}
 \varphi(x,x^i) =\pi_1(x,x^i) +i\pi_2(x,x^i) ,
\end{eqnarray}
whose the covariant derivative is given by 
${\cal D}_N \varphi=
 \partial_N \varphi +iA_N \varphi -i \varphi A_N$.
For the fermionic part ${\cal O}_F$, we obtain
\begin{eqnarray}
 {\cal O}_F&\!\!\!=\!\!\!&
  -i\bar{\Psi}_l\Gamma^N {\cal D}_N \Psi^l
   +2\sqrt{2}\bar{\Psi}_1\varphi^*\Psi^2
    +2\sqrt{2}\bar{\Psi}_2\varphi\Psi^1 
\nonumber
\\
 &&+\left(
  -\sqrt{2}i\bar{\Psi}_1\phi(1-i\Gamma^7)\Psi^1
   +\sqrt{2}i\bar{\Psi}_2\phi(1+i\Gamma^7)\Psi^2
   +\textrm{H.c.}\right) .
 \label{ferop}
\end{eqnarray}
Here the 6D  fermions are defined 
in terms of two-component Weyl fermions as
\begin{eqnarray}
 \Psi^1 = {1\over 2}
  \left(\begin{array}{c}
    \lambda_\alpha-\chi_\alpha \\
    \bar{\lambda}^{\dot{\alpha}}
     +\bar{\chi}^{\dot{\alpha}} \\
   -i\lambda_{5\beta}+\lambda_{6\beta} \\
    i\bar{\lambda}_5^{\dot{\beta}}
   +\bar{\lambda}_6^{\dot{\beta}}
  \end{array}
 \right) ,~~~
 \Psi^2 ={1\over 2}
  \left(\begin{array}{c}
   -i\lambda_{5\alpha}-\lambda_{6\alpha} \\
    i\bar{\lambda}_5^{\dot{\alpha}}
    -\bar{\lambda}_6^{\dot{\alpha}} \\
   \lambda_\beta+\chi_\beta \\
   \bar{\lambda}^{\dot{\beta}}
    -\bar{\chi}^{\dot{\beta}} \\
   \end{array}
  \right) , 
\\
  \bar{\Psi}_1
  ={1\over 2}\left(
   \begin{array}{c}
    \lambda^\alpha+\chi^\alpha \\
   \bar{\lambda}_{\dot{\alpha}}
    -\bar{\chi}_{\dot{\alpha}} \\
   i\lambda_5^\beta +\lambda_6^\beta \\
   -i\bar{\lambda}_{5\dot{\beta}}
   +\bar{\lambda}_{6\dot{\beta}} \\
   \end{array}\right)^T ,~~~
\bar{\Psi}_2
  ={1\over 2}\left(
  \begin{array}{c}
   i\lambda_5^\alpha -\lambda_6^\alpha\\
   -i\bar{\lambda}_{5\dot{\alpha}}
    -\bar{\lambda}_{6\dot{\alpha}} \\
    \lambda^\beta -\chi^\beta \\
   \bar{\lambda}_{\dot{\beta}}
     +\bar{\chi}_{\dot{\beta}} \\
  \end{array}\right)^T .
\end{eqnarray}
These 6D fermions satisfy the symplectic Majorana condition
\begin{eqnarray}
  \Psi^l=\epsilon^{lk}C_6\bar{\Psi}_k^T , 
 \label{6dsymp}
\end{eqnarray}
with $l,k=1,2$.
The notation of spinors, gamma matrices and 
charge conjugation matrix is shown in appendix~\ref{spinor}.

\subsection{Orbifold Parity and mode expansion}

We choose radii of two extra dimensions 
as $R_5 > R_6$
and impose the orbifold parity $Z_2$
in each of $x^5$ and $x^6$ directions.
The theory has four fixed lines 
at $x^5=0, \pi R_5$ and $x^6=0, \pi R_6$
and four fixed points at $(x^5,x^6)=(0,0)$, $(\pi R_5,0)$,
$(0,\pi R_6)$, $(\pi R_5, \pi R_6)$.
The boundary conditions of fields are 
specified by the orbifold parity.

The boundary condition for the vector multiplet are given 
as follows:
For the direction of $x^5$,
\begin{eqnarray}
   V   (x,-x^5,x^6)
   &\!\!\!=\!\!\!&P_1
  V   (x,x^5,x^6)P_1^\dag , 
 ~
   V   (x,\pi R_5-x^5,x^6)
   =P_2
  V   (x,\pi R_5+x^5,x^6)P_2^\dag ,
\nonumber
\\
   \Sigma_5   (x,-x^5,x^6)
   &\!\!\!=\!\!\!& -P_1
  \Sigma_5   (x,x^5,x^6)P_1^\dag ,
 ~
   \Sigma_5   (x,\pi R_5-x^5,x^6)
   =-P_2
  \Sigma_5   (x,\pi R_5+x^5,x^6)P_2^\dag ,
\nonumber
\\
    \Sigma_6  (x,-x^5,x^6)
   &\!\!\!=\!\!\!&P_1
   \Sigma_6   (x,x^5,x^6)P_1^\dag ,
 ~
   \Sigma_6   (x,\pi R_5-x^5,x^6)
   =P_2
  \Sigma_6   (x,\pi R_5+x_5,x_6)P_2^\dag ,
\nonumber
\\
   \Phi   (x,-x^5,x^6)
   &\!\!\!=\!\!\!&-P_1
  \Phi   (x,x^5,x^6)P_1^\dag ,
 ~
   \Phi   (x,\pi R_5-x^5,x^6)
   =-P_2
  \Phi   (x,\pi R_5+x^5,x^6)P_2^\dag .
\nonumber
\\
 \label{parityg5}
\end{eqnarray}
For the direction of $x^6$,
\begin{eqnarray}
   V   (x,x^5,-x^6)
   &\!\!\!=\!\!\!&P_3
  V   (x,x^5,x^6)P_3^\dag , 
 ~
   V   (x,x^5,\pi R_6-x^6)
   =P_4
  V   (x,x^5,\pi R_6+x^6)P_4^\dag ,
\nonumber
\\
   \Sigma_5   (x,x^5,-x^6)
   &\!\!\!=\!\!\!& P_3
  \Sigma_5   (x,x^5,x^6)P_3^\dag ,
 ~
   \Sigma_5   (x,x^5,\pi R_6-x^6)
   =P_4
  \Sigma_5   (x,x^5,\pi R_6+x^6)P_4^\dag ,
\nonumber
\\
    \Sigma_6  (x,x^5,-x^6)
   &\!\!\!=\!\!\!&-P_3
   \Sigma_6   (x,x^5,x^6)P_3^\dag ,
 ~
   \Sigma_6   (x,x^5,\pi R_6-x^6)
   =-P_4
  \Sigma_6   (x,x^5,\pi R_6+x^6)P_4^\dag ,
\nonumber
\\
   \Phi   (x,x^5,-x^6)
   &\!\!\!=\!\!\!&-P_3
  \Phi   (x,x^5,x^6)P_3^\dag ,
 ~
   \Phi   (x,x^5,\pi R_6-x^6)
   =-P_4
  \Phi   (x,x^5,\pi R_6+x^6)P_4^\dag .
\nonumber
\\
 \label{parityg6}
\end{eqnarray}
These have the periodicity
\begin{eqnarray}
   G   (x,x^5+2\pi R_5,x^6)
   &\!\!\!=\!\!\!&P_2 P_1
  G   (x,x^5,x^6)P_1^\dag P_2^\dag , 
\nonumber
\\
   G   (x,x^5,x^6+2\pi R_6)
   &\!\!\!=\!\!\!&P_4 P_3
  G   (x,x^5,x^6)P_3^\dag P_4^\dag ,
\nonumber
\end{eqnarray}
where $G=V,\Sigma_5, \Sigma_6,\Phi$.
The existence of fixed lines and fixed points
explicitly breaks gauge invariance in 6D.
At these lines and points,
at most the restricted gauge invariance is preserved \cite{Hall:2003yc}.
The gauge transformation parameters obey
the same boundary conditions as
the corresponding gauge fields.
The consistency condition of these 
restricted gauge invariance gives a constraint to
the selection of boundary conditions.
For example, as seen from properties of the structure constants 
shown in Appendix~\ref{lam},
breaking of gauge group such as 
SU(6)$\to$SO(6) or independently SO(6)$\to$SU(3)$\times$U(1) 
can be achieved by boundary conditions.
On the other hand, breaking of gauge group 
such as SU(6)$\to$SO(6)$\to$SU(3)$\times$U(1)
(for example, SU(6)$\to$SO(6) in the direction of $x^6$ 
and SO(6)$\to$SU(3)$\times$U(1) in the direction of $x^5$)
by Neumann and Dirichlet boundary conditions
is seen to be incompatible with restricted gauge invariance.
It was explicitly shown that an example of  boundary conditions 
to break restricted gauge symmetry
gives rise to breakdown of Ward-Takahashi identity
and unitarity violation~\cite{Sakai:2006qi}.
Imposing boundary conditions with the orbifold parity
ensures the restricted gauge invariance.

\begin{figure}[htbp]
 \begin{center}
 \includegraphics[width=10.5cm]{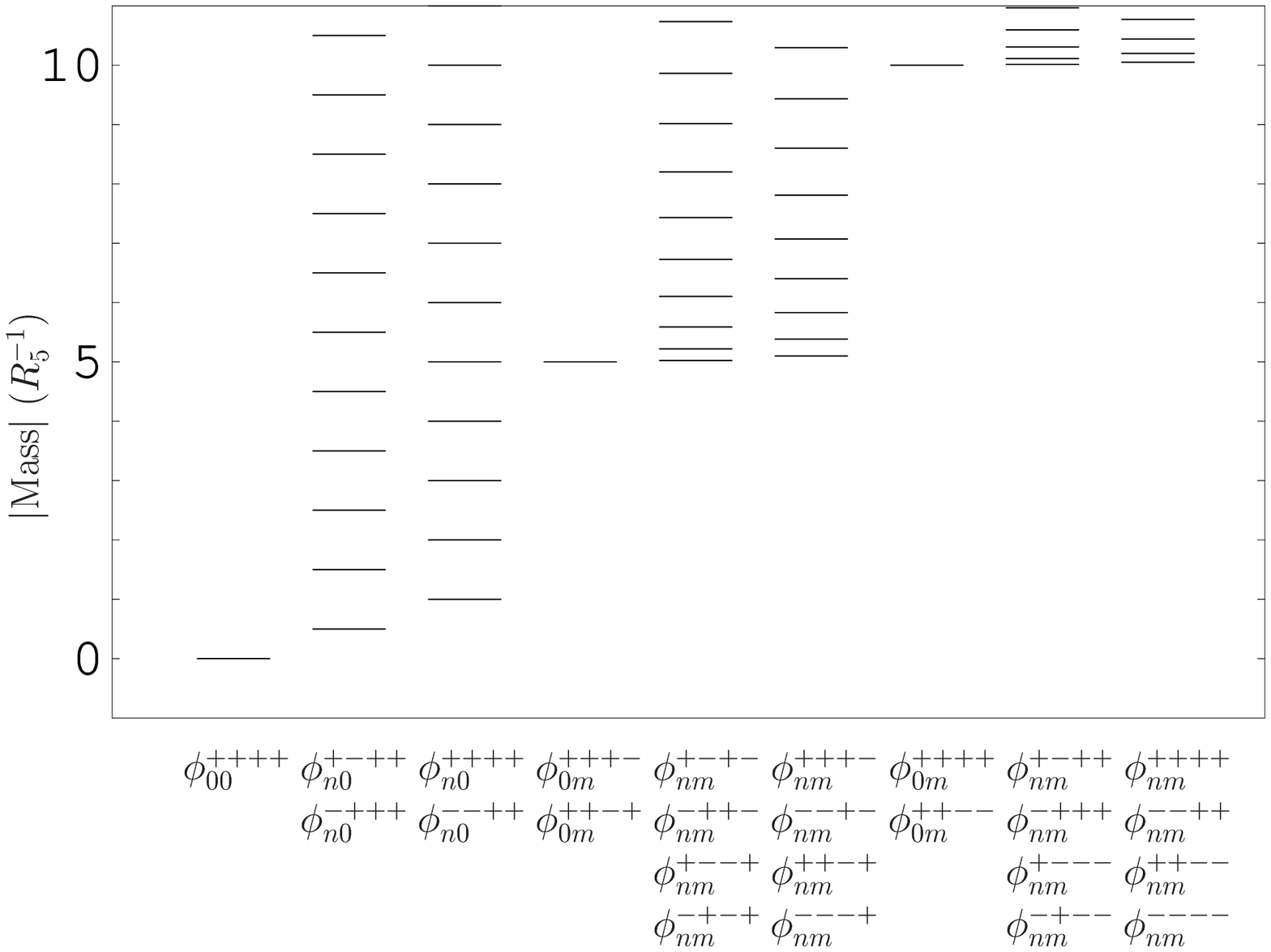} 
\caption{Absolute values of masses of 4D modes
in the unit of $R_5^{-1}$, where
$R_6=R_5/10$ is chosen.
 \label{figmass}}
\end{center}  
\end{figure}

We introduce mode functions in which the property of 
orbifold parity is manifest.
For $(++++)$ where four signs indicate 
the orbifold parity at $x^5=0,\pi R_5$, $x^6=0,\pi R_6$ in order,
a generic 6D massless field $\phi_B(x,x^5,x^6)$ is expanded 
in terms of 4D mode as 
\begin{eqnarray}
 &&\phi_B(x,x^5,x^6)
 =\phi_{00}^{++++}(x)
  +\sum_{n=1}^\infty
   \sqrt{2}\phi_{n0}^{++++}(x)
   \cos\left({nx^5\over R_5}\right)
\nonumber
\\
 &&\qquad
 +\sum_{n=1}^\infty
   \sqrt{2}
   \phi_{0n}^{++++}(x)\cos\left(nx^6\over R_6\right)
  +\sum_{n=1}^\infty \sum_{m=1}^\infty
     2\phi_{nm}^{++++}(x)
   \cos\left({nx^5\over R_5}\right)
   \cos\left({mx^6\over R_6}\right) .
\nonumber
\\
 && \qquad\qquad\qquad\qquad\qquad\qquad \qquad\qquad\qquad
  \label{modepppp}
\end{eqnarray}
The mode expansions for all possible parities are shown in 
Appendix~\ref{xpan}.
As mentioned above,
$\phi$ has dimensionality 
$\left[\phi\right]=\left[\textrm{mass}\right]$.
Each 4D mode has the mass squared tabulated in Table \ref{tabmass}.
\begin{table}[h]
\begin{center}
\caption{4D modes and masses squared.\label{tabmass}}
\begin{tabular}{llllll}
  \hline\hline
  $\phi_{00}^{++++}$ &  0 &  
~~ $\phi_{n0}^{+-++}$, $\phi_{n0}^{-+++}$ & 
$(n-1/2)^2 /R_5^2$ & ~~ 
  $\phi_{n0}^{++++}$, $\phi_{n0}^{--++}$ & $n^2 /R_5^2$  \\
 && ~~
 $\phi_{0m}^{+++-}$, $\phi_{0m}^{++-+}$ & $(m-1/2)^2 /R_6^2$ & ~~ 
 $\phi_{0m}^{++++}$, $\phi_{0m}^{++--}$ & $m^2 /R_6^2$ \\ \hline 
 \multicolumn{3}{l}{
  $\phi_{nm}^{+-+-}$, $\phi_{nm}^{-++-}$,    
 $\phi_{nm}^{+--+}$, $\phi_{nm}^{-+-+}$} & 
 \multicolumn{3}{l}{$(n-1/2)^2 /R_5^2 
 +(m-1/2)^2 /R_6^2$} \\
 \multicolumn{3}{l}{ 
  $\phi_{nm}^{+++-}$, $\phi_{nm}^{--+-}$, 
  $\phi_{nm}^{++-+}$, $\phi_{nm}^{---+}$} &
 \multicolumn{3}{l}{$n^2 /R_5^2 
 +(m-1/2)^2 /R_6^2$}   \\
 \multicolumn{3}{l}{
 $\phi_{nm}^{+-++}$, $\phi_{nm}^{-+++}$, 
  $\phi_{nm}^{+---}$, $\phi_{nm}^{-+--}$} & 
\multicolumn{3}{l}{$(n-1/2)^2 /R_5^2
 +m^2 /R_6^2$}  \\
 \multicolumn{3}{l}{
  $\phi_{nm}^{++++}$, $\phi_{nm}^{--++}$, 
  $\phi_{nm}^{++--}$, $\phi_{nm}^{----}$} & 
 \multicolumn{3}{l}{$n^2 /R_5^2
 +m^2 /R_6^2$} \\ \hline\hline
\end{tabular}
\end{center}
\end{table}
Spectrum for small Kaluza-Klein (KK) 
numbers is shown in Fig.~\ref{figmass}.
The first KK mode appears at the energy scale $1/(2R_5)$.
These fields have the parities $(+-++)$ and $(-+++)$.
When the energy scale is beyond $1/R_5$, 
the fields with the parity $(--++)$ as well as $(++++)$ 
become dynamical.
Below the energy scale $1/(2R_6)$, only the fields 
with the even parity with respect to the $x^6$ direction appear.
These boundary conditions are $(++++)$, $(+-++)$, $(-+++)$ 
and $(--++)$.

On fixed lines, 
unbroken gauge group corresponds to the gauge transformation
parameters which are nonzero under boundary conditions.
Confined fields with representation 
of unbroken gauge group may be placed on the fixed lines.
Let $(\phi (x,x^5))^{i_1\cdots i_t}_{j_1\cdots j_s}$ 
be a field with $t$ fundamental representation indices  
and $s$ fundamental
conjugate representation indices with respect to SU($N$)
on the fixed line $x^6=0$ or $x^6=\pi R_6$.
Throughout this paper, 
we work with $N=1$ supersymmetry 
for 5D theory on fixed lines,
which corresponds to $N=2$ in the 4D language.
For $\phi$, the field $\phi^c$ is paired as its complex conjugate. 
Their boundary conditions are given by
\begin{eqnarray}
 (\phi(x,-x^5))^{i_1\cdots i_t}_{j_1\cdots j_s}
 &\!\!\!=\!\!\!&
 \eta_1(\hat{P}_1)^{i_1}_{k_1}
 \cdots 
  (\hat{P}_1)^{i_t}_{k_t}
  (\phi(x,x^5))^{k_1\cdots k_t}_{l_1\cdots l_s}
  (\hat{P}_1^\dag)^{l_1}_{j_1}
 \cdots 
  (\hat{P}_1^\dag)^{l_t}_{j_t}  ,
\nonumber
\\
 (\phi^c(x,-x^5))_{i_1\cdots i_t}^{j_1\cdots j_s}
 &\!\!\!=\!\!\!&
 -\eta_1(\hat{P}_1)^{j_1}_{l_1}
 \cdots 
  (\hat{P}_1)^{j_s}_{l_s}
  (\phi^c(x,x^5))_{k_1\cdots k_t}^{l_1\cdots l_s}
  (\hat{P}_1^\dag)^{k_1}_{i_1}
 \cdots 
  (\hat{P}_1^\dag)^{k_t}_{i_t} .
\nonumber
\\
 (\phi(x,\pi R_5-x^5))^{i_1\cdots i_t}_{j_1\cdots j_s} ,
 &\!\!\!=\!\!\!&
 \eta_2(\hat{P}_2)^{i_1}_{k_1}
 \cdots 
  (\hat{P}_2)^{i_t}_{k_t}
  (\phi(x,\pi R_5+x^5))^{k_1\cdots k_t}_{l_1\cdots l_s}
  (\hat{P}_2^\dag)^{l_1}_{j_1}
 \cdots 
  (\hat{P}_2^\dag)^{l_t}_{j_t}  ,
\nonumber
\\
 (\phi^c(x,\pi R_5-x^5))_{i_1\cdots i_t}^{j_1\cdots j_s}
 &\!\!\!=\!\!\!&
 -\eta_2(\hat{P}_2)^{j_1}_{l_1}
  \cdots 
  (\hat{P}_2)^{j_s}_{l_s}
  (\phi^c(x,\pi R_5+x^5))_{k_1\cdots k_t}^{l_1\cdots l_s}
  (\hat{P}_2^\dag)^{k_1}_{i_1}
 \cdots 
  (\hat{P}_2^\dag)^{k_t}_{i_t} ,
\nonumber
\\
 \label{tensorp}
\end{eqnarray}
where parity matrices with a hat over a symbol such as $\hat{P}_1$ 
indicate that they are projected parity matrices 
corresponding to the reduction of 6D unified gauge group to SU($N$).
Here the parity of sign is $\eta_1=\pm 1$ and
$\eta_2=\pm 1$. 
The boundary conditions for confined fields $\phi(x,x^6)$ on fixed 
lines $x^5=0$ or $x^5=\pi R_5$
are defined in a parallel way.
As a concrete example, for the parity matrices
\begin{eqnarray}
 \hat{P}_1=\left(\begin{array}{cc}
   {\bf 1}_5 & \\
    & -1 \\
  \end{array}\right),\qquad~
 \hat{P}_2=\left(\begin{array}{cc}
   {\bf 1}_2 & \\
    & -{\bf 1}_4 \\
  \end{array}\right) ,
 \label{p5124}
\end{eqnarray}
the boundary conditions of several small representations in SU(6),
the fundamental representation $\phi^i$, 
the two-rank antisymmetric representation $\phi^{ij}$ and
the third-rank antisymmetric representation $\phi^{ijk}$, are
tabulated in Table~\ref{use}.
All components can have zero modes.
The origin of zero modes depends on 
$\hat{P}_1$, $\hat{P}_2$, $\eta_1$ and $\eta_2$.
\begin{table}[h]
\begin{center}
\caption{Boundary conditions of {\bf 6}, {\bf 15} and {\bf 20} for
the parity matrices (\ref{p5124}).
Here two signs in parentheses indicate 
the parities at $x^5=0$ and $x^5=\pi R_5$
in order.\label{use}}
\begin{tabular}{ll|cccc} \hline\hline
Field  & Component  & \multicolumn{4}{l}{
Boundary condition}
\\ \hline
{\bf 6} $\phi^i$ &1,2 &  $(++)$ & $(-+)$ & $(+-)$ & $(--)$  \\ 
&3,4,5 &$(+-)$ & $(--)$ & $(++)$ & $(-+)$ \\
&6 & $(--)$ & $(+-)$ & $(-+)$ & $(++)$  \\ \hline
{\bf 15} $\phi^{ij}$
&12,34,35,45 &  $(++)$ & $(-+)$ & $(+-)$ & $(--)$  \\ 
&13,14,15,23,24,25 &$(+-)$ & $(--)$ & $(++)$ & $(-+)$ \\
&16,26 & $(--)$ & $(+-)$ & $(-+)$ & $(++)$  \\ 
&36,46,56 & $(-+)$ & $(++)$ & $(--)$ & $(+-)$  \\ \hline
{\bf 20} $\phi^{ijk}$
&123,124,125,345 &  $(+-)$ & $(--)$ & $(++)$ & $(-+)$  \\ 
&126,346,356,456 &$(--)$ & $(+-)$ & $(-+)$ & $(++)$ \\
&134,135,145,234,235,245 & $(++)$ & $(-+)$ & $(+-)$ & $(--)$  \\ 
&136,146,156,236,246,256 & $(-+)$ & $(++)$ & $(--)$ & $(+-)$  \\ \hline
\multicolumn{2}{c|}{$\eta_1,\eta_2$} & $1,1$ & $-1,1$ & $1,-1$ & $-1,-1$  \\ \hline\hline
\end{tabular}
\end{center}
\end{table}
For example, the two components $1,2$ of {\bf 6}
has zero mode for $\eta_1=\eta_2=1$,
while the three components $3,4,5$ of {\bf 6}
has zero mode for $\eta_1=-\eta_2=1$.
Properties of the boundary conditions of 
the representations given in Table~\ref{use}
will be used in explicit models
analyzed later.

\subsection{Couplings on fixed lines and points}

Fields with the positive parity 
at fixed lines and fixed points of orbifold
have nonzero values at the lines and points.
These fields are coupled to fields confined on the fixed line and point.
For  a left-handed chiral superfield of 
fundamental representation denoted as $\Phi_l(x,x^5)$
and its complex conjugate $\Phi_l^c(x,x^5)$ confined
on the fixed line $x^6=0$,
the 6D fields have the 
interaction~\cite{ArkaniHamed:2001tb}\cite{Hebecker:2001ke}
\begin{eqnarray}
 S_5 =\int d^6 x ~
  \delta(x^6)
\left({1\over 2} 
  \left[\Phi_l^\dag e^{2V}\Phi_l
   +\Phi_l^c e^{-2V}\Phi_l^c{}^\dag\right]_D
 +2\textrm{Re}\left[
   \left(\Phi^c_l (\partial_5+\Sigma_5)\Phi_l\right)
  \right]_F\right) .
 \label{actions5}
\end{eqnarray}
The 5D fields $\Phi_l$ and $\Phi_l^c$ have the $\Lambda$-transformation
\begin{eqnarray}
 \Phi_l\to\Phi_l'= 
  e^{-\Lambda_l}  \Phi_l ,~ \quad
 \Phi_l^c\to \Phi_l^{c'}=\Phi_l^c
 e^{\Lambda_l} ,
\end{eqnarray}
where $\Lambda(x,x^5,x^6=0)=\Lambda_l(x,x^5)$.
The component action of (\ref{actions5}) is given in 
Appendix~\ref{ap:sf}. 
The term $\int d^6x \delta(x^6) 2\textrm{Re}
\left[\Phi^c_l (\partial_5+\Sigma_5)\Phi_l\right]_F$ 
is invariant under $\Lambda$-transformation by itself.
Similar terms appear
also for other representations
following properties of tensors in Eq.(\ref{tensorp}).
The interaction of Eq.(\ref{actions5}) is a gauge interaction.
The gauge coupling constant appears in these equations
when zero mode of 6D field is canonically normalized.

The 5D massless scalar field is expanded in terms of 4D mode as
\begin{eqnarray}
 \phi_l(x,x^5)={1\over \sqrt{\pi R_5}}\left[
   \phi_0^{++}(x)
  +\sum_{n=1}^{\infty}\sqrt{2}\phi_n^{++}(x)\cos\left(
  {nx^5\over R_5}\right)\right] ,
 \label{5dmasses}
\end{eqnarray}
for the boundary condition $(++)$
in the orbifold property (\ref{tensorp}).
The dimensionality of the field $\phi_l$ is defined 
as 
$\left[\phi_l\right]=\left[\textrm{mass}\right]^{3/2}$
which is different from the dimensionality of 
6D fields and 4D fields confined on 
fixed points.
In analogy to Table~\ref{tabmass},
5D confined fields have the masses 
0, $(n-{1\over 2})/R_5$, $n/R_5$
depending on boundary conditions.

We next consider interactions at fixed points.
For a left-handed chiral superfield of 
fundamental representation denoted as $\Phi_p(x)$ 
confined on the fixed point $(x^5,x^6)=(0,0)$, 
the interactions of 6D fields with the fixed point field are
\begin{eqnarray}
 S_4=\int d^6x  ~\delta(x^5)\delta(x^6)
 \bigg({1\over 2}\left[\Phi_p^\dag e^{2V}
   \Phi_p\right]_D + 2\textrm{Re}\left[f(\Phi_p)\right]_F\bigg) ,
 \label{actions4}
\end{eqnarray}
where the superpotential is denoted as $f(\Phi_p)$.
The $\Lambda$-transformation of $\Phi_p$ is given by
\begin{eqnarray}
 \Phi_p\to \Phi_p'=e^{-\Lambda_p} \Phi_p ,
\end{eqnarray}
where $\Lambda(x,x^5=0,x^6=0)=\Lambda_p(x)$.

For the other fixed lines and points,
$x^5=0$, $(x^5,x^6)=(\pi R_5,0)$,
$(0,\pi R_6)$, $(\pi R_5,\pi R_6)$, 
the interactions of the 6D fields 
with fields of various representations 
on the fixed line and point can be obtained
by the suitable replacements of delta functions and the operator 
$(\partial_5+\Sigma_5)$.

\subsection{Localization and decoupling of fields on fixed lines \label{ldff}}

We have written a 5D massless superfield for a field
confined on a fixed line.
A 5D massive superfield can be defined consistently 
with supersymmetry
on the fixed line $x^6=0$ by adding 
to Eq.(\ref{actions5})
\begin{eqnarray}
 \int d^6 x \delta(x^6)\,
    2\textrm{Re}\left[M\epsilon(x^5) 
 \Phi_l^c \Phi_l\right]_F .
\end{eqnarray}
A parallel discussion can be made for 5D massive superfields
on other fixed lines.
Here the sign function is given by
\begin{eqnarray}
   \epsilon(x^5)=\left\lbrace
  \begin{array}{ccl}
      1 &\textrm{for} & 2z \pi R_5 < x^5< (2z+1)\pi R_5 , \\
     -1 & \textrm{for}& (2z-1) \pi R_5 <x^5 < 2z \pi R_5 , \\
      0 & \textrm{for}& x^5=z \pi R_5 .  \\
   \end{array}
  \right.
\end{eqnarray}
where $z$ is an integer.

From the 5D action obtained 
with Eqs.(\ref{s5after}), (\ref{ap:fheq}) and (\ref{ap:fheqc}),
the scalar components of $\Phi_l$ and $\Phi_l^c$
are seen to obey
\begin{eqnarray}
  \left(\partial_5-M\epsilon\right)
 \left(\partial_5+M\epsilon\right)\phi_l
   =-M_n^2\phi_l , \quad~
 \left(\partial_5+M\epsilon\right)
\left(\partial_5-M\epsilon\right)\phi_l^c
=-M_n^2\phi_l^c ,
 \label{kkeq}
\end{eqnarray}
where
$M_n$ indicates the KK mass.
The fermionic components have the same KK mass as $M_n$.
With the mode expansion
\begin{eqnarray}
 \phi_l(x,x^5)=\sum_{n=0}^{\infty} \phi_{n}(x)
f_n(x^5) ,\quad~
 \phi_l^c(x,x^5)=\sum_{n=0}^{\infty} \phi_{n}^c(x) 
 g_n(x^5) ,
\end{eqnarray}
the equation (\ref{kkeq}) is equivalent to
\begin{eqnarray}
 \left(\partial_5 +M\epsilon\right)f_n(x^5)= M_n g_n(x^5) ,\quad ~
  \left(-\partial_5 +M\epsilon\right)g_n(x^5)= M_n f_n(x^5) .
\end{eqnarray}
The function $f_0$ and $g_0$ with $n=0$ are zero mode eigenfunction.
The corresponding 4D fields $\phi_{0}$ and $\phi_{0}^c$ 
have zero mass
$m_0=0$.
For zero mode, the equations of motion for $f_0$ and $g_0$
is the two independent equations
\begin{eqnarray}
  \left(\partial_5 +M\epsilon\right)f_0(x^5)= 0 ,\quad ~
  \left(-\partial_5 +M\epsilon\right)g_0(x^5)= 0 .
\end{eqnarray}
From these equations, it is seen that
the solutions have exponential forms
which would lead to localization.

For the parity $(++)$ for $f_0$,
the mode function is
\begin{eqnarray}
 f_0^{(++)}=
   \sqrt{M \over \sinh(\pi R_5 M)}
   \times 
 \left\lbrace \begin{array}{c}
e^{-M(|x^5|-\pi R_5/2)} , ~~~~~~~     M>0 , \\
e^{M(|x^5-\pi R_5|-\pi R_5/2)} , ~~~~  M<0 , \\
\end{array}
\right.
  \label{f0mp}
\end{eqnarray}
for $0\leq x^5\leq \pi R_5$.
The normalization constant is fixed by
$\int_0^{\pi R_5} dx^5 (f_0^{(++)})^2=1$.
For $M>0$, $\phi_l^{++}$ is localized at $x^5=0$
and for $M<0$, it is localized at $x^5=\pi R_5$.
For the parity $(--)$ for $f_0$, $f_0^{(--)}=0$. 
For the parity $(++)$ for $g_0$,
the mode function is
\begin{eqnarray}
 g_0^{(++)}=
   \sqrt{M\over \sinh(\pi R_5 M)}
  \times
 \left\lbrace
  \begin{array}{c}
   e^{-M(|x^5-\pi R_5|-\pi R_5/2)} , ~~~~     M>0 , \\
   e^{M(|x^5|-\pi R_5/2)} , ~~~~~~~~~~~~  M<0 , \\
 \end{array}
 \right.
  \label{g0mp}
\end{eqnarray}
for $0\leq x^5\leq \pi R_5$. 
The normalization constant is fixed by
$\int_0^{\pi R_5} dx^5 (g_0^{(++)})^2=1$.
For $M>0$, $\phi_l^{c++}$ is localized at $x^5=\pi R_5$
and for $M<0$, it is localized at $x^5=0$.
For the parity $(--)$ for $g_0$, $g_0^{(--)}=0$. 
These mean that
the fields $\phi_l^{++}$ and $\phi_l^{c++}$
are localized at the opposite fixed points.
The product $f_0^{(++)}g_0^{(++)}$ is written as
\begin{eqnarray}
  f_0^{(++)}g_0^{(++)}
 ={M\over \sinh(\pi R_5 M)} ,
 \label{fgfixedp}
\end{eqnarray}
on the fixed line $0\leq x^5\leq \pi R_5$
for arbitrary $M$ (including $M=0$).

The $n$-th eigenfunction $f_n$ 
is related to the function $g_n$ with the opposite
parities.
For the parity $(++)$ for $f_n$ and $(--)$ for $g_n$,
the functions are
\begin{eqnarray}
 f_n^{(++)} =
 \sqrt{2\over \pi R_5}\cos \left({n\over R_5}x^5\right) ,\quad~
  g_n^{(--)} =
 \sqrt{2\over \pi R_5}\sin \left({n\over R_5}x^5+\Omega_n\right) ,
\end{eqnarray}
where $\tan \Omega_n =-MR_5/n$.
The corresponding 4D modes $\phi_{n}^{(++)}$ and 
$\phi_{n}^{c(--)}$
have the mass squared
$M_n^2= M^2+(n/R_5)^2$ .
The functions 
$f_n^{(--)}$ and $g_n^{(++)}$ are
\begin{eqnarray}
 f_n^{(--)} =
 \sqrt{2\over \pi R_5}\sin \left({n\over R_5}x^5-\Omega_n\right) ,\quad~
  g_n^{(++)} =
 \sqrt{2\over \pi R_5}\cos \left({n\over R_5}x^5\right) .
\end{eqnarray} 
The 4D modes $\phi_{n}^{(--)}$ and $\phi_{n}^{c(++)}$
have the mass squared $M_n^2= M^2+(n/R_5)^2$.
The other eigenfunctions are given by
$f_n^{(+-)} = f_{n+1/2}^{(++)}$,
$g_n^{(-+)} = g_{n+1/2}^{(--)}$,
$f_n^{(-+)}=f_{n+1/2}^{(--)}$ and  $g_n^{(+-)}=g_{n+1/2}^{(++)}$.
The 4D modes $\phi_{n}^{(+-)}$, $\phi_{n}^{c(-+)}$,
$\phi_{n}^{(-+)}$ and $\phi_{n}^{c(+-)}$
have the mass squared
$M_n^2= M^2+((n+1/2)/R_5)^2$.
The 4D KK modes of fixed-line fields with 5D masses 
are heavier than the KK modes of
the corresponding 5D massless fields.

\subsubsection*{Fixed-point couplings for heavy fields}

We have seen that zero mode arises from 
fixed-line fields with and without 5D mass terms.
In contrast, 4D heavy modes yield if 
4D interactions are added.

Here for the 5D superfield $\Phi_l^c(x,x^5)$ 
with the scalar component $\phi_l^c$ of dimensionality
$\left[\phi_l^c\right]=\left[\textrm{mass}\right]^{3/2}$
and the 4D superfield $\Phi_p(x)$ with 
the scalar component $\phi_p$ of
dimensionality $\left[\phi_p\right]=\left[\textrm{mass}\right]$,
we consider the fixed-point coupling
\begin{eqnarray}
 \delta(x^5)\delta(x^6)(
   2\textrm{Re}[\sqrt{m_{\textrm{\scriptsize dec}}} \Phi_l^c \Phi_p]_F
 +{1\over 2}[\Phi_p^\dag e^{2V}\Phi_p]_D) ,
 \label{bradec}
\end{eqnarray} 
which is still supersymmetric.
If  $\Phi_l^c$ has a 5D mass,
the mass is taken as positive so that the coupling (\ref{bradec})
is made on the fixed point where $\Phi_l^c$ is localized.
After integration of auxiliary fields,
the supersymmetric mass terms are obtained as
\begin{eqnarray}
  {\cal L}_{\textrm{\scriptsize mass}}
   =\delta(x^5)\delta(x^6)\left(-m_{\textrm{\scriptsize dec}}
  \delta(x^5)\phi_p\phi_p^*
  -m_{\textrm{\scriptsize dec}} \phi_l^c \phi_l^{c*} 
   +(\textrm{fermions})\right) .
\end{eqnarray}
For a large $m_{\textrm{\scriptsize dec}}$,
the 4D mode $\phi_p(x)$, $\phi_l^{c\,++}(x)$ and their superpartners 
become heavy.

\vspace{4ex}

One could introduce Chern-Simons terms on 5D fixed lines.
Such terms are relevant to anomalies in 5D localized on
fixed points~\cite{ArkaniHamed:2001is}.
We will not treat this issue further in the paper.

\section{High energy behavior of gauge couplings \label{myg}}

Employing the action defined above, 
we will consider 6D SU(6) models
where zero mode is composed of
right-handed neutrino and their superpartners
as well as the field content of MSSM.
Before moving on to explicit models,
we give the approach to 
examine high energy behavior of gauge couplings.
The characteristic scales of theory are
the radii $R_5$ and $R_6$ and the cutoff of 6D theory $M_s$.
Since we assume $R_5>R_6$, it is needed to 
deal with effective field theory at energy scales 
from $1/(2R_5)$ to $1/(2R_6)$.
Here $1/(2R_5)$ is the minimum mass of KK mode obtained
from Table~\ref{tabmass} and from 
discussion of 5D massless fields below (\ref{5dmasses})
and of 5D massive fields,
whereas $1/(2R_6)$ is the mass when the $x^6$ direction
appears.
Our approach is based on 4D KK effective
Lagrangian~\cite{Bhattacharyya:2006ym}\cite{Uekusa:2007im}.
Although higher-dimensional theory is non-renormalizable,
it can be assumed that the contributions from the KK
states with masses larger than the scale of interest are 
decoupled from the theory.
In this approximation the corrections to the gauge couplings
can be calculated.
The first KK excitation occurs at
the scale $1/(2R_5)$.
Up to this scale the gauge coupling evolution 
has a contribution only from zero mode.
Between $1/(2R_5)$ and $1/R_5$
(to avoid complexity the discussion is presented
for the masses listed in 
Table~\ref{tabmass} or for the moment we assume
that the absolute values of 5D masses $|M|$ is zero or
larger than $1/(2R_6)$), 
the running is still logarithmic but 
beta functions are modified due to the first KK excitation.
Whenever a KK threshold is crossed, 
beta functions are renewed.
Gauge couplings are described as functions of the energy,
depending on the number of KK states.

We consider gauge field Lagrangian
\begin{eqnarray}
 {\cal L}_{\textrm{\scriptsize gauge}}=-{1\over 4g^2} (F_{MN}^a)^2 ,
 \label{lgauge}
\end{eqnarray}
from the action (\ref{actionbf}).
The gauge field for the parity $(++++)$ is decomposed as in
Eq.(\ref{modepppp}),
\begin{eqnarray}
 &\!\!\!\!\!\!&  A_m^a(x,x^i)=
       A_m^a{}_{(00)}^{++++}(x) 
  +\sum_{n=1} \sqrt{2} A_m^a{}_{(n0)}^{++++}(x) 
   \cos\left({n x^5\over R_5}\right)
\nonumber
\\
 &\!\!\!\!\!\!&~
 +\sum_{n=1} \sqrt{2} A_m^a{}_{(0n)}^{++++}(x) 
   \cos\left({n x^6\over R_6}\right)
 + \sum_{n=1}\sum_{m=1} 
    2 A_m^a{}_{(nm)}^{++++}(x) 
   \cos\left({n x^5\over R_5}\right)
 \cos\left({m x^6\over R_6}\right) .
  \label{gaugepppp}
\end{eqnarray} 
The mode expansions for $(+-++)$, $(--++)$ and $(-+++)$ 
are obtained similarly to
(\ref{modepmpp}), (\ref{modemmpp}) and (\ref{modemppp}).
The 5-component and 6-component of gauge field corresponding to
(\ref{gaugepppp})
have the parities $(--++)$ and $(++--)$, respectively.
They are given by
\begin{eqnarray}
   A_5^a(x,x^i)&\!\!\!=\!\!\!&\sum_{n=1}
   \sqrt{2} A_5^a{}_{(n0)}^{--++}(x) 
  \sin\left({n x^5\over R_5}\right)
 +\sum_{n=1}\sum_{m=1}
   2A_5^a{}_{(nm)}^{--++}(x) 
  \sin\left({n x^5\over R_5}\right)
 \cos\left({m x^6\over R_6}\right) ,
\nonumber
 \\
   A_6^a(x,x^i)&\!\!\!=\!\!\!&\sum_{m=1}
   \sqrt{2} A_6^a{}_{(0m)}^{++--}(x) 
  \sin\left({m x^6\over R_6}\right)
 +\sum_{n=1}\sum_{m=1}
   2A_6^a{}_{(nm)}^{++--}(x) 
  \cos\left({n x^5\over R_5}\right)
 \sin\left({m x^6\over R_6}\right) .
\nonumber
\\
\end{eqnarray}
We keep only the mode $(++)$ with respect to $x^6$,
since we have interest in states with masses less than $1/(2R_6)$.
Substituting the KK mode expansion into the Lagrangian (\ref{lgauge}),
we obtain the 4D gauge part Lagrangian
\begin{eqnarray}
 && \int_0^{\pi R_5} dx^5 \int_0^{\pi R_6}dx^6~ 
 {\cal L}_{\textrm{\scriptsize gauge}}
\nonumber
\\
 &\!\!=\!\!&-{\pi^2 R_5 R_6\over 4g^2} 
   (F_{mn}^a {}_{(0)})^2
\nonumber
\\
&&
 -{\pi^2 R_5 R_6\over 4g^2} 
 \sum_{n=1}\left((F_{mn}^a {}_{(n)})^2
  +2 f^{abc}F_{mn}^a{}_{(0)}  A^{mb}{}_{(n)} 
  A^{nc}{}_{(n)}
  +2F_{m5}^a {}_{(n)}
   F^{m5a} _{(n)}\right) .
 \label{lag4d}
\end{eqnarray}
For notational simplicity,
we have defined
 $A_m^a{}_{(0)}(x)\equiv A_m^a{}_{(00)}^{++++}(x)$,
$A_m^a{}_{(n)}(x)\equiv A_m^a{}_{(n0)}^{++++}(x)$,
$A_5^a{}_{(n)}(x)\equiv A_5^a{}_{(n0)}^{--++}(x)$.
The field strengths are given by
\begin{eqnarray}
  F_{mn}^a{}_{(0)}&\!\!=\!\!&
 \partial_m A_n^a{}_{(0)}- \partial_n A_m^a{}_{(0)}
 +f^{abc}A_m^b{}_{(0)}A_n^c{}_{(0)} ,
\nonumber
\\
  F_{mn}^a{}_{(n)}&\!\!=\!\!&
 \partial_m A_n^a{}_{(n)}- \partial_n A_m^a{}_{(n)}
 +f^{abc}A_m^b{}_{(m)}A_n^c{}_{(r)}\delta_{nmr}^+
 +f^{abc}(A_m^b{}_{(0)}A_n^c{}_{(n)}-A_n^b{}_{(0)}A_m^c{}_{(n)}) ,
\nonumber
\\
  F_{m5}^a {}_{(n)} &\!\!=\!\!&
  \partial_m A_5^a {}_{(n)}
   +(\textrm{${n\over R_5}$})A_m^a {}_{(n)}
   +f^{abc} A_m^b {}_{(m)} A_5^c{}_{(r)}
    \delta_{nmr}^-
   +f^{abc}A_m^b{}_{(0)}A_5^c{}_{(n)} , 
\nonumber
\\
  &&\label{fm5}
\end{eqnarray}
with
 $\delta_{nmr}^{\pm}
 =(\delta_{n,m+r}+\delta_{m,n+r}\pm\delta_{r,n+m})/\sqrt{2}$.
The $n$-th KK gauge field begins to play a dynamical role 
at energy scales higher than $(n/R_5)$. 
We treat the summation over KK modes in the Lagrangian (\ref{lag4d})
as scale-dependent.
This is explicitly written as $\sum_{n=1}^m$ 
at the energy range $m/R_5 \le E< (m+1)/R_5$.
At scales less than $1/(2R_5)$,
the Lagrangian describes  only zero mode
as the summation is simply zero.
The $n$-th KK scalar $A_5^a{}_{(n)}$ has the same mass as that of 
the $n$-th KK gauge field, 
as more explicitly seen after gauge fixing.
The summation over the KK modes is treated
similarly to the KK gauge field case.
 
The 4D gauge couplings are obtained as
\begin{eqnarray}
 {1\over g_{4D}^2}={\pi^2 R_5 R_6\over g^2} + \cdots
={(\pi R_5 M_s)(\pi R_6 M_s)\over \hat{g}^2}
  + \cdots ,
\label{g4d}
\end{eqnarray}
where a dimensionless coupling constant is 
defined as $\hat{g}=M_s g$. 
Here the ellipses denote possible contributions of
fixed-line and fixed-point kinetic terms.
If dimensionless gauge coupling constants with 
the origins in 6D, 5D and 4D are
of the same order,
the first terms in Eq.(\ref{g4d}) would dominate due to 
the volume 
suppression $1>(R_6M_s)^{-1}>(R_5M_s)^{-1}$.
However, in the expression (\ref{g4d}), there is subtlety.
If $M_s=10^{18}$GeV, $1/R_6=4\times 10^{17}$GeV,
$1/R_5=4\times 10^{16}$GeV and
$g_{4D}\sim 1$ are chosen, 
the dimensionless gauge coupling constants
are of order of 
$\hat{g}\sim (100\pi)\times (10\pi)\gg 1$.
It seems unnatural
that $\hat{g}$ is larger than 1
since it is made dimensionless with the 
cutoff $M_s$. 
To obtain $\hat{g}\sim 1$ naturally,
taking into account geometric loop factor may be 
viable~\cite{Chacko:1999mi}.
We here simply ignore the contributions of fixed-line
and fixed-point terms.

The 4D Lagrangian (\ref{lag4d})
is invariant under a standard 4D
gauge transformation.
The infinitesimal gauge transformation law for $A_m^a {}_{(0)}$
is given by  
\begin{eqnarray}
  \delta A_m^a {}_{(0)}
  =\partial_m \alpha^a 
    +f^{abc} A_m^b {}_{(0)} \alpha^c  
\end{eqnarray}
where $\alpha^a(x)$ is an infinitesimal parameter
dependent on 4D coordinates.
The KK gauge field and scalar are transformed as 
adjoint matter fields,
\begin{eqnarray}
 \delta A_m^a {}_{(n)}&\!\!=\!\!&f^{abc}A_m^b{}_{(n)} \alpha^c ,
 \label{eq:trsfm}
\\
 \delta A_5^a {}_{(n)}&\!\!=\!\!&f^{abc}A_5^b {}_{(n)}\alpha^c .
 \label{eq:trsf5}
\end{eqnarray}
In addition to the standard gauge invariance,
the Lagrangian (\ref{lag4d})
is also invariant under another gauge transformation
with the infinitesimal local parameter $\alpha^a_{(n)} (x)$, 
\begin{eqnarray}
  &&\delta_{(n)} A_m^a {}_{(0)}=0 ,
\\
  &&\delta_{(n)} A_m^a {}_{(n)}
  =\partial_m \alpha^a_{(n)} 
    +f^{abc} A_m^b {}_{(m)} \alpha^c_{(r)} \delta_{nmr}^+  
    +f^{abd} A_m^b {}_{(0)} \alpha^c_{(n)} ,
\\
 &&\delta_{(n)} A_5^a {}_{(n)}
  =-({n\over R_5})\alpha_{(n)}^a
   +f^{abc}A_5^b {}_{(m)}\alpha^c_{(r)}\delta_{nmr}^- .
\end{eqnarray}
For these two kinds of gauge invariance,
a version of the generalized Lorenz gauge 
convenient for the background field method
can be chosen~\cite{Uekusa:2007im}.
It is straightforward to
include part of fermion and scalar fields
as well as the other boundary conditions 
$(+-++)$, $(--++)$ and $(-+++)$.

We now give gauge coupling constants
in such a way to track the number of KK modes.
For scales less than the mass of the first KK mode,
the gauge coupling constant 
is given in terms of 
$\alpha^{-1}\equiv 4\pi g^{-2}(\pi^2 R_5 R_6)$
by
\begin{eqnarray}
 \alpha^{-1}(\mu)= 
 \alpha^{-1}(M)+{b\over 2\pi} \log {M\over \mu} ,
 ~\quad \textrm{for}~~ \mu < {1\over 2R_5} ,
 \label{alpha0}
\end{eqnarray}
where $\mu$ and $M$ are the scale of interest and 
another scale. 
In Eq.(\ref{alpha0}),  $b$ is obtained only from zero mode part as
\begin{eqnarray}
  b= -\!\sum_{\textrm{\scriptsize
   gauge}}
 \! {11\over 3}C_2(G)
     +\sum_{\textrm{\scriptsize Weyl}}{2\over 3}C(r)
     +\!\!\!\!\sum_{\textrm{\scriptsize
   \begin{tabular}{c} 
    complex \\ scalar
    \end{tabular}}} \!\!\!\!{1\over 3}C(r) ,
\end{eqnarray}
with the quadratic Casimir operator for the adjoint representation
$C_2(G)=N$ and 
the coefficient in the trace of the product of
two generator matrices $C(r)={1\over 2}$.
In the factor $11/3$ for gauge, 
the contributions of the corresponding ghost
fields have been included. 
As scales cross the mass of the first KK excitation, 
the first KK mode becomes dynamical.
Then $\alpha^{-1}$ is given by
\begin{eqnarray}
  \alpha^{-1}(\mu)=\alpha^{-1}(M)
   +{b\over 2\pi} \log {M\over \mu}
   +{b_{(1)}\over 2\pi}
    \log {1\over 2\mu R_5} ,
 ~\quad \textrm{for}~~
   {1 \over 2R_5} \le \mu < {1\over R_5} ,
  \label{al1}
\end{eqnarray}
where
\begin{eqnarray}
  b_{(1)}= -\!\sum_{\textrm{\scriptsize
 gauge}} \!{11\over 3}C(r)
     +\sum_{\textrm{\scriptsize Dirac}}{4\over 3}C(r)
     +\!\!\!\!\sum_{\textrm{\scriptsize
   \begin{tabular}{c} 
    complex \\ scalar
    \end{tabular}}} \!\!\!\!{1\over 3}C(r) ,
  \label{b1aa}
\end{eqnarray}
When the second KK mode becomes dynamical,
$\alpha^{-1}$ is given by
\begin{eqnarray}
  \alpha^{-1}(\mu)=\alpha^{-1}(M)
   +{b\over 2\pi} \log {M\over \mu}
   +{b_{(1)}\over 2\pi}
    \log {1\over 2\mu R_5} +{b_{(2)}\over 2\pi}
    \log {1\over \mu R_5} ,
  \label{al2}
\end{eqnarray}
for $1/R_5 \le \mu < 3/(2R_5)$,
where
\begin{eqnarray}
  b_{(2)}= -\!\!\sum_{\textrm{\scriptsize
 gauge}}\! {11\over 3}C_2(G)
     +\sum_{\textrm{\scriptsize Dirac}}{4\over 3}C(r)
     +\!\!\!\!\sum_{\textrm{\scriptsize
   \begin{tabular}{c} 
    complex \\ scalar
    \end{tabular}}} \!\!\!\!{1\over 3}C(r) .
 \label{b2aa}
\end{eqnarray}
In the equations (\ref{b1aa}) and (\ref{b2aa}),
the coefficients of gauge part are different from
each other.
The gauge part of $b_{(1)}$ appears from coset and it gives $C(r)$,
whereas the gauge part of $b_{(2)}$ is 
the contribution of group and it gives $C_2(G)$ which is 
the same as that of zero mode.
If every mass is given in $n/(2R_5)$,
a generic $\alpha^{-1}$ is given by
\begin{eqnarray}
 \alpha^{-1}(\mu)=\alpha^{-1}(M)
+{b\over 2\pi}\log{M\over \mu}
  +\sum_{s=1,2}{b_{(s)}\over 2\pi}
   \left[(n+1-s)\log{1\over  2\mu R_5}
   +\log ((2n-s)!!)\right] ,
\end{eqnarray}
for $(n-1/2)/R_5 \leq \mu< n/R_5$ and
\begin{eqnarray}
 \alpha^{-1}(\mu)=\alpha^{-1}(M)
+{b\over 2\pi}\log{M\over \mu}
  +\sum_{s=1,2}{b_{(s)}\over 2\pi}
    \left[n\log{1\over 2\mu R_5}
  +\log((2n-2+s)!!)\right],
\end{eqnarray}
for $n/R_5 \leq \mu< (n+1/2)/R_5$.

\section{Model with Higgs as 5D multiplets
\label{5DHm}}

We consider two cases where
Higgs fields are 5D chiral superfields and
they are a part of the 6D gauge multiplet.
In this section, we consider a model where Higgs fields are 
5D chiral superfields as a version with 
gaugino-mediated supersymmetry breaking of the model~\cite{Hall:2002qw}.

The starting parity matrices are given by
\begin{eqnarray}
 P_1=P_3={\bf 1}_6,\quad~
 P_2=\left(
  \begin{array}{cc}
   {\bf 1}_{2}& \\
         &-{\bf 1}_{4} \\
  \end{array}\right) ,\quad~
 P_4=\left(
  \begin{array}{ccccc|c}
    {\bf 1}_{5} & \\
     &-1 \\ 
  \end{array}\right) ,
 \label{parity1}
\end{eqnarray}
for the boundary conditions (\ref{parityg5}) and 
(\ref{parityg6}). 
The vector superfield $V(A_m,\lambda)$ 
has the boundary conditions
\begin{eqnarray}
 \left(\begin{array}{cc|ccc|c}
   _{(++++)} & _{(++++)} & _{(+-++)} & _{(+-++)} & _{(+-++)} & _{(+-+-)} \\
   _{(++++)} & _{(++++)} & _{(+-++)} & _{(+-++)} & _{(+-++)} & _{(+-+-)} \\ \hline
   _{(+-++)} & _{(+-++)} & _{(++++)} & _{(++++)} & _{(++++)} & _{(+++-)} \\
   _{(+-++)} & _{(+-++)} & _{(++++)} & _{(++++)} & _{(++++)} & _{(+++-)} \\
   _{(+-++)} & _{(+-++)} & _{(++++)} & _{(++++)} & _{(++++)} & _{(+++-)} \\ \hline
   _{(+-+-)} & _{(+-+-)} & _{(+++-)} & _{(+++-)} & _{(+++-)} & _{(++++)} \\
 \end{array}\right) .
 \label{vparity5dh}
\end{eqnarray}
The zero mode $(++++)$ is in the blocks 
of square matrices of rank 2, 3 and 1.
Among these zero mode components,
the blocks of the square matrices of 
rank 2 and 3 have the unbroken gauge group 
SU(3)$\times$SU(2)$\times$U(1)
which is identified as 
SU(3)$_\textrm{\scriptsize C}$$\times$SU(2)$_\textrm{\scriptsize
W}$$\times$U(1)$_Y$.
This  gives the correct normalization
for hypercharges.
The SU(5) relation for the three MSSM gauge
coupling constants is obtained
and
the successful prediction of the MSSM for $\sin^2\theta_w$
is recovered.
The other block has extra U(1) gauge group 
which we write as U(1)${}'$.
In the up-left corner
the components of square matrix of rank 5 
have the positive parity
in the $x^6$ direction and all of them 
contributes to dynamics at scales less than $1/(2R_5)$.
For each fixed line and point,
an unbroken gauge group of restricted gauge symmetry
and the number of supersymmetry in the 4D language are
tabulated in Table~\ref{tab:susy5dh}.
\begin{table}[htbp]
\begin{center}
\caption{Gauge groups and supersymmetries $N$ 
on fixed lines and
points.
\label{tab:susy5dh}}
 \begin{tabular}{lll|lll} \hline\hline
   Location & Gauge group     & $N$
 & Location & Gauge group     & $N$ \\ \hline
$x^6=0$ & SU(6) & 2 
& $(0,0)$ & SU(6) & 1 \\ 
   $x^6=\pi R_6$ & SU(5)$\times$U(1)${}'$ & 2   
&   $(0,\pi R_6)$ &  SU(5)$\times$U(1)${}'$ & 1 \\
   $x^5=0$    & SU(6) & 2  
& $(\pi R_5,0)$ & SU(4)$\times$SU(2)$\times$U(1)  & 1 \\
$x^5=\pi R_5$ & SU(4)$\times$SU(2)$\times$U(1)  & 2  
& $(\pi R_5,\pi R_6)$ &  
SU(3)$\times$SU(2)$\times$U(1)$\times$U(1)$'$ &1 
 \\ \hline \hline
 \end{tabular}
\end{center}
\end{table} 
The gauge supersymmetry and supersymmetry at each fixed point
are given by the intersection of those on the adjacent fixed 
lines; for example, SU(3)$\times$SU(2)$\times$U(1)$\times$U(1)$'$
$=$ SU(5)$\times$U(1)${}'$ $\cap$ SU(4)$\times$SU(2)$\times$U(1).

The charges for U(1) and U(1)$'$
are defined in a similar way to case of the 4D SU(5).
The three independent diagonal matrices
invariant under SU(3)
are 
\begin{eqnarray}
 \textrm{diag}(1,-1,0,0,0,0) , \quad
 \textrm{${1\over \sqrt{15}}$}\,\textrm{diag}(1,1,1,1,1,-5) ,\quad
 \textrm{${1\over \sqrt{15}}$}\,\textrm{diag}(3,3,-2,-2,-2,0) .
 \label{diag3}
\end{eqnarray}
In the notation of Appendix~\ref{lam},
they are $\lambda_3$, $\lambda_{35}$ and
$( \textrm{${\sqrt{5}\over 3}$}\lambda_8
 +\textrm{${\sqrt{10}\over 3}$}\lambda_{15}
 +\textrm{${\sqrt{6}\over 6}$}\lambda_{24})$,
respectively.
For the electric charge 
after SU(2)$_\textrm{\scriptsize W}$$\times$U(1)$_Y$ is 
broken to U(1)$_{\textrm{\scriptsize em}}$,
the charge matrix of U(1)$_{\textrm{\scriptsize em}}$ is 
\begin{eqnarray}
 Q &\!\!\!=\!\!\!&
   T_3+Y
 =\textrm{${1\over 2}$}\lambda_3
  +\textrm{${1\over 2}$}
  \textrm{${\sqrt{15}\over 3}$}( \textrm{${\sqrt{5}\over 3}$}\lambda_8
 +\textrm{${\sqrt{10}\over 3}$}\lambda_{15}
 +\textrm{${\sqrt{6}\over 6}$}\lambda_{24})
 = \textrm{diag}(1,0,\textrm{$-{1\over 3}$},\textrm{$-{1\over
  3}$},\textrm{$-{1\over 3}$},0) .
 \label{u1em}
\end{eqnarray}
From the other independent diagonal matrix in Eq.(\ref{diag3}),
we take the charge matrix for U(1)$'$ as
\begin{eqnarray}
 Q_{\textrm{\scriptsize x}}&\!\!\! =\!\!\!&3\sqrt{15}\,p\lambda_{35} 
=3p\,\textrm{diag}(1,1,1,1,1,,-5) ,
 \label{u1prime}
\end{eqnarray}
where $p$ is a constant.
For the fixed lines with unbroken SU(5)$\times$U(1)${}'$,
the U(1)${}'$ is not quantized, whereas
the U(1)$_Y$ in SU(5) is quantized.

In the gauge multiplet,
the superfield $\Sigma_5(\sigma_5,\lambda_5)$ has the boundary 
condition, 
\begin{eqnarray}
 \left(\begin{array}{cc|ccc|c}
   _{(--++)} & _{(--++)} & _{(-+++)} & _{(-+++)} & _{(-+++)} & _{(-++-)} \\
   _{(--++)} & _{(--++)} & _{(-+++)} & _{(-+++)} & _{(-+++)} & _{(-++-)} \\ \hline
   _{(-+++)} & _{(-+++)} & _{(--++)} & _{(--++)} & _{(--++)} & _{(--+-)} \\
   _{(-+++)} & _{(-+++)} & _{(--++)} & _{(--++)} & _{(--++)} & _{(--+-)} \\
   _{(-+++)} & _{(-+++)} & _{(--++)} & _{(--++)} & _{(--++)} & _{(--+-)} \\ \hline
   _{(-++-)} & _{(-++-)} & _{(--+-)} & _{(--+-)} & _{(--+-)} & _{(--++)} \\
 \end{array}\right) .
 \label{5parity5dh}
\end{eqnarray}
The boundary conditions of $\Sigma_5$ also has
the positive parity in the $x^6$ direction 
in the blocks of square matrices of rank 5 and 1.
Thus $\Sigma_5$ has dynamical components below $1/(2R_6)$.
The superfields $\Phi$ and $\Sigma_6$
also propagate in 6D,
in addition to 
$V$ and $\Sigma_5$.
The components for $\Sigma_6$ and $\Phi$
are seen to be non-dynamical until $1/(2R_6)$.
The boundary conditions 
for $\Sigma_6$ and $\Phi$ are
explicitly given in Appendix~\ref{ap:par}.

We take confined fields on fixed lines and points
as matter.
The field content is shown in Table~\ref{tab:h5}
with a consideration 
similar to that in obtaining explicit boundary conditions
given in Table~\ref{use}.
\begin{table}[htbp]
\begin{center}
\caption{Field content added to
the gauge multiplet $(V,\Sigma_5,\Sigma_6,\Phi)(x,x^5,x^6)$ in the model with Higgs multiplets on a fixed line.
Quantum numbers are shown for gauge groups of
the restricted gauge symmetry
on each fixed line or point.
For $(\Psi_\imath, \Psi_\imath^c)$, 
the parity of sign should be regarded
as not $\eta_1,\eta_2$ but the correspondent
in the $x^6$ direction. 
Here $\imath=1,2$, $\jmath=1,2,3$. \label{tab:h5}}
 \begin{tabular}{lcclcc} \hline\hline
   Location & Field     & \multicolumn{2}{l}{Quantum number}
 & $\eta_1$ & $\eta_2$ \\  \hline
   $x^6=\pi R_6$  
 & $(H_1,H_1^c)(x,x^5)$   & $(5^*,2p)$ &
  \scriptsize{$\in$ SU(5)$\times$U(1)${}'$} & 1&1   \\
 & $(H_2,H_2^c)(x,x^5)$ & $(5,-2p)$ && 1&1 \\
 & $(T_\imath,T_\imath^c)(x,x^5)$   & $(10,p)$ && 1& $-1$  \\ 
 & $(T'_\imath, T_\imath^{'c})(x,x^5)$  & $(10,p)$ && 1&1 \\ \hline
   $(0,\pi R_6)$ & $T_3(x)$   & $(10,p)$ &
 \scriptsize{$\in$ SU(5)$\times$U(1)${}'$} 
 & \multicolumn{2}{c}{---}  \\ 
 & $N_\jmath(x)$  & $(1,5p)$ &&   \multicolumn{2}{c}{---} \\ \hline
   $x^5=0$   & $(\Psi_\jmath, \Psi_\jmath^c)(x,x^6)$    & $6^*$
& \scriptsize{$\in$ SU(6)} 
  &1&1  \\  \hline
   $(\pi R_5,0)$ & $S(x)$  
& $(1,1,0)$ & \scriptsize{$\in$ SU(4)$\times$SU(2)$\times$U(1)} &
  \multicolumn{2}{c}{---} \\
\hline \hline
 \end{tabular}
\end{center}
\end{table}
For all fields on fixed lines in this model, 5D masses are zero.
In our notation, a conjugated field has the opposite transformation
property with the non-conjugated field, and 
we specify the transformation property
of a hypermultiplet by that of the non-conjugated chiral superfield;
for instance, $H_1$ and $H_1^c$ 
transforms as $(5^*,2p)$ and $(5,-2p)$
under SU(5)$\times$U(1)${}'$, respectively.
The superfields $T$, $T'$ and $\Psi$ has Yukawa coupling 
at the fixed point $(x^5,x^6)=(0,\pi R_6)$.
In order to avoid the fermion mass relation
for the first two generations,
$T_\imath$, $T_\imath^c$, $T_\imath'$ and $T_\imath^{'c}$
propagate in 5D.
For the third generation, $T_3$ is set at 
the intersection fixed point $(x^5,x^6)=(0,\pi R_6)$
to give the approximate mass relation for $m_\tau/m_b$ in SU(5),
for example, if down-type quarks and charged leptons have
small mixing angles in flavor space. 
The superfield $S$ confined at 
$(x^5,x^6)=(\pi R_5,0)$
develops $\theta\theta F_s$.
This field is directly coupled to the gauge multiplet only.
It originates gaugino-mediated supersymmetry breaking.

The charges of U(1)${}'$ for confined fields are assigned so as to
allow the Yukawa interactions
\begin{eqnarray}
  TTH_2, \quad~ T\Psi H_1 ,
\end{eqnarray}
where the ${\bf 5}^*$-part of the whole ${\bf 6}^*$
of $\Psi$ contributes to the coupling. 
From Eq.(\ref{u1prime}),
$H_1$, $H_2$ and $T$ have the U(1)${}'$ charges 
$2p$, $-2p$ and $p$, respectively as shown in Table~\ref{tab:h5},
where we have
 assumed $H_1$ and $H_2$ have the opposite U(1)${}'$ charges.
The zero modes $(++++)$ of the vector multiplet
and $(++)$ of matter multiplets contribute to 
4D chiral anomaly.
The SU(3)$\times$SU(2)$\times$U(1) part is 
clearly anomaly free
because zero mode is composed of
right-handed neutrinos and their superpartners
as well as the field content of MSSM.
The part of U(1)$'$ should be checked.
The zero modes of $H_\imath$, $T_\imath$, $T_\imath'$, 
$\Psi_\jmath$ ($\imath=1,2$, 
$\jmath=1,2,3$) and their charge conjugate as well as $T_3$ 
contribute
\begin{eqnarray}
 0+3\times 10\times p +3\times 5\times (-3p)&\!\!\!=\!\!\!& -3 (5p) 
 \qquad~~\textrm{to}~ \textrm{Tr}\,Q_{\textrm{\scriptsize x}},
\\
 0+3\times 3\times p +3\times 1\times (-3p)&\!\!\!=\!\!\!& 0 
 \qquad\qquad\quad~\textrm{to}~ \textrm{Tr}_{\textrm{\scriptsize
 SU(2)}}\, Q_{\textrm{\scriptsize x}},
\\
 0+3\times (p+2p) +3\times (-3p)&\!\!\!=\!\!\!& 0
 \qquad\qquad\quad~ \textrm{to}~ \textrm{Tr}_{\textrm{\scriptsize
 SU(3)}}\, Q_{\textrm{\scriptsize x}},
\\
 0+3\times 10\times p^3 +3\times 5\times (-3p)^3
  &\!\!\!=\!\!\!& -3 (125p^3) 
 \quad~ \textrm{to}~ \textrm{Tr}\,Q_{\textrm{\scriptsize x}}^{3} .
\end{eqnarray}
The SU(5) singlets $N_\jmath~(\jmath=1,2,3)$  
with the U(1)${}'$ charge $5p$ confined on
the fixed point $(x^5,x^6)=(0, \pi R_6)$
cancel the contributions given above to the anomalous terms.

The gauge symmetry for U(1)$'$ is 
broken via the coupling
\begin{eqnarray}
 \delta(x^5)\delta(x^6-\pi R_6)\,
  2\textrm{Re}\left[X(B\bar{B}-\Lambda_\textrm{\scriptsize x}^2)
 \right]_F , \label{upb}
\end{eqnarray}
where $X$, $B$ and $\bar{B}$ confined on the fixed point
have the quantum number
$(1,0)$, $(1,10p)$ and $(1,-10p)$ under SU(5)$\times$U(1)$'$,
respectively. A dimensionful constant is 
denoted as $\Lambda_\textrm{\scriptsize x}$.
The U(1)$'$ is broken by the vacuum expectation value
\begin{eqnarray}
 \langle B\rangle=\langle\bar{B}\rangle =
 \Lambda_\textrm{\scriptsize x} .
\end{eqnarray}
Adding the coupling
$\delta(x^5)\delta(x^6-\pi R_6)
2\textrm{Re}\left[\bar{B}N_\jmath N_\jmath\right]_F $ to
Eq.(\ref{upb})
induces Majorana mass of right-handed neutrino.

\subsection{Mass spectrum}

Here we give quantum numbers for components of zero modes
and KK modes of the 6D gauge multiplet and 
the fields given in Table~\ref{tab:h5}.
In the 6D gauge multiplet, 
the vector superfield $V$ has zero mode.
They give one vector and one two-component Weyl fermion.
Each 4D chiral superfield has one complex scalar and
one two-component Weyl fermion.
Each pair of 5D superfields, collectively ($\Phi_l,\Phi_l^c$), 
contributes to one complex scalar and one two-component Weyl fermion.
Because $S$ is a spurion, it does not
have zero mode nor KK modes.
The massless mode is the right-handed neutrino and its superpartner
as well as the field content of 
MSSM with two Higgs fields.
Representations under SU(3)$_\textrm{\scriptsize C}$
$\times$SU(2)$\textrm{\scriptsize W}$$\times$U(1)$_Y$ are 
tabulated in Table~\ref{tab:rep}.
\begin{table}[h]
\begin{center}
 \caption{Quantum numbers for 
SU(3)$_\textrm{\scriptsize C}$$\times$SU(2)$_\textrm{\scriptsize W}$$\times$U(1)$_Y$ 
of zero mode for superfields given in Table~\ref{tab:h5}
and the gauge multiplet. The notation for 
zero-mode quarks and leptons is also shown.
The superfields
$\Sigma_5,\Sigma_6,\Phi,
H_\imath^c,T_\imath^c,T_\imath'{}^c,\Psi_j^c,S$ do not have zero mode. 
Here $\imath=1,2$, $\jmath=1,2,3$. \label{tab:rep}}
\begin{tabular}{ll|l} \hline\hline
 $V(x,x^5,x^6)$ & $(1,3,0)\oplus(8,1,0)$ & Zero-mode \\ 
 $H_1(x,x^5)$  & 
    $(1,2,-1/2)$ & quark and\\
  $H_2(x,x^5)$ & 
   $(1,2,1/ 2)$  & lepton \\ \hline 
  $T_\imath(x,x^5)$ & $(3,2,1/6)$ 
    & ${1\over \sqrt{\pi R_5}}$$q_\imath(x)$ \\ 
  $T_\imath'(x,x^5)$  & $(3^*,1,-2/3)\oplus(1,1,1)$ 
   & ${1\over \sqrt{\pi R_5}}$$u_\imath^c(x) 
 \oplus {1\over \sqrt{\pi R_5}}$$e_\imath^c(x)$ \\
 $T_3(x)$ & $(3,2,1/6)\oplus(3^*,1,-2/3)\oplus(1,1,1)$ 
 & $q_3(x) \oplus u_3^c(x) \oplus e_3^c(x)$ \\
  $\Psi_\jmath(x,x^6)$  
 & $(3^*,1,1/3)\oplus(1,2,-1/2)$ 
 & ${1\over \sqrt{\pi R_6}}$$d_\jmath^c(x)
                 \oplus{1\over \sqrt{\pi R_6}}$$l_\jmath(x)$ 
 \\
 $N_\jmath(x)$  &  $(1,1,0)$ & $\nu_\jmath^c(x) $ \\ \hline
\hline
\end{tabular}
\end{center}
\end{table}
The two lower generation down-type quarks and
charged leptons are included in different multiplets,  
$d_\imath\in T_\imath$ and $e_\imath^c\in T_\imath'$.
The third generation down-type quarks and charged leptons are
included in the identical multiplet, $d_3\oplus e_3^c \in T_3$.

For fields with the masses $(n-\textrm{${1\over 2}$})/R_5$ 
and $n/R_5$, representations are shown
in Table~\ref{tab:rep2}.
\begin{table}[h]
\begin{center}
  \caption{Quantum numbers
for SU(3)$_\textrm{\scriptsize C}$$\times$
SU(2)$_\textrm{\scriptsize W}$$\times$U(1)$_Y$ 
of 4D mode with the masses 
$(n-\textrm{${1\over 2}$})/R_5$ 
and $n/R_5$.
The superfields
$\Sigma_6,\Phi,T_3,\Psi_\jmath,N_\jmath,S$ do not have modes with
these masses.  \label{tab:rep2}}
 \begin{tabular}{l|l|l}
 \hline\hline
 Superfield & Quantum number for 
 $(n-\textrm{${1\over 2}$})/R_5$
& Quantum number for $n/R_5$
 \\ \hline
 $V+\Sigma_5$ & $(3,2,-5/ 6)\oplus 
(3^*,2,5/ 6)$  
 & $(1,3,0)\oplus(8,1,0)$ 
  \\ 
 $H_1+H_1^c$ & $(3^*,1,1/3)$ 
 & $(1,2,-1/2)$ 
\\ 
  $H_2+H_2^c$ & $(3,1,-1/3)$  
 & $(1,2,1/2)$  
    \\  
$T_\imath+T_\imath^c$
 & $(3^*,1,-2/3)\oplus(1,1,1)$  
& $(3,2,1/6)$  
\\ 
$T'_\imath+T^{'c}_\imath$
 & $(3,2,1/6)$  
& $(3^*,1,-2/3)\oplus(1,1,1)$  
\\ \hline\hline
 \end{tabular}
\end{center}
\end{table}
The massive modes of $V+\Sigma_5$ have 
one vector, one Dirac fermion and one complex scalar.
Each pair of 5D superfields $(\Phi_l,\Phi_l^c)$ has
two complex scalars and one Dirac fermion.

\subsection{Gauge coupling correction}

Following the approach shown in Section~\ref{myg},
for the mass spectrum in the previous section,
we examine high energy behavior of gauge coupling constants.
At scales less than the unification scale,
gauge coupling constants obey Eq.(\ref{alpha0}) at one loop.
In Eq.(\ref{alpha0}),  the coefficients for zero mode 
$b$ are
$b_3=-3$, $b_2=1$, $b_1=33/5$.
As scales cross the mass of the first KK excitation, 
the first KK mode becomes dynamical.
From the first KK-mode contribution (\ref{al1}),
the second KK-mode contribution (\ref{al2})
and Table~\ref{tab:rep2}, we obtain $b_{(1)}$ and $b_{(2)}$ 
for each gauge group as
\begin{eqnarray}
  b_{(1)3}&\!\!=\!\!& -4+n_h 
   +3n_T =4, \qquad\qquad b_{(2)3}= -6+3n_T =0, 
\\
  b_{(1)2}&\!\!=\!\!& -6 +3n_T =0 , 
 \qquad\qquad\qquad~ b_{(2)2}= -4 +n_h+3n_T =4 
\\
  b_{(1)1}&\!\!=\!\!& -10+{2\over 5}n_h
  +3n_T = -{16\over 5} , \quad~
 b_{(2)1}= {3\over 5}n_h+3n_T = {36\over 5} ,
\end{eqnarray}
where $n_h$ stands for the number of pair of Higgs superfields, 
$n_h=2$ and $n_T$ stands for the number of 
the pairs $((T_\imath,T_\imath^c)+(T'_\imath,T_\imath^{'c}))$, 
$n_T=2$.
The coefficients $b_{(1)}$ and $b_{(2)}$
are not independent of gauge group.
In this method, even if 
gauge coupling constants coincide
at $\mu=M_X$,
the lines of the three gauge couplings do not become 
a single  line
beyond $\mu=M_X$ unlike 4D unified model. 
On the other hand,
the sum of $b_{(1)}$ and $b_{(2)}$ is independent of
gauge group.
We find
\begin{eqnarray}
 b_{(1)3}+b_{(2)3}=b_{(1)2}+b_{(2)2}
=b_{(1)1}+b_{(2)1}=
 -10+n_h+6n_T=4.
\end{eqnarray}
This equality is 
because the sum of the modes with $(n-\textrm{$1\over 2$})/R_5$ 
and $n/R_5$ 
spans over every component needed in
a multiplet under SU(5) representation.

A remarkable feature is that
$b_{(1)}+b_{(2)}$ is positive.
It is completely different from  that of the 4D SU(5) unified model.
This is originated from more components of Dirac fermions and 
complex scalars which contributes to positive part 
in the coefficients of $\beta$ function
(\ref{b1aa}) and (\ref{b2aa})
compared to the 4D case.
If one considered a distinct model with $n_T=0$,
the sum of $b_{(1)}$ and $b_{(2)}$ would be obtained as
$b_{(1)}+b_{(2)}=-8$ for $n_h=2$.
This is the same as in the 4D SU(5) unified model
beyond the unification scale.
The inclusion of matter fields with KK modes
gives rise to strong couplings
at high energies.

\begin{figure}[htbp]
 \begin{center}
 \includegraphics[width=10.5cm]{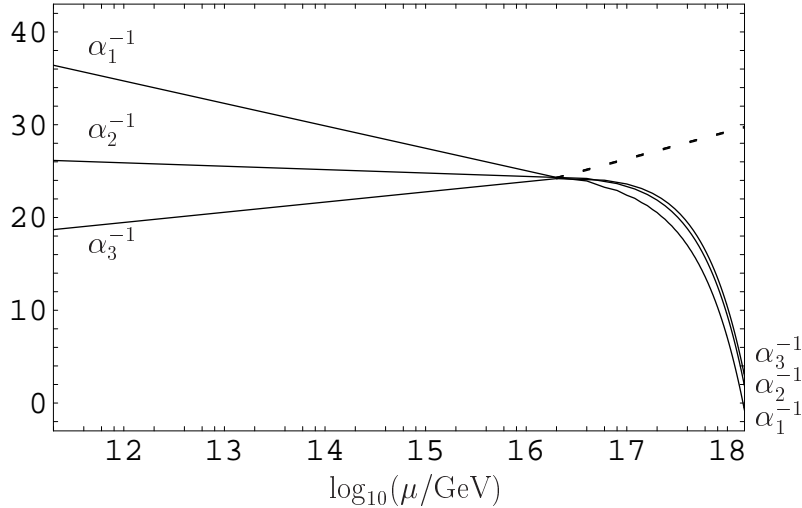} 
\caption{Energy dependence of gauge couplings. 
The input parameters are given by
$\alpha^{-1}_{\textrm{\scriptsize em}}=127.906$,
$\sin^2\theta_w=0.2312$ and
$\alpha_3=0.1187$ at $M_Z=91.1876$GeV.
$M_{\textrm{\scriptsize SUSY}}=M_Z$ and
$1/(2R_5)=2\times 10^{16}$GeV are chosen.
The broken line indicates gauge couplings
in minimal 4D supersymmetric
grand unified model where 
$X,Y$ gauge bosons, two triplet Higgs fields
and their superpartners in addition to standard model particles and 
superparticles lead to the coefficient of the $\beta$ function 
$b_1=b_2=b_3= -8$.
 \label{gcfig}}
\end{center}  
\end{figure}

To evaluate energy dependence of gauge coupling constants,
successful SU(5) relations are available.
For zero mode, the gauge couplings
$\alpha^{-1}_{\textrm{\scriptsize em}}=127.906$ 
and $\alpha_s=0.1187$
at $M_Z=91.1876\textrm{GeV}$ lead to
the weak mixing angle
\begin{eqnarray}
\sin^2\theta_w={5\over 5b_1+3b_2-8b_3}
\left((b_1-b_2){\alpha_{\textrm{\scriptsize em}}\over\alpha_s}
+(b_2-b_3){3\over 5}\right)=0.231 ,
\nonumber
\end{eqnarray}
which is compatible with experiments.
As another version of zero mode renormalization group equations,
the unified scale is written as
\begin{eqnarray}
M_X=M_Z\,\textrm{exp}
\left[{16\pi\over 5}
 {\alpha_{\textrm{\scriptsize em}}^{-1}(M_Z)\over
 (b_1-b_2)}
 \left({3\over 8}-\sin^2\theta_w(M_Z)\right)\right]
 =1.99\times 10^{16}\textrm{GeV} .
\end{eqnarray}
In order to solve the doublet-triplet splitting 
by boundary conditions,
we choose  $1/(2R_5)=M_X$.
In Figure~\ref{gcfig},
$\alpha_{\textrm{\scriptsize em}}$, $\alpha_3$
and $\sin^2\theta_w$
at $M_Z$ are taken as input parameters
and the radius is chosen as 
$1/(2R_5)=2\times 10^{16}\textrm{GeV}$.
The couplings $\alpha^{-1}$ approach to zero 
above $10^{18}\textrm{GeV}$.
In order for this model with rapidly growing strong coupling 
at high energies to be valid,
the scale $1/R_6$ has the upper bound.

\subsection{Yukawa couplings}

The values of the running quark masses $m(\mu)$
below the electroweak scale $\Lambda_W$
can be evaluated by
using the formula $\mu(d/d\mu) m(\mu)=\gamma(\alpha_s)m(\mu)$
where $\gamma(\alpha_s)$ denotes the anomalous dimension of 
the quark mass operator.
On the other hand, at high energies far from $\Lambda_W$, 
evolution equations of 
Yukawa coupling constants $y(\mu)$
are needed for the quark masses. 
With $1/(2R_5)$ as the unification scale 
to solve the doublet-triplet splitting, we have seen that
$1/R_6$ should be less than the scale where $\alpha^{-1}$ is zero.
In this light, 
the values of the parameters $R_5$, $R_6$ and $g$ are almost fixed
although a variation is still possible.
It should be checked whether
the values of derived Yukawa coupling constants are compatible with 
experiments and whether 
effects of extra dimensions are useful for generating hierarchical
numbers.

We examine Yukawa coupling
constants at $\mu= 1/(2R_5)=M_X$
where the unified gauge group is broken by boundary conditions. 
At energies $\Lambda_W\leq
\mu\leq M_X$,
renormalization group equations are governed by
contributions of zero mode, but still
with extra-dimensional effects
as seen from the factors $(1/\sqrt{\pi R_5})$ and
$(1/\sqrt{\pi R_6})$ in Table~\ref{tab:rep}.
For experimental data, 
we adopt the values of the running Yukawa coupling 
for quarks and charged leptons shown in 
Table~\ref{tab:mq} by quoting Ref.\cite{Fusaoka:1998vc}.
The values of quark and lepton masses
are sensitive to the value of $\tan\beta=v_2/v_1$.
We choose $\tan\beta= 10$,
avoiding too large or too small $\tan\beta$
which can give rise to the burst of 
Yukawa coupling constants.
As the numerical results of 
Yukawa coupling constants are not sensitive to the value of
$M_{\textrm{\scriptsize SUSY}}$,
the scale of the supersymmetry breaking 
is taken as $M_{\textrm{\scriptsize SUSY}}=M_Z$.
\begin{table}[h]
\begin{center}
 \caption{Yukawa coupling constants at $M_Z$ and $M_X$.
For convenience, instead of the Yukawa coupling constants $y(\mu)$,
the values of
$m(\mu)=y(\mu)v\sin\beta/\sqrt{2}$ for up-type quarks
and $m(\mu)=y(\mu)v\cos\beta/\sqrt{2}$ for down-type quarks and 
charged leptons are listed in unit of MeV,
where $v=\sqrt{2}\Lambda_W$, 
$\Lambda_W=174.1$GeV,
 $\tan\beta=v_2/v_1=10$ and
$M_{\textrm{\scriptsize SUSY}}=M_Z$.
 \label{tab:mq}}
\begin{tabular}{c|ccccccccc}
  \hline\hline  
$\mu$ &  $m_u(\mu)$ & $m_c(\mu)$ & $m_t(\mu)$ & $m_d(\mu)$ 
 & $m_s(\mu)$ & $m_b(\mu)$ 
 & $m_e(\mu)$ & $m_{\mu}(\mu)$ & $m_{\tau}(\mu)$ \\ \hline
$M_Z$ & 2.33 & 677 & 181000 & 4.69 & 93.4
& 3000 & 0.4868 & 102.751 & 1746.7 \\ 
$M_X$ & 1.04 & 302 & 129000 & 1.33 & 26.5
& 1000 & 0.3250 & 68.598 & 1171.4 \\ \hline\hline
\end{tabular}
\end{center}
\end{table}

The Yukawa couplings arise as superpotential terms
of the fixed point coupling (\ref{actions4})
where delta functions should be read as
$\delta(x^5)\delta(x^6-\pi R_6)$.
The up-type Yukawa superpotential $f_u$ is given by
\begin{eqnarray}
 f_u= 
   \left(\begin{array}{c}
   T_1 \\ T_2 \\ T_3 \\
   \end{array}\right)^{\!\!\!T\,ij}\!\!
 \left(\begin{array}{ccccc}
    Y_{u11}'
 & Y_{u12}' & 0 & 0 & 0 \\
  Y_{u21}'
& Y_{u22}' & 0 & 0 & 0 \\
Y_{u31}' & Y_{u32}' & Y_{u31}
& Y_{u32} &Y_{u33} \\
  \end{array}\right) \!\!
 \left(\begin{array}{c}
   T_1' \\ T_2' \\ T_1 \\ T_2 \\ T_3 \\
  \end{array}\right)^{\!\!lm} \!\!\! H_2^n \epsilon_{ijlmn} ,
 \label{fu}
\end{eqnarray}
where $i,j,l,m,n=1,\cdots,5$.
The Yukawa couplings are taken for 
boundary values of fields rather than they are for zero mode.
In terms of the dimensionless Yukawa coupling constants
$y_{u\jmath\imath}'$ and 
$y_{u3\jmath}$ ($\jmath=1,2,3$, 
$\imath=1,2$)
normalized 
with the cutoff of 6D theory $M_s$, 
the Yukawa coupling constant matrix is written as
\begin{eqnarray}
 \left(\begin{array}{ccccc}
    {1\over M_s^{3/2}}y_{u11}'
 & {1\over M_s^{3/2}}y_{u12}' & 0 & 0 & 0 \\
  {1\over M_s^{3/2}}y_{u21}'   
& {1\over M_s^{3/2}}y_{u22}' & 0 & 0 & 0 \\
{1\over M_s} y_{u31}'& {1\over M_s}y_{u32}' & {1\over M_s}y_{u31} 
& {1\over M_s}y_{u32} &{1\over \sqrt{M_s}}y_{u33} \\
  \end{array}\right) .
\end{eqnarray}
When the Higgs field is replaced with the vacuum expectation value 
$\langle H_2^2\rangle=v_2/\sqrt{\pi R_5}$,
the zero mode leads to 
the 4D up-type quark 
mass term at $\mu=M_X$
in the notation  
given in Table~\ref{tab:rep}, 
\begin{eqnarray}
 \left(\begin{array}{c}
   u_1 \\ u_2 \\ u_3 \\
  \end{array}\right)^T
  \left(\begin{array}{ccc}
    \epsilon_5^3\, y_{u11}' v_2
  & \epsilon_5^3\, y_{u12}' v_2 
  &\epsilon_5^2\, y_{u31} v_2 \\
\epsilon_5^3\,y_{u21}' v_2 
 & \epsilon_5^3\, y_{u22}' v_2
 &\epsilon_5^2\, y_{u32} v_2 \\
 \epsilon_5^2\, y_{u31}' v_2 
 & \epsilon_5^2\, y_{u32}' v_2
 & \epsilon_5\,y_{u33} v_2 \\
 \end{array}\right)  
 \left(\begin{array}{c}
   u_1^c \\ u_2^c \\ u_3^c \\
  \end{array}\right) 
 +\textrm{H.c.}, 
 \label{yu}
\end{eqnarray}
where $\epsilon_5=1/\sqrt{\pi R_5 M_s}$.
The mass matrix in Eq.(\ref{yu}) in general has 
nonzero off-diagonal components in flavor space.
It should be diagonalized into mass eigenstates.
As an example to estimate the size of Yukawa couplings,
we here compare diagonal components in Eq.(\ref{yu}) with the values 
shown in 
Table~\ref{tab:mq}.
For $1/(2R_5)=2\times 10^{16}$GeV,
$\tan\beta=10$ and $M_s=10^{18}$GeV,
we obtain
the Yukawa coupling constants as
\begin{eqnarray}
 y_{u11}' \approx 0.003 , \quad~
 y_{u22}' \approx 0.9 , \quad~
 y_{u33} \approx 2 .
\end{eqnarray}
The mass hierarchy is generated  as
\begin{eqnarray}
{m_u\over m_t}(=8.06\times 10^{-6})&\!\!\!\ll\!\!\!&
  {y_{u11}'\over y_{u33}}(\approx 0.0015) ,
\\
{m_c\over m_t}(=0.00234)&\!\!\!\ll\!\!\!& 
 {y_{u22}'\over y_{u33}}(\approx 0.45).
\end{eqnarray}
due to the suppression 
$\epsilon_5\approx 0.1$.

The down-type and charged-lepton-type 
Yukawa couplings are generated from the superpotential 
\begin{eqnarray}
 f_d&\!\!\!=\!\!\!&
 \left(\begin{array}{c}
  T_1 \\ T_2 \\ T_3 \\ 
  \end{array}\right)^{\!\!\!T\,ij}
  \left(\begin{array}{ccc}
   {1\over M_s^{3/2}}\,y_{d11} &{1\over M_s^{3/2}}\, y_{d12}
  &{1\over M_s^{3/2}}\, y_{d13}\\
  {1\over M_s^{3/2}}\, y_{d21} & {1\over M_s^{3/2}}\,y_{d22}
 & {1\over M_s^{3/2}}\,y_{d23}\\
  {1\over M_s}\, y_{d31} &{1\over M_s}\, y_{d32}
 &{1\over M_s}\, y_{d33}\\
  \end{array}
\right)
 \left(\begin{array}{c}
  \Psi_1 \\ \Psi_2 \\ \Psi_3 \\
 \end{array}\right)_{\!\! i} H_{1\,j}
\nonumber
\\
 &&
 +\left(\begin{array}{c}
  T_1' \\ T_2' \\ 
  \end{array}\right)^{\!\!T\,ij}
  \left(\begin{array}{ccc}
  {1\over M_s^{3/2}}\, y_{e11} & {1\over M_s^{3/2}}\,y_{e12} 
&{1\over M_s^{3/2}}\, y_{e13}\\
   {1\over M_s^{3/2}}\,y_{e21} & {1\over M_s^{3/2}}\,y_{e22} 
& {1\over M_s^{3/2}}\, y_{e23}\\
  \end{array}
\right)
 \left(\begin{array}{c}
  \Psi_1 \\ \Psi_2 \\ \Psi_3 \\
 \end{array}
  \right)_{\!\! i} H_{1\,j} ,
 \label{fd}
\end{eqnarray}
where the Yukawa coupling constants 
$y_{d\imath\jmath}$, $y_{d3\jmath}$ and 
$y_{e\imath\jmath}$
($\jmath=1,2,3$, $\imath=1,2$)
are 
normalized to be dimensionless
with the cutoff of 6D theory $M_s$.
From this equation,  with the vacuum expectation value
$\langle H_{1,\,j=2}\rangle=v_1/\sqrt{\pi R_5}$ and the notation of 
zero-mode given in Table~\ref{tab:rep}, the 4D down-type
 quark and charged lepton mass terms are 
\begin{eqnarray}
&& \left(\begin{array}{c}
   d_1 \\ d_2 \\ d_3 \\
  \end{array}\right)^T
  \left(\begin{array}{ccc}
    \epsilon_5^2\epsilon_6\, y_{d11} v_1 
  & \epsilon_5^2\epsilon_6\, y_{d12} v_1 
  & \epsilon_5^2\epsilon_6\, y_{d13} v_1 \\
    \epsilon_5^2\epsilon_6\, y_{d21} v_1 
  & \epsilon_5^2\epsilon_6\, y_{d22} v_1 
  & \epsilon_5^2\epsilon_6\, y_{d23} v_1 \\
    \epsilon_5\epsilon_6\, y_{d31} v_1 
  & \epsilon_5\epsilon_6\, y_{d32} v_1
  & \epsilon_5\epsilon_6\,y_{d33} v_1 \\
 \end{array}\right)  
 \left(\begin{array}{c}
   d_1^c \\ d_2^c \\ d_3^c \\
  \end{array}\right)
\nonumber
\\
&& + \left(\begin{array}{c}
   e_1^c \\ e_2^c \\ e_3^c \\
  \end{array}\right)^T
  \left(\begin{array}{ccc}
    \epsilon_5^2\epsilon_6\, y_{e11} v_1 
  & \epsilon_5^2\epsilon_6\, y_{e12} v_1 
  & \epsilon_5^2\epsilon_6\, y_{e13} v_1 \\
    \epsilon_5^2\epsilon_6\, y_{e21} v_1 
  & \epsilon_5^2\epsilon_6\, y_{e22} v_1 
  & \epsilon_5^2\epsilon_6\, y_{e23} v_1 \\
   \epsilon_5\epsilon_6\, y_{d31} v_1 
  & \epsilon_5\epsilon_6\, y_{d32} v_1
  & \epsilon_5\epsilon_6\,y_{d33} v_1 \\
 \end{array}\right)  
 \left(\begin{array}{c}
   e_1 \\ e_2 \\ e_3 \\
  \end{array}\right) 
  +\textrm{H.c.},
 \label{yd}
\end{eqnarray}
where $\epsilon_6=1/\sqrt{\pi R_6 M_s}(>\epsilon_5)$.
The down-type quarks and charged leptons have
the common Yukawa coupling constants 
in
$\sum_{\jmath=1}^3y_{d3\jmath}(d_3d_\jmath^c+e_3^ce_\jmath)+\textrm{H.c.}$.
Because the two matrices for down-type quarks and charged leptons are 
distinct,
the fermion mass relations are avoided.
The relation between $m_b$ and $m_\tau$ also depends
on the form of the matrices.
The difference of $m_b$ and $m_\tau$ in data might
yield as a result of diagonalization of the matrices with the
common and uncommon components.
As an example of estimation,
we compare diagonal components in Eq.(\ref{yd}) 
with data at $M_X$ in Table~\ref{tab:mq}.
For $R_6=R_5/10$,
we obtain the Yukawa coupling constants as
\begin{eqnarray}
 y_{d11}\approx 0.01 ,\quad~
 y_{d22} \approx 0.2 ,\quad~
 y_{d33} \approx 0.1 ,\quad~
 y_{e11} \approx 0.003 ,\quad~
 y_{e22} \approx 0.6 .
\end{eqnarray}
The mass hierarchy is generated as
\begin{eqnarray}
{m_d\over m_b}(=0.00133)&\!\!\!\ll\!\!\!&
  {y_{d11}\over y_{d33}}(\approx 0.1) ,
\\
{m_s\over m_b}(=0.0265)&\!\!\!\ll\!\!\!& 
 {y_{d22}\over y_{d33}}(\approx 2) ,
\\
{m_e\over m_\tau}(=2.774\times 10^{-4})&\!\!\!\ll\!\!\!&
  {y_{e11}\over y_{d33}}(\approx 0.03) ,
\\
{m_\mu\over m_\tau}(=0.0585)&\!\!\!\ll\!\!\!& 
 {y_{e22}\over y_{d33}}(\approx 6) ,
\end{eqnarray}
due to the suppression 
$\epsilon_5\approx 0.1$ and $\epsilon_6\approx 0.3$. 
We have chosen $\tan\beta=10$.
For such a moderate value,
the suppression factor $\epsilon_6$
provides a part of
the suppression for $m_b/m_t$.
The effect of $v_2>v_1$ is partly compensated by
the effect of $R_6< R_5$.
The mass matrices could also be compared
with data taking into account the Cabibbo-Kobayashi-Maskawa matrix.
For example, in the base where up-type quarks are diagonal,
the down-type quark mass matrix 
may be \cite{Fusaoka:1998vc}
(as each component is written in absolute value)
\begin{eqnarray}
 M_d(M_X)=m_b(M_X)
  \left(\begin{array}{ccc}
  0.0026 & 0.0054 & 0.0025 \\
  0.0054 & 0.0263 & 0.0310 \\
  0.0025 & 0.0310 & 0.9990 \\
 \end{array}\right) .
 \label{mdb}
\end{eqnarray}
Then the corresponding Yukawa matrix is written as
\begin{eqnarray}
 \left(\begin{array}{ccc}
  y_{d11} &  y_{d12} &  y_{d13} \\
  y_{d21} &  y_{d22} &  y_{d23} \\
  y_{d31}  &  y_{d32} &  y_{d33} \\
       \end{array}\right)
\approx
 y_{d33} \left(
 \begin{array}{ccc}
   0.1 & 0.1 & 0.1 \\
   0.1 & 1 & 1 \\
   0.001 & 0.01 & 1 \\
 \end{array}\right) 
\end{eqnarray} 
Similarly to estimation from diagonal components of mass matrices, 
small numbers are generated via 
the suppression factor $\epsilon_5$ and $\epsilon_6$.

\subsection{Proton stability}

As in the 4D minimal SU(5) unified model,
extra fields which become dynamical at $\mu=1/(2R_5)=M_X$
can give rise to baryon number violating processes.
Contributions of
$X$, $Y$ gauge boson exchange give 
dimension-six baryon-number violating
operators.
Their amplitudes are suppressed 
by $M_X$ which is large in supersymmetric case.
As supersymmetric contributions, 
dimension-four and dimension-five
proton decay operators can yield
unless some selection rules are applied.

\begin{figure}[h]
 \begin{center}
 \includegraphics[width=5cm]{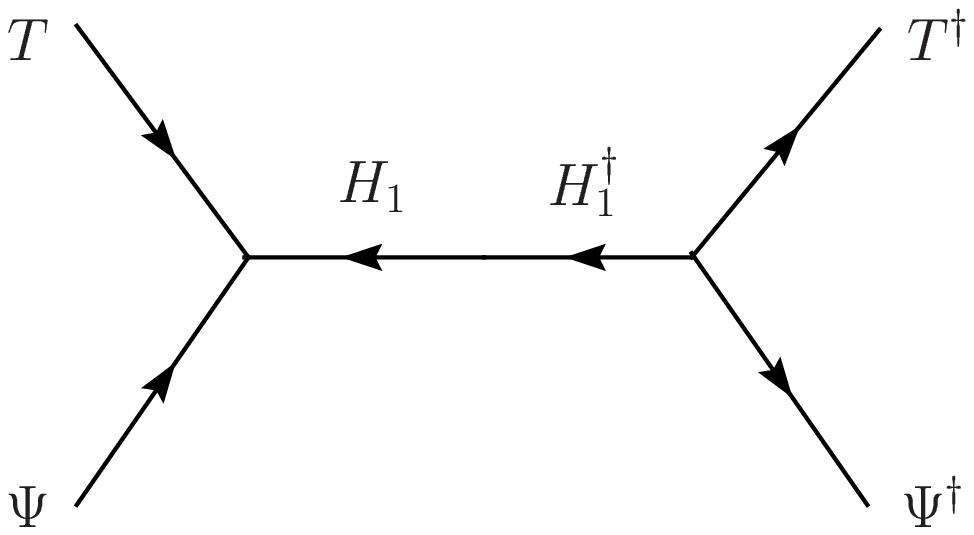} \qquad~ 
 \includegraphics[width=5cm]{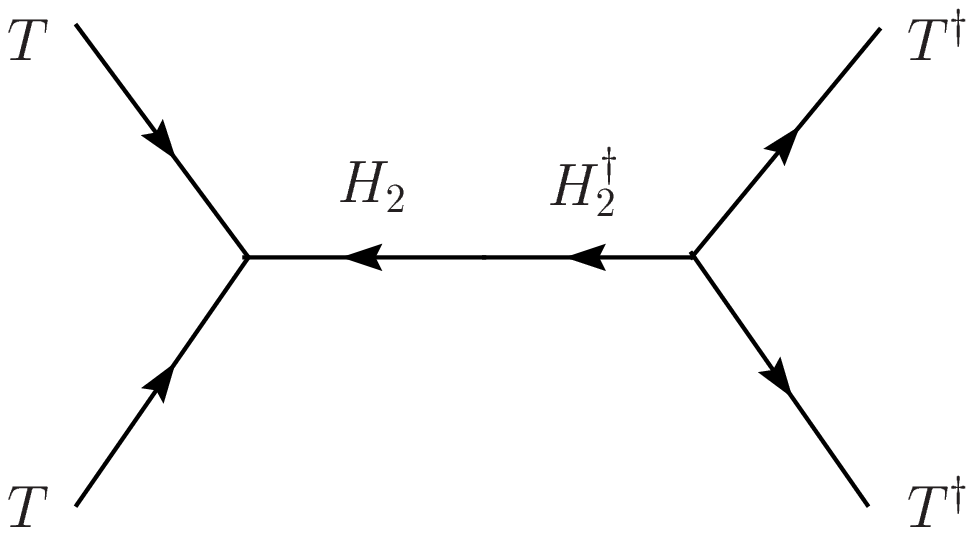} 
\caption{Dimension-six baryon number violating process.
 \label{dim6}}
\end{center}  
\end{figure}

For the field content given in Table~\ref{tab:rep},
the superfields 
$T^{ij}$, $\Psi_\imath$, $H_{1i}$, $H_1^{ci}$, 
$H_2^i$ and $H_{2i}^c$ ($i,j=1,\cdots, 5$)
give the gauge-singlet 
lower dimension operators
on the fixed point like the Yukawa couplings,
\begin{eqnarray}
&&
 \Psi_\imath H_2^i ,~~ \Psi_\imath H_1^{c\, i} ,~~ H_2^i H_{1i} ,~~
  H_2^i H_{2i}^c ,~~  H_{1i} H_1^{c\,i} ,~~
  H_{2i}^c H_1^{c\,i} , ~~
 T^{ij} H_{1i} H_{2j}^c,
 \label{5op1}
\\
&& T^{ij} \Psi_\imath H_{1j},\quad ~ T^{ij}T^{kl}H_2^m \varepsilon_{ijklm},
 \label{ytt}
\\
 && T^{ij} \Psi_\imath H_{2j}^c , \quad~
 T^{ij} T^{kl} H_1^{c\,m}\varepsilon_{ijklm}
 \label{dim5}
\\
&&
 T^{ij}T^{kl}T^{mn}\Psi_n \varepsilon_{ijklm} ,~~
 T^{ij}T^{kl}T^{mn}H_{1n}\varepsilon_{ijklm} ,~~
 T^{ij}T^{kl}T^{mn}H_{2n}^{c}\varepsilon_{ijklm} ,
 \label{5op2}
\end{eqnarray}
where $T^{ij}$ collectively denotes $T_\imath^{ij}$, $T_\imath'{}^{ij}$ and $T_3^{ij}$ ($\imath=1,2$)
and for $\Psi$, 
$\bf{5}^*$-part $\Psi_{\jmath i}$ ($\jmath=1,2,3$) is coupled to other
fields. 
The operators (\ref{ytt}) gives the Yukawa couplings
as shown in Eqs.(\ref{fu}) and (\ref{fd}).
If the operators (\ref{dim5}) exist,
they would give dimension-five 
baryon number violation processes by colored higgsino exchange.
If $\Psi_\imath H_2^i$ and $H_2^i H_{1i}$ exist, 
they would generate large masses for the weak-doublet Higgs fields,
which gives rise to recurrence of the doublet-triplet splitting problem.
These dangerous operators are avoided by R invariance
as a subgroup of SU(2)$_\textrm{\scriptsize R}$
inherent in 6D $N=2$ theory.
In Table~\ref{r5dh},
R charges of superfields are shown.
\begin{table}[h]
\begin{center}
\caption{R-charges in 5D Higgs model.\label{r5dh}}
\begin{tabular}{cccccccccc} \hline\hline
  $\theta$ & $V$, $\Sigma_5$, $\Sigma_6$ & $\Phi$ & $H_1$, 
 $H_2$ & $H_1^c$, $H_{2}^c$ 
& $T$, $\Psi$, $N$  & $S$ & $X$ & $B$ & $\bar{B}$\\ \hline
 $-1$ & 0 & 2 & 0 & 2 & 1 
&0 &2 &2 &$-2$ \\ \hline \hline
\end{tabular} 
\end{center}
\end{table}
With this charge assignment, 
the 5D mass operators $H_2H_2^c$ and $H_1H_1^c$
as well as the Yukawa couplings
$T\Psi H_1$ and $TTH_2$ 
are only allowed in Eqs.(\ref{5op1})-(\ref{5op2}).
While R invariance prohibits operators of dimension four and five,
$T\Psi H_1$ and $TTH_2$ give
dimension-six baryon number violation processes 
by colored Higgs exchange.
The diagrams for these dimension-six operators 
are shown in Fig.~\ref{dim6}, 
where
two groups of superfields of different chirality are 
connected by a chirality non-flip superpropagator.
The lifetime is proportional to $\tau\propto (1/(2R_5))^{4}$, 
which for $1/(2R_5)=M_X$ is compatible  with the observation.

\subsection{Supersymmetry breaking}

So far we have worked with the supersymmetries
$N=2$ in 6D bulk (4D $N=4$),
$N=1$ on 5D fixed lines (4D $N=2$) and
$N=1$ on 4D fixed points (4D $N=1$).
In this section, we consider breaking 4D $N=1$ supersymmetry
to no supersymmetry.

The supersymmetry breaking term is effectively 
generated on the 4D fixed point
where the gauge multiplet is directly coupled,
\begin{eqnarray}
 \int d^6 x \,\delta (x^5-\pi R_5)\delta (x^6) 
 \, 2\textrm{Re}\left[
  \textrm{Tr} {1\over 4kg^2}
   SW^\alpha W_\alpha \right]_F .
 \label{sww}
\end{eqnarray}
From the configuration shown in Table~\ref{tab:h5},
the superfield $S$ is spatially separated from 
all fields except for the 6D gauge multiplet 
($V,\Sigma_5,\Sigma_6,\Phi$).
The 6D fields 
$\Sigma_5$, $\Sigma_6$ and $\Phi$ do not have zero mode.
As long as leading zero mode part
of supersymmetry breaking effect is concerned,
the model is similar to the model of 
Ref.\cite{Kaplan:1999ac} where 
the standard-model gauge fields and gauginos only propagate in bulk.
At $\mu=M_X$
gauginos obtain masses through their direct couplings
(\ref{sww}) as
\begin{eqnarray}
 && 
 2\textrm{Re}  \left[
 {1\over M_s}S W^\alpha W_\alpha \right]_F
 \supset
M_{1/2} \lambda^\alpha\lambda_\alpha ,\qquad~
 M_{1/2}=
 {F_S\over M_s}
\end{eqnarray}
via $S=F_S \theta\theta$.
All other supersymmetry breaking masses 
are suppressed by 
a Yukawa factor of the 
spatial separation of the source and matter
and by loop factors $\alpha^2$.
Their masses are generated via running.
For squarks, sleptons and higgsino whose masses
are denoted as $\tilde{m}$ collectively, one-loop renormalization group
equations have the 
form~\cite{Inoue:1982pi}\cite{Inoue:1983pp}\cite{Buchmuller:2005ma}
\begin{eqnarray}
  {d\tilde{m}^2\over  dt}
  = -c_g g^2 M_{1/2}^2 + c_m g^2 \textrm{Tr} Y \tilde{m}^2
   +c_y y^2 m_{\tilde{h}}^2 , 
\end{eqnarray}
where $c_g$ and $c_y$ are positive coefficients and
$c_m$ are not necessarily positive and
the higgsino mass is denoted as $m_{\tilde{h}}$.
The renormalization group flow
starts to grow with the gaugino contribution with a negative sign
in the direction of low energies from the unification scale.
This do not lead to new flavor violation
because the sources of flavor violation 
are aligned with the Yukawa matrices.
At one loop approximation,
gaugino masses are proportional to gauge couplings squared
\begin{eqnarray}
 {M_1\over \alpha_1}= {M_2\over \alpha_2}= {M_3\over \alpha_3}
 = {M_{1/2}\over \alpha_X} .
\end{eqnarray}
Since $\alpha_1$, $\alpha_2$, $\alpha_3$ and $\alpha_X$
are of the same order,
the gaugino masses $M_1$, $M_2$, $M_3$ and $M_{1/2}$
are of the same order.
That charged superparticles and the lightest Higgs have lower bounds
gives the minimum of gaugino mass at the unification scale.
For $M_{1/2}=350\textrm{GeV}$ 
($M_1=144\textrm{GeV}$,
$M_2=287\textrm{GeV}$, $M_3= 1009\textrm{GeV}$) and
$M_s=10^{18}\textrm{GeV}$,
$\sqrt{F_S}=5.9\times 10^{10}\textrm{GeV}$.
Since $F_S$ is generated at $\mu=M_s$,
it would be natural that $F_S\sim M_s$.
In order to produce a small $F_S$,
one may apply a mechanism in 
Ref.\cite{ArkaniHamed:1999pv}\cite{Schmaltz:2000gy}
by utilizing the two extra dimensions.
Following this idea,
it may also be possible to address naturalness
for $\mu$-term by setting fields confined on fixed lines.
Here we do not treat this issue further.

All soft supersymmetry breaking masses 
are of the same order at the electroweak scale.
It is possible that neutralino or slepton has
the mass $\lesssim 200$\,GeV.
The lightest superparticle (LSP) can be gravitino 
$m_{3/2}\sim F_S/ M_{Pl}$ with the Planck scale $M_{Pl}$  as in case of 
Ref.\cite{Buchmuller:2005ma}. 
Like models given in Refs.\cite{Kaplan:1999ac}\cite{Chacko:1999mi},
the number of input parameters might be taken as four
where $\tan\beta$ is treated as a free parameter.fixed value of tan$\beta=10$, 
The quantity tan $\beta$ affects 
the third generation in running.
Charged lepton mass matrix includes $\tan\beta$ in off diagonal
components. 
If tan$\beta$ is increased,
it means a large mixing. 
Consequently
stau mass 
is decreased.
For a large tan$\beta$,
an input $M_{1/2}$ must be large 
in order not to make stau too light.
Further consideration about these issues and 
high energy behavior of gauge couplings 
would restrict the model.

\vspace{4ex}

Before concluding the model with Higgs fields as 5D superfields, 
we mention other choice of parity matrices.
Similar results are also obtained for
another choice of parity matrices 
\begin{eqnarray}
 P_1=P_3={\bf 1}_6 ,\quad~
 P_2=\left(
  \begin{array}{cc}
   {\bf 1}_{3}& \\
         &-{\bf 1}_{3} \\
  \end{array}\right) ,\quad~
 P_4=\left(
  \begin{array}{ccccc|c}
    {\bf 1}_{5} & \\
     &-1 \\ 
  \end{array}\right) .
 \label{anotherp}
\end{eqnarray}
The difference between Eq.(\ref{anotherp}) and Eq.(\ref{parity1})
is the numbers of the components
 $V^{(--+-)}$, $\Sigma_5^{(+++-)}$,
  $\Sigma_6^{(---+)}$, $\Phi^{(++-+)}$ and
$V^{(-++-)}$, $\Sigma_5^{(+-+-)}$,
  $\Sigma_6^{(-+-+)}$, $\Phi^{(+--+)}$.
These affect dynamics at scales beyond $1/(2R_6)$.
The boundary conditions for Eq.(\ref{anotherp}) are shown in 
Appendix~(\ref{ap:par}).

\section{Model with Higgs in gauge multiplet}

From here on we consider a model  
where Higgs fields are components of the 6D 
gauge multiplets,
which is a 6D version of the model given in~\cite{Burdman:2002se}.
The parity matrices are given by
\begin{eqnarray}
 P_1=\left(\begin{array}{cc}
    {\bf 1}_5& \\
    & -1 \\
   \end{array}\right) ,\quad~
 P_2=\left(\begin{array}{cc}
    {\bf 1}_2 & \\
    & -{\bf 1}_4 \\
    \end{array}\right) ,\quad~
 P_3=P_4={\bf 1}_6 ,
 \label{parity2}
\end{eqnarray}
for the equations (\ref{parityg5}) and 
(\ref{parityg6}).
The vector superfield $V(A^m,\lambda)$ 
has the boundary conditions
\begin{eqnarray}
  \left(\begin{array}{cc|ccc|c}
  _{(++++)} &_{(++++)}& _{(+-++)}& _{(+-++)}& _{(+-++)}& _{(--++)} \\
  _{(++++)} &_{(++++)}& _{(+-++)}& _{(+-++)}& _{(+-++)}& _{(--++)} \\ \hline
  _{(+-++)} &_{(+-++)}& _{(++++)}& _{(++++)}& _{(++++)}& _{(-+++)} \\
  _{(+-++)} &_{(+-++)}& _{(++++)}& _{(++++)}& _{(++++)}& _{(-+++)} \\
  _{(+-++)} &_{(+-++)}& _{(++++)}& _{(++++)}& _{(++++)}& _{(-+++)} \\ \hline
  _{(--++)} &_{(--++)}& _{(-+++)}& _{(-+++)}& _{(-+++)}& _{(++++)} \\
\end{array}\right) .
 \label{vparitygh}
\end{eqnarray}
Similarly to the model with 5D Higgs
in Section~\ref{5DHm},
the zero mode $(++++)$ is in the blocks 
of square matrices of rank 2, 3 and 1.
They have
the unbroken gauge group 
SU(3)$_\textrm{\scriptsize C}$$\times$SU(2)$_\textrm{\scriptsize
W}$$\times$U(1)$_Y$$\times$U(1)${}'$.
The U(1)${}_{\textrm{\scriptsize em}}$
and U(1)${}'$ charge matrices are given by
Eqs.(\ref{u1em}) and (\ref{u1prime}), respectively.
As in the previous section,
this gives the correct normalization for hypercharges.
The SU(5) relation for MSSM gauge coupling constants
is maintained.
The difference from the case with 5D Higgs is that
the the boundary conditions
$(--++)$ and $(-+++)$ appear.
The whole block 
has all the positive parity 
in the $x^6$ direction.  
They becomes dynamical when 
the extra dimension $x^5$ is visible.
For each fixed line and point,
unbroken gauge groups of restricted gauge symmetry
and the number of supersymmetry in the language of 4D are
tabulated in Table~\ref{tab:susy}.
\begin{table}[htbp]
\begin{center}
\caption{Gauge groups and supersymmetries $N$ 
on fixed lines and
points.
\label{tab:susy}}
 \begin{tabular}{lll|lll} \hline\hline
   Location & Gauge group     & $N$
 & Location & Gauge group     & $N$ \\ \hline
$x^6=0$ & SU(6) & 2 
& $(0,0)$ & SU(5)$\times$U(1)${}'$ & 1 \\ 
   $x^6=\pi R_6$ & SU(6) & 2   
&   $(0,\pi R_6)$ &  SU(5)$\times$U(1)${}'$ & 1 \\
   $x^5=0$    & SU(5)$\times$U(1)${}'$ & 2  
& $(\pi R_5,0)$ & SU(4)$\times$SU(2)$\times$U(1)  & 1 \\
$x^5=\pi R_5$ & SU(4)$\times$SU(2)$\times$U(1)  & 2  
& $(\pi R_5,\pi R_6)$ &  
SU(4)$\times$SU(2)$\times$U(1) &1 
 \\ \hline \hline
 \end{tabular}
\end{center}
\end{table}
In the gauge multiplet,
the superfield $\Sigma_5(\sigma_5,\lambda_5)$ has the boundary 
condition, 
\begin{eqnarray}
  \left(\begin{array}{cc|ccc|c}
  _{(--++)} &_{(--++)}& _{(-+++)}& _{(-+++)}& _{(-+++)}& _{(++++)} \\
  _{(--++)} &_{(--++)}& _{(-+++)}& _{(-+++)}& _{(-+++)}& _{(++++)} \\ \hline
  _{(-+++)} &_{(-+++)}& _{(--++)}& _{(--++)}& _{(--++)}& _{(+-++)} \\
  _{(-+++)} &_{(-+++)}& _{(--++)}& _{(--++)}& _{(--++)}& _{(+-++)} \\
  _{(-+++)} &_{(-+++)}& _{(--++)}& _{(--++)}& _{(--++)}& _{(+-++)} \\ \hline
  _{(++++)} &_{(++++)}& _{(+-++)}& _{(+-++)}& _{(+-++)}& _{(--++)} \\
\end{array}\right) . 
  \label{5paritygh}
\end{eqnarray}
The superfield $\Sigma_5$ has 
two SU(2) doublet as zero mode.
These zero modes are identified as the two MSSM Higgs fields.
In addition, the whole block of $\Sigma_5$ also has 
the positive parity in the $x^6$ direction.
They becomes dynamical when 
the extra dimension $x^5$ is visible.
The superfields $\Sigma_6$ and $\Phi$ have
the negative parity in the $x^6$ direction.
They are non-dynamical up to the scale $1/(2R_6)$.
The boundary conditions for $\Sigma_6$ and $\Phi$
are given in Appendix~\ref{ap:par}.

We take confined fields on a fixed line as matter.
The field content is shown in Table~\ref{model2}
with a consideration similar to that in obtaining
explicit boundary conditions given in Table~\ref{use}.
\begin{table}[htbp]
\begin{center}
\caption{Field content added to
the gauge multiplet $(V,\Sigma_5,\Sigma_6,\Phi)(x,x^5,x^6)$ in the model with Higgs fields in the gauge multiplet.
Quantum numbers are shown for gauge groups of
the restricted gauge symmetry
on each fixed line or point.
Here $\jmath=1,2,3$. \label{model2}}
 \begin{tabular}{lcclcc} \hline\hline
   Location & Field     & \multicolumn{2}{l}{Quantum number}
 & $\eta_1$ & $\eta_2$ \\  \hline
   $x^6=\pi R_6$  
 & $(N_\jmath,N_\jmath^c)(x,x^5)$   & 6 &
  \scriptsize{$\in$ SU(6)} & $-1$&$1$   \\
 & $(E_\jmath,E_\jmath^c)(x,x^5)$ & 15 && 1&1 \\
 & $(D_\jmath,D_\jmath^c)(x,x^5)$   & ~15$^*$ && $-1$& 1  \\ 
 & $(U_\jmath,U_\jmath^c)(x,x^5)$  & 20 && 1&$-1$ \\ \hline
   $(\pi R_5,0)$ & $S(x)$  
& $(1,1,0)$ & \scriptsize{$\in$ SU(4)$\times$SU(2)$\times$U(1)} &
  \multicolumn{2}{c}{---} \\
\hline \hline
 \end{tabular}
\end{center}
\end{table}
Quantum numbers are shown for charge-unconjugated.
For all fields on the fixed line 
(not $S$ on a fixed point), negative 5D masses $M<0$ are chosen 
so that $N_\jmath$, $E_\jmath$, $D_\jmath$ and $U_\jmath$ 
are localized at $x^5=\pi R_5$.
The decoupling of redundant components 
will be taken into account
so that zero mode field content is the same as
the model of 5D Higgs with the identical unbroken gauge group. 
It is 4D gauge anomaly free.

Gauge symmetry breaking of 
U(1)${}'$ is simultaneously treated with 
the decoupling of redundant components of fields
in next section.

\subsection{Mass spectrum}

In this section, following~\cite{Burdman:2002se} we 
start with reviewing that
the fields given in Table~\ref{model2} include
redundant components,
that it can decouple with fixed-line interactions 
described in Section~\ref{ldff}
and that this decoupling is responsible for
flavor mixing.
After this decoupling is performed,
we give quantum numbers for components of zero mode
and KK modes of the 6D gauge multiplet and the fields given
in Table~\ref{model2}.

The boundary conditions for the matter superfields are given by
\begin{eqnarray}
 N_\jmath &\!\!\!=\!\!\!& \left[ N_{\bar{l}'}^{(--)}\right]{}_\jmath
  \oplus \left[N_{\bar{d}'}^{(-+)}
  \oplus N_{n}^{(++)}\right]{}_\jmath ,
 \label{fieldn}
\\
 E_\jmath&\!\!\!=\!\!\!& \left[E_e^{(++)}\right]{}_\jmath 
 \oplus \left[E_u^{(++)}
   \oplus E_{\bar{d}}^{(-+)}\right]{}_\jmath
 \oplus \left[E_q^{(+-)} \oplus E_{\bar{l}}^{(--)}
 \right]{}_\jmath ,
\\
 D_\jmath&\!\!\!=\!\!\!& \left[D_{\bar{e}}^{(-+)}\right]{}_\jmath
 \oplus 
 \left[D_{\bar{u}}^{(-+)} \oplus D_d^{(++)}\right]{}_\jmath
   \oplus \left[D_{\bar{q}}^{(--)} 
\oplus D_l^{(+-)}\right]{}_\jmath,
\\
 U_\jmath  &\!\!\!=\!\!\!& \left[U_{u'}^{(++)} 
  \oplus U_{\bar{e}'}^{(-+)} \right]{}_\jmath
 \oplus \left[U_{e'}^{(++)}
    \oplus U_{\bar{u}'}^{(-+)}\right]{}_\jmath
   \oplus \left[U_{q'}^{(+-)} +
U_{\bar{q}'}^{(--)}\right]_\jmath ,
 \label{fieldu}
\end{eqnarray}
In Table~\ref{model2},
the parities of sign 
have been chosen so that $N_\jmath$, 
$E_\jmath$, $D_\jmath$ and $U_\jmath$ 
have
the $(++)$ components 
$N_{n\,\jmath}^{(++)}$, $E_{e\,\jmath}^{(++)}$, 
$D_{d\,\jmath}^{(++)}$
and $U_{u'\,\jmath}^{(++)}$.
Here the subscript Roman indices label
SU(3)$_C$ $\times$ SU(2)$_W$ $\times$ 
U(1)$_Y$ $\times$ U(1)$'$ quantum numbers 
shown in Table~\ref{qnt}.
The quantum numbers are represented for 
for left-handed superfields.
\begin{table}[htbp]
\begin{center}
\caption{
Quantum numbers and labels. \label{qnt}}
\begin{tabular}{l|l|l|l} \hline \hline
  $q$  $(3,2,1/6,6p)$ 
& $q'$  $(3,2,1/6,-9p)$ 
& $l$  $(1,2,-1/2, 12p)$ 
& $l'$  $(1,2,-1/2, -3p)$   \\
  $u$  $(3^*,1,-2/3,6p)$ 
& $u'$  $(3^*,1,-2/3,-9p)$ 
& $e$  $(1,1,1, 6p)$    
& $e'$  $(1,1,1, -9p)$    \\
  $d$  $(3^*,1,1/3,12p)$ 
&  $d'$  $(3^*,1,1/3,-3p)$ 
& $n$  $(1,1,0, -15p)$  &  \\
\hline\hline
\end{tabular}
\end{center}
\end{table}
The labels with a bar over a symbol 
are used to indicate each opposite quantum number such as 
$\bar{q}$ for $(3^*,2,-1/6,-6p)$.
In Eqs.(\ref{fieldn})--(\ref{fieldu}),
the square brackets stand for
the classification of multiplets 
under SU(4)$\times$SU(2)$\times$U(1),
the unbroken gauge group 
at $x^5=\pi R_5$ where $N_\jmath$, $E_\jmath$, $D_\jmath$ and $U_\jmath$ 
are localized.
The charge-conjugated fields are decomposed as
\begin{eqnarray}
 N^c_\jmath &\!\!\!=\!\!\!& \left[N_{l'}^{c(++)}
  \oplus N_{d'}^{c(+-)}\right]{}_\jmath
  \oplus \left[N_{\bar{n}}^{c(--)} \right]{}_\jmath ,
\\
 E^c_\jmath&\!\!\!=\!\!\!& \left[E_{\bar{e}}^{c(--)} 
\oplus E_{\bar{u}}^{c(--)}
   \oplus E_{\bar{q}}^{c(-+)}\right]{}_\jmath \oplus 
 \left[E_l^{c(++)}
   \oplus E_d^{c(+-)}\right]{}_\jmath ,
\\
 D^c_\jmath&\!\!\!=\!\!\!& \left[D_e^{c(+-)} 
 \oplus D_u^{c(+-)}
   \oplus D_q^{c(++)} \right]{}_\jmath\oplus 
 \left[D_{\bar{l}}^{c(-+)}
   \oplus D_{\bar{d}}^{c(--)} \right]{}_\jmath ,
\\
 U^c_\jmath  &\!\!\!=\!\!\!& \left[U_{\bar{u}'}^{c(--)} 
 \oplus U_{\bar{e}'}^{c(--)}
   \oplus U_{\bar{q}'}^{c(-+)}\right]{}_\jmath\oplus 
\left[U_{e'}^{c(+-)} \oplus 
 U_{u'}^{c(+-)}
    \oplus U_{q'}^{c(++)} \right]{}_\jmath,
\end{eqnarray}
where 
square brackets denote the classification of multiplets 
under SU(5)$\times$U(1)${}'$,
the unbroken gauge group 
at $x^5=0$ where $N_\jmath^c$, $E_\jmath^c$, $D_\jmath^c$ and $U_\jmath^c$ 
are localized.

The parity $(++)$ components are
\begin{eqnarray}
 (3,2,\textrm{${1\over 6}$}) && D_{q\,\jmath}^{c\,(++)}, U_{q'\,\jmath}^{c\,(++)} 
 \quad~
 (1,2,-\textrm{${1\over 2}$})~~ N_{l'\,\jmath}^{c\,(++)}, E_{l\,\jmath}^{c\,(++)} \\
 (3^*,1,\textrm{${1\over 3}$})&& D_{d\,\jmath}^{(++)} 
 \quad\qquad\qquad~
 (1,1,0)\quad~ N_{n\,\jmath}^{(++)} \\
 (3^*,1,-\textrm{${2\over 3}$})&& U_{u'\,\jmath}^{(++)}, E_{u\,\jmath}^{(++)} 
 \qquad~
 (1,1,1)\quad~ E_{e\,\jmath}^{(++)}, U_{e'\,\jmath}^{(++)} 
\end{eqnarray}
The three numbers in the parentheses denote
the quantum numbers for SU(3)$_\textrm{\scriptsize C}$,
SU(2)$_\textrm{\scriptsize W}$ 
and U(1)$_Y$ in order.
For all of 5D superfields, negative 5D masses $M<0$ have been chosen 
so that $N,E,D,U$ are localized at $x^5=\pi R_5$.
Redundant components as massless modes 
are decoupled with fixed-point couplings shown in 
Section~\ref{ldff}.
The SU(2)-singlet superfields 
[$E_{\bar{d}}^{(-+)} \oplus E_{u}^{(++)}$] 
and [$U_{\bar{u}'}^{(-+)}\oplus U_{e'}^{(++)}$] are made heavy through
their couplings to additional matters 
of $(6,1)$ and $(4^*,1)$, respectively
(in SU(4) $6^*=6$).
The other fields localized at $x^5=0$ include
SU(2) doublet (++) components. They are
made heavy for linear combinations.
This linear combination 
makes mixing between generations,
while the original Yukawa couplings are 
diagonal in flavor space since
Higgs fields are part of the gauge multiplet.

Now we specify part of SU(2) doublet.
For quark sector, 
[$D_q^{c(++)}\oplus D_d^{c(+-)} \oplus D_e^{c(+-)}
$]$_\jmath\equiv D_{T\,\jmath}^c$ and 
[$U_{q'}^{c(++)} \oplus 
U_{u'}^{c(+-)} \oplus U_{e'}^{c(+-)}
$]$_\jmath\equiv U_{T'\,\jmath}^c$
have the quantum number $(10,6p)$ and $(10,9p)$ for
SU(5)$\times$U(1)${}'$, respectively.
The fixed-point couplings with 
U(1)${}'$ symmetry breaking are chosen as 
\begin{eqnarray}
  &&\delta(x^5)\delta(x^6-\pi R_6)\bigg(
 [(U_{T'\,\imath}^c \mathbf{M}_{1\imath\jmath}
 + D_{T\,\imath}^c
  \mathbf{M}_{2\imath\jmath}\bar{\cal B})
 \bar{T}_\jmath
 + {\cal X}({\cal B}\bar{\cal B}-\Lambda_{x}^2)]_{\theta\theta}
  +\textrm{H.c.}\bigg) .
 \label{branec}
\end{eqnarray}
with the 4D K\"ahler terms.
For the fixed-point fields,
the quantum numbers 
with respect to SU(5)$\times$U(1)${}'$
are given by
\begin{eqnarray}
 \bar{T}_\imath ~ (10^*,9p) ,\quad~
  {\cal B} ~ (1,15p) ,\quad~ 
 \bar{\cal B} ~ (1,-15p), \quad~ 
 {\cal X} ~(1,0) 
\end{eqnarray}
and $\mathbf{M}_{1\imath\jmath}$, 
$\mathbf{M}_{2\imath\jmath}$ and $\Lambda_x$ are constants.
From the equation of motion for ${\cal X}$, the vacuum expectation
values are developed
\begin{eqnarray}
 \langle {\cal B}\rangle 
=\langle \hat{\cal B} \rangle
  =\Lambda_x .
\end{eqnarray}
This breaks U(1)${}'$ gauge symmetry. 
After the U(1)${}'$ symmetry breaking,
linear combinations with mixing between generations yields.
The elements of the linear combinations are 
$U_{T'1}^c$, $U_{T'2}^c$, $U_{T'3}^c$, 
$D_{T1}^c$, $D_{T2}^c$ and $D_{T3}^c$.
These fields $U_{T'\,\jmath}^c$ and $D_{T\,\jmath}^c$ 
are transformed into
another basis $H_{T\,\jmath}$ and $T_\jmath$, 
where $H_{T\,\jmath}$ would be made
heavy and $T_\jmath$ has quark doublet
as its $(++)$ component.
The transformation between the two basis is given by
\begin{eqnarray}
  U_{T'}^c =H_T U_1 +T U_3 , \quad~
D_T^c = H_T U_2 +T U_4 .
 \label{tutu}
\end{eqnarray}
Here the $3\times 3$ unitary matrices $U_\ell$, $\ell=1,\cdots, 4$ are
written as components of the $6\times 6$ unitary matrix 
\begin{eqnarray}
 U=\left(\begin{array}{cc}
   U_1 & U_2 \\
   U_3 & U_4 \\ \end{array}\right) ,
\end{eqnarray}
which mixes the 6 elements.
Due to linear combinations with complex numbers,
CP phases are included.
After U(1)${}'$ is broken,
relevant part of the fixed-point coupling (\ref{branec}) becomes
\begin{eqnarray}
 (U_{T'}^c, D_T^c) \left(\begin{array}{c}
    \mathbf{M}_1 \\ \mathbf{M}_2 \Lambda_x \\
  \end{array}\right)
 \bar{T} 
 =(H_T, T) \left(\begin{array}{c}
    U_1\mathbf{M}_1  + U_2 \mathbf{M}_2 \Lambda_x \\
  U_3\mathbf{M}_1  + U_4 \mathbf{M}_2 \Lambda_x \\
  \end{array}\right)
 \bar{T} . \label{qhq}
\end{eqnarray}
For the non-heavy part 
$U_3\mathbf{M}_1  + U_4 \mathbf{M}_2 \Lambda_x=0$,
the matrix $\mathbf{M}_2$ is written 
in terms of other variables
as $\mathbf{M}_2=-\Lambda_x^{-1}U_4^{-1}U_3 \mathbf{M}_1$.
Eq.(\ref{qhq}) reduces to
\begin{eqnarray}
 (U_{T'}^c  \mathbf{M}_1 +D_T^c\mathbf{M}_2 \Lambda_x)
 \bar{T} 
 =H_T (U_1-U_2 U_4^{-1} U_3)\mathbf{M}_1 
\bar{U}^\dag 
 (\bar{U}\bar{T}) .
\end{eqnarray}
We choose $U$, $\mathbf{M}_1$ and $\bar{U}$ 
so that $(U_1-U_2 U_4^{-1} U_3)\mathbf{M}_1 \bar{U}^\dag$ is
diagonal and the diagonal components are large.
Thus $H_T$ becomes non-dynamical.
From Eq.(\ref{qhq}),
zero mode is obtained as
\begin{eqnarray}
 U_{q'}^{c(++)} =q(x) g_{0U}^{(++)}U_3 , \quad~
 D_q^{c(++)}=   q(x) g_{0D}^{(++)}U_4  ,
   \label{udq}
\end{eqnarray}
where $q(x)$ is SU(2)-doublet quark,
$g_{0U}^{(++)}$ and $g_{0D}^{(++)}$
denote the mode functions
given in (\ref{g0mp}) with $M_U$,
the 5D mass of $(U,U^c)$ and  $M_D$,
the 5D mass of $(D,D^c)$, respectively.

For lepton sector,
[$N_{L'}^{c\,(++)}+N_{D'}^{c\,(+-)}
$]$_\jmath\equiv N_{F'\,\jmath}^c$ and 
[$E_{L}^{c\,(++)}+E_{D}^{c\,(+-)}
$]$_\jmath\equiv E_{F\,\jmath}^c$
have the quantum number
$(5^*,-3p)$ and  $(5^*,12p)$, respectively.
The linear combinations are made as
\begin{eqnarray}
 (N_{F'}^c, E_F^c)
 &\!\!\!=\!\!\!&(H_F, F)
 \left(\begin{array}{cc}
   U_5 & U_6 \\
   U_7 & U_8 \\
 \end{array}\right)
 \label{FU8}
\end{eqnarray}
After the mass term of $H_F$ is generated,
the zero mode is obtained as
\begin{eqnarray}
 N_{L'}^{c(++)}= l(x)g_{0N}^{(++)} U_7, \quad~
 E_L^{c(++)}= l(x)g_{0E}^{(++)} U_8 ,
   \label{nel}
\end{eqnarray}
where $l(x)$ is SU(2)-doublet lepton,
$g_{0N}^{(++)}$ and $g_{0E}^{(++)}$
denote the mode functions
given in (\ref{g0mp}) with $M_N$,
the 5D mass of $(N,N^c)$
and  $M_E$,
the 5D mass of $(E,E^c)$, respectively.

For the decoupling discussed above,
we give quantum numbers for zero mode and KK mode.
The vector superfield $V$ has zero mode of
one vector and one two-component Weyl fermion.
The superfield $\Sigma_5$ has zero mode
of two Higgs multiplets which are 
made out of two complex scalars and two two-component Weyl 
fermions.
Each pair of 5D superfields 
contributes to one complex scalar and one two-component Weyl fermion.
For massless fields, group representations are 
tabulated in Table~\ref{tab:repgh}.
\begin{table}[h]
\begin{center}
 \caption{Quantum numbers for 
SU(3)$_\textrm{\scriptsize C}$$\times$SU(2)$_\textrm{\scriptsize W}$$\times$U(1)$_Y$ 
of zero mode for superfields in the model
with Higgs in the 6D gauge multiplet.
The notation for 
zero-mode quarks and leptons is also shown.
Here $\imath,\jmath=1,2,3$. \label{tab:repgh}}
\begin{tabular}{l|l|l|l} \hline\hline
 $V(x,x^5,x^6)$  & Quark and
 & $\Sigma_5(x,x^5,x^6)$    
      & Quark and \\
  $(1,3,0) \oplus(8,1,0)$ & lepton 
 & $(1,2,1/ 2) \oplus(1,2,-1/2)$ & lepton
 \\ \hline
  $U_\jmath(x,x^5)$  $(3^*,1,-2/3)$ 
  &   $u_\jmath^c(x)f_{0U\jmath}^{(++)}(x^5)$ 
  & $U_\jmath^c(x,x^5)$   $(3,2,1/6)$ 
    & $q_\imath(x)g_{0U\jmath}^{(++)}(x^5)
U_{3\imath\jmath}$  \\
$D_\jmath(x,x^5)$   $(3^*,1,1/3)$ 
     & $d_\jmath^c(x)f_{0D\jmath}^{(++)}(x^5)$  
  & $D_\jmath^c(x,x^5)$   $(3,2,1/6)$ 
    & $q_\imath(x)g_{0D\jmath}^{(++)}(x^5)
U_{4\imath\jmath}$  \\
$N_\jmath(x,x^5)$  $(1,1,0)$ 
 &    $\nu_\jmath^c(x)f_{0N\jmath}^{(++)}(x^5)$ 
  &$N_\jmath^c(x,x^5)$   $(1,2,-1/2)$ 
   & $l_\imath(x)g_{0N\jmath}^{(++)}(x^5)
U_{7\imath\jmath}$  \\
$E_\jmath(x,x^5)$  $(1,1,1)$ 
    & $e_\jmath^c(x)f_{0E\jmath}^{(++)}(x^5)$ 
  & $E_\jmath^c(x,x^5)$   $(1,2,-1/2)$ 
    & $l_\imath(x)g_{0E\jmath}^{(++)}(x^5)
U_{8\imath\jmath}$  \\ \hline\hline
\end{tabular}
\end{center}
\end{table}
The down-type quarks and charged leptons have 
different origin of multiplets.
Thus unfavorable SU(5) mass relations are avoided.

For the fields with the masses $(n-\textrm{${1\over 2}$})/R_5$ 
and $n/R_5$, representations are shown
in Table~\ref{tab:rep2}.
\begin{table}[h]
\begin{center}
  \caption{Quantum numbers
for SU(3)$_\textrm{\scriptsize C}$$\times$SU(2)$_\textrm{\scriptsize W}$$\times$U(1)$_Y$ 
of the 4D mode with the masses 
$(n-\textrm{${1\over 2}$})/R_5$ 
and $n/R_5$.
 \label{tab:rep2gh}}
 \begin{tabular}{lll}
 \hline\hline
 Superfield & Quantum number & Mass \\ \hline
 $V+\Sigma_5$ &
 $(3,2,-5/ 6)\oplus 
(3^*,2,5/ 6)\oplus (3,1,-1/3)\oplus (3^*,1,1/3)$  
& $(n-\textrm{${1\over 2}$})/R_5$
 \\
 & $(1,3,0)\oplus(8,1,0)\oplus (1,2,1/2)
 \oplus (1,2,-1/2)$ & $n/R_5$
  \\ 
 \hline\hline
 \end{tabular}
\end{center}
\end{table}
For massive mode, $V+\Sigma_5$ have 
one vector, one Dirac fermion and one complex scalar.
We omit massive modes of superfields 
shifted by 5D masses.

\subsection{Gauge coupling correction \label{ggh}}

As 5D mass and $1/R_6$ are large,
the contribution on gauge couplings are generated from
$V$ and $\Sigma_5$ as well as zero modes.
In Eqs.(\ref{alpha0}),
(\ref{al1}) and (\ref{al2}), we obtain the coefficients $b$ as
\begin{eqnarray}
  b_3&\!\!\!=\!\!\!& -3,  ~~~~ b_{(1)3}=-6, ~~~~~~b_{(2)3}=-6 ,
\\
  b_2&\!\!\!=\!\!\!&1, ~~~~~~\, b_{(1)2}=-6, ~~~~~~b_{(2)2}=-6 ,
\\
 b_1&\!\!\!=\!\!\!&\textrm{$33\over 5$}, 
  ~~~~~ b_{(1)1}=-\textrm{$54\over 5$},
  ~~~~~ b_{(2)1}=-\textrm{$6\over 5$}.
\end{eqnarray}
As in the model with Higgs fields as 5D supermultiplets,
the sum of $b_{(1)}$ and $b_{(2)}$ is independent
of gauge group.
The following relation is satisfied:
\begin{eqnarray}
 b_{(1)3}+b_{(2)3}=b_{(1)2}+b_{(2)2}=b_{(1)1}+b_{(2)1}=-12 .
\end{eqnarray}
\begin{figure}[htbp]
 \begin{center}
 \includegraphics[width=10.5cm]{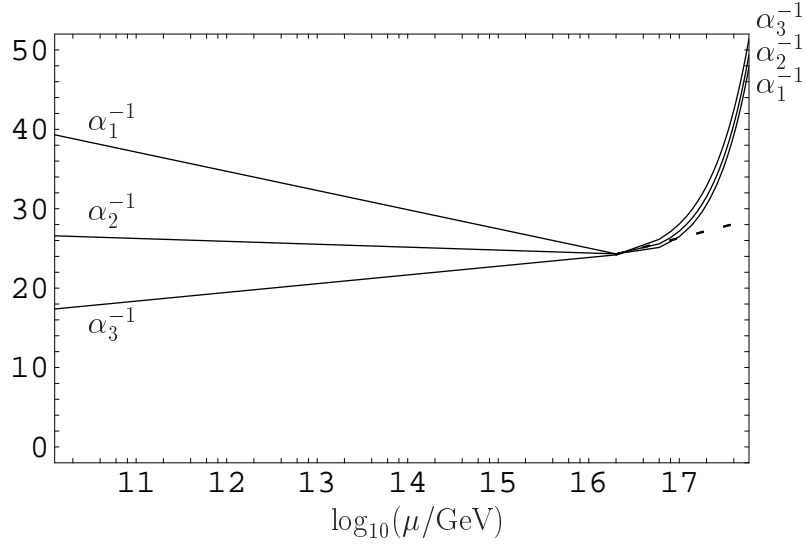} 
\caption{Energy dependence of gauge couplings. 
The input parameters are given by $\alpha^{-1}_{em}=127.906$,
$\sin^2\theta_w=0.2312$ and
$\alpha_3=0.1187$ at $M_Z=91.1876$GeV.
$M_{\textrm{\scriptsize SUSY}}=M_Z$ and
$1/(2R_5)=2\times 10^{16}$GeV are chosen.
The broken line indicates gauge couplings
in minimal 4D supersymmetric
grand unified model where 
$X,Y$ gauge bosons, two triplet Higgs fields
and their superpartners in addition to MSSM particles and 
lead to the coefficient of the $\beta$ function 
$b_1=b_2=b_3= -8$.
 \label{gcfig}}
\end{center}  
\end{figure}
The coupling constants strengthen
asymptotic freedom compared to the minimal 4D 
unified model.
This is because in the coefficients of $\beta$ function (\ref{b1aa}) 
and (\ref{b2aa}), negative contributions by the vector multiplet
dominate.

As seen from Tables~\ref{tab:rep2} and \ref{tab:rep2gh},
the quantum numbers
\begin{eqnarray}
  (3,1,-1/3)\oplus (3^*,1,1/3)
  \oplus (1,2,1/2)\oplus (1,2,-1/2) .
\end{eqnarray} 
appear commonly in all of the 4D unified model and
the models with Higgs fields in 5D and in 6D.
On the other hand, 
a way of counting in the coefficients of $\beta$ function
depends on the models.
In the 4D unified model, they are regarded as 
one complex scalar and one two-component Weyl fermion.
In counting of KK mode contributions 
in the model with 5D Higgs, they are regarded as
two complex scalar and one Dirac fermion.
In counting of KK mode contributions
in the model with Higgs fields in the 6D gauge multiplet,
they are regarded as one vector, one Dirac fermion,
one complex scalar.
These give rise to the difference of
the coefficients of $\beta$ function.

Like the model with 5D Higgs,
for gauge couplings, successful SU(5) relations
are available.
Energy dependence of gauge coupling constants are 
shown in Fig.~\ref{gcfig}, where the parameters 
are the same as that of the model with 5D Higgs.
Beyond the unification scale, 
gauge couplings become asymptotic free more rapidly
compared to the 4D unified model.
Requiring that this model to be valid
may put an upper bound to
the scale $1/R_6$.

\subsection{Yukawa couplings \label{ygh}}

As in the model with Higgs fields as 5D supermultiplets,
$R_5$, $R_6$ and $g$ are restricted also
in the model with Higgs fields in the 6D gauge multiplet.
Unlike the model with 5D Higgs, the Yukawa couplings 
are obtained from the gauge interaction
$2\textrm{Re}\left[\Phi_l^c\Sigma_5\Phi_l\right]_F$
in (\ref{actions5}) with properties 
of components shown in Table~\ref{use}.
The neutrino Yukawa interactions arise from 
$N^c\Sigma_5 N$.
For the superfield $N$, the component with the quantum number
$(1,1,0,-15p)$ has the parity $(++)$.
For $N^c$, $N^c_{l'}$ has the parity $(++)$.
Then the interaction which induces the Yukawa coupling is
\begin{eqnarray}
 N_{l'}^{c(++)}\Sigma_{5 ~(1,2,{1\over 2},18p)} N_n^{(++)} .
\end{eqnarray} 
The superfield $\Sigma_5\,{}_{(1,2,{1\over 2},18p)}$ 
has the parity $(++++)$. 
This zero mode is 
the Higgs superfield for which the vacuum expectation value
$v_2$ is developed.
For $E$, $E_{u}$ and $E_{e}$
have the parity $(++)$.
For $E^c$, $E^c_{l}$ has the parity $(++)$.
The operator ($E^{c(++)}_L E_U^{(++)}$)
has the quantum number $(3^*,2,-{7\over 6},18p)$ and 
cannot be coupled to any component of $\Sigma_5$.
The Yukawa interaction arises from the term with the component $E_e$
\begin{eqnarray}
    E_l^{c(++)}\Sigma_{5 ~ (1,2,-{1\over 2},-18p)}^{(++++)} E_e^{(++)} .
\end{eqnarray}
The superfield $\Sigma_5\,{}_{(1,2,-{1\over 2},18p)}$ is
the other component with the parity $(++++)$ among components 
of $\Sigma_5$.
This zero mode is the Higgs superfield for which the vacuum
expectation value $v_1$ is developed.
For $U^c$, the component with the parity $(++)$ is
$U_{q'}^{c\,(++)}$.
For $U$, $U_{u'}^{(++)}$
and $U_{e'}^{(++)}$ have the parity $(++)$.
The operator $(U_{q'}^{c\,(++)}U_{e'}^{(++)})$ does not have
its coupling to $\Sigma_5$.
The possible interaction is
\begin{eqnarray}
 U_{q'}^{c(++)}\Sigma_{5 ~ (1,2,{1\over 2},18p)}^{(++++)}
 U_{u'}^{(++)} .
 \label{yukawau}
\end{eqnarray}
For $D^c$ and $D$, $D_{q}^c$ and
$D_{d}$ have the parity $(++)$.
The interaction is
\begin{eqnarray}
 D_{q}^{c(++)} \Sigma_{5~ (1,2,-{1\over 2},-18p)}^{(++++)} D_d^{(++)} .
 \label{yukawad}
\end{eqnarray}

For quark sector,
the Yukawa interactions are given by
\begin{eqnarray}
 \int d^6x \delta(x^6-\pi R_6)
  \, 2\textrm{Re}\left[U_{q'\,\jmath}^c \Sigma_5 U_{u'\,\jmath} +
   D_{q\,\jmath}^c\Sigma_5 D_{d\jmath}\right]_F ,
 \label{yukawa6}
\end{eqnarray} 
where generation indices are denoted as $\jmath=1,2,3$.
The interactions are diagonal in flavor space.
From Eq.(\ref{udq}),
the fields $U_{q'}^c$ and $D_q^c$ is written in terms 
of $q$.
With the notation of zero modes given in 
Table~\ref{tab:repgh} and 
the product of mode functions 
(\ref{fgfixedp}),
Eq.(\ref{yukawa6}) leads to 
the zero mode Yukawa interaction 
\begin{eqnarray}
 &&\int d^6 x \delta(x^6-\pi R_6)
 \left(q_\imath g_{0U\jmath}^{(++)} U_{3\imath\jmath}
  \Sigma_{5(1,2,\textrm{${1\over 2}$})}
  u_\jmath^c f_{0U\jmath}^{(++)}
+ q_\imath g_{0D\jmath}^{(++)} U_{4\imath\jmath}
  \Sigma_{5(1,2,\textrm{$-{1\over 2}$})}
  d_\jmath^c f_{0D\jmath}^{(++)}
\right) 
\nonumber
\\
 &\!\!\!=\!\!\!&
 \int d^4 x 
 \left({\pi R_5 M_{U\jmath}\over 
\sinh(\pi R_5 M_{U\jmath})}
   q_\imath U_{3\,\imath\jmath}  
\Sigma_{5(1,2,\textrm{${1\over 2}$})}u_\jmath^c 
+ 
  {\pi R_5 M_{D\jmath}\over \sinh(\pi R_5 M_{D\jmath})}
   q_\imath U_{4\, \imath\jmath} 
\Sigma_{5(1,2,\textrm{$-{1\over 2}$})}  d_\jmath^c 
  \right) ,
\end{eqnarray}
with Hermitian conjugate terms added.
From this equation, quark masses at the energy scale $1/2R_5$ 
are obtained as
\begin{eqnarray}
 (\textrm{diag}(m_u, m_c, m_t))_{\jmath_1\jmath_2}
 &\!\!\!=\!\!\!& 
 g_{4D} v_2
 (V_{uL}^\dag)_{\jmath_1 \imath}
 U_{3\, \imath\jmath}
{\pi R_5 M_{U\jmath}\over \sinh(\pi R_5M_{U\jmath})}
   (V_{uR})_{\jmath \jmath_2} 
\\
  (\textrm{diag}(m_d, m_s, m_b))_{\jmath_1\jmath_2}
 &\!\!\!=\!\!\!&
   g_{4D} v_1
 (V_{dL}^\dag)_{\jmath_1 \imath}
 U_{4\, \imath\jmath}
{\pi R_5 M_{D\jmath}\over \sinh(\pi R_5M_{D\jmath})}
   (V_{dR})_{\jmath\jmath_2} .
\end{eqnarray}
where $V_{uL}$, $V_{uR}$, $V_{dL}$ and $V_{dR}$
are mixing matrices for quarks.
When the kinetic terms for Higgs fields are canonically
normalized, $g_{4D}$ appears.
The Cabibbo-Kobayashi-Maskawa matrix is obtained as
$V_{\textrm{\scriptsize CKM}}=V_{uL}^\dag V_{dL}$.
From these equations,
5D masses of fields confined on fixed lines are estimated.
For simplicity we choose
\begin{eqnarray}
 U_3 = V_{uL} V_{dL}^\dag ,\quad~
 U_4 = V_{dL} V_{dR}^{\dag} .
\end{eqnarray}
For $\textrm{tan}\beta=10$,
$\alpha_X^{-1}=24.3$ and
$v=\sqrt{2}\times 174.1$GeV,
we obtain the 5D masses of up-type and down-type quarks
for each generation as
\begin{eqnarray}
 M_{U1}= -10 ,~~
   M_{U2}=-6 ,~~ M_{U3}=-0.9 ,~~
 M_{D1}= -8 ,~~
  M_{D2}=-6 ,~~ M_{D3}=-3 , 
\end{eqnarray}
in unit of $1/(2R_5)$
by comparing data shown in Table~\ref{tab:mq}.
The mass hierarchy is generated due to 
exponential factor.
Since $|M_{U3}|<1/(2R_5)$ as well as $|M_{U\jmath}|, 
|M_{D\jmath}| \leq 1/(2R_6)$ 
for $R_6=R_5/10$,
KK modes of matter affect 
gauge coupling correction analyzed in Section~\ref{ggh}.
Since the effect of matter on $\beta$ function are
positive for $b$, 
the inclusion of the KK modes would 
weaken enhancement of $\alpha^{-1}(\mu)$.

For lepton sector
the masses and mixing angles are obtained in a parallel way.
From Eq.(\ref{nel}),
the lepton masses at the energy scale $1/(2R_5)$ are 
obtained as
\begin{eqnarray}
 (\textrm{diag}(m_{\nu_e}, m_{\nu_\mu}, 
  m_{\nu_\tau}))_{\jmath_1\jmath_2}
 &\!\!\!=\!\!\!& 
 g_{4D} v_2
 (V_{nL}^\dag)_{\jmath_1 \imath}
 U_{7\, \imath\jmath}
{\pi R_5 M_{N\jmath}\over \sinh(\pi R_5M_{N\jmath})}
   (V_{nR})_{\jmath \jmath_2} 
\\
  (\textrm{diag}(m_e, m_\mu, m_\tau))_{\jmath_1\jmath_2}
 &\!\!\!=\!\!\!&
   g_{4D} v_1
 (V_{eL}^\dag)_{\jmath_1 \imath}
 U_{8\, \imath\jmath}
{\pi R_5 M_{E\jmath}\over \sinh(\pi R_5M_{E\jmath})}
   (V_{eR})_{\jmath\jmath_2} .
\end{eqnarray}
where $V_{nL}$, $V_{nR}$, $V_{eL}$ and $V_{eR}$
are mixing matrices for quarks.
Small Dirac neutrino masses are obtained 
via exponential suppression.
A small mass such as $10^{-3}$eV is 
generated by 5D negative mass of the order of $10/R_5$.
For $\tan\beta$, $\alpha_X^{-1}$ and $v$ given above,
we obtain the 5D masses of charged leptons 
for each generation as
\begin{eqnarray}
  M_{E1}=-9, ~~ M_{E2}=-5 ,~~ M_{E3}=-3 ,
\end{eqnarray}
in unit of $1/(2R_5)$ for $U_8=V_{eL}V_{eR}^\dag$.
The mass hierarchy is generated due to 
exponential factor.
Since  $|M_{E\jmath}| \leq 1/(2R_6)$ for $R_6=R_5/10$,
these KK modes also affect 
gauge coupling correction analyzed in Section~\ref{ggh}.
Since these effects on $\beta$ function is
positive for $b$, 
the inclusion of the KK modes would 
weaken enhancement of $\alpha^{-1}(\mu)$.

\subsection{Proton stability}

As in the model with Higgs superfields
as 5D supermultiplets, 
baryon-number violating operators appear
also in the model with Higgs superfields in the 6D gauge multiplet.
As superpotentials on fixed points, 
there can be operators to generate 
dimension-five baryon number violating processes such as 
\begin{eqnarray}
  U_{T'}^c U_{T'}^c U_{T'}^c E_F^c \approx
    (T U_3)(T U_3)(T U_3)(F U_8) ,
 \label{uuue}
\end{eqnarray} 
where Eqs(\ref{tutu}) and (\ref{FU8}) have been used
and $\approx$ denotes
equivalence except for decoupled heavy fields. 
The operator (\ref{uuue})
corresponds to the first operator in Eq.(\ref{5op2})
in the model with Higgs superfields as 5D supermultiplets.
Such operators are forbidden by R invariance
with charges assigned in Table~\ref{rgh}.
\begin{table}[h]
\begin{center}
\caption{R-charges in model with Higgs in gauge multiplet.\label{rgh}}
\begin{tabular}{cccccccc} \hline\hline
  $\theta$ & $V$, $\Sigma_5$, $\Sigma_6$ & $\Phi$ 
& $D$, $D^c$, $U$, $U^c$, $E$, $E^c$, $N$, $N^c$ 
& $\bar{T}$ & $S$ & ${\cal X}$ & ${\cal B}$, $\bar{\cal B}$    \\ \hline
 $-1$ & 0 & 2 & 1 & 1 & 0 & 2 & 0   \\ \hline
\end{tabular} 
\end{center}
\end{table}
In addition, 
the dimension-six proton decay by 
colored Higgs exchange is absent
unlike the model with Higgs fields as 5D chiral superfields,
while the dimension-six proton decay by $X$ and $Y$ boson 
exchange would exist.

\subsection{Supersymmetry breaking}

The two MSSM Higgs 
fields which are part of the 6D gauge multiplet
propagate in 6D.
One may expect that the Higgs superfields
would be directly coupled to supersymmetry breaking source as
\begin{eqnarray}
 && 2\textrm{Re}\left[
 {1\over M}S W^\alpha W_\alpha \right]_F
\nonumber
\\
 && +{1\over 2}\left[
  \left(\left({1\over M}S^\dag + {1\over M^2} S^\dag S\right)
 H_1 H_2 + \textrm{H.c.}\right)
 +{1\over M^2} S^\dag S
  \left( H_1^\dag H_1 +H_2^\dag H_2
\right) \right]_D ,
 \label{higgsino}
\end{eqnarray}
at $(x^5,x^6)=(\pi R_5,0)$.
Such couplings appeared
in gaugino mediation with the standard model gauge fields and 
Higgs fields and their superpartners in bulk~\cite{Chacko:1999mi}.
If the coupling
\begin{eqnarray}
S^\dag S \textrm{Tr}(\Sigma_5^\dag \Sigma_5)
=S^\dag S (H_1^\dag H_1+H_2^\dag H_2) +\cdots ,
\end{eqnarray}
is allowed, the coupling seems to lead to Higgs mass terms.
At the fixed point $(\pi R_5,0)$, 
restricted gauge symmetry has SU(4)$\times$ SU(2)$\times$U(1). 
The field $\Sigma_5$ is written as
\begin{eqnarray}
 \Sigma_5\bigg|_{(\pi R_5,0)}=\left(\begin{array}{c|c}
  \Sigma_{2\cdot 2}& \Sigma_{2\cdot 4} \\ \hline
   \Sigma_{4\cdot 2}& \Sigma_{4\cdot 4} \\
 \end{array}\right) .
 \label{sig42}
\end{eqnarray}
If the coupling
\begin{eqnarray}
S^\dag \textrm{Tr}(\Sigma_{2\cdot 4}\Sigma_{4\cdot 2}) 
 =S^\dag H_1 H_2 +\cdots 
\end{eqnarray}
is allowed, 
this seems $\mu$ term.
However, the operators $\textrm{Tr}(\Sigma_5^\dag \Sigma_5)$ and 
$\textrm{Tr}(\Sigma_{2\times 4} \Sigma_{4\times 2})$ are 
gauge non-invariant
due to $\partial_5$-factor of the transformation (\ref{sig5})
\begin{eqnarray}
 \Sigma_5\to 
 \Sigma_5'=e^{-\Lambda}(\partial_5+\Sigma_5)e^{\Lambda} .
\end{eqnarray}
Thus in this model, Higgs masses and $\mu$ and $B\mu$ terms
are not generated by contact interactions.
At $\mu=M_X$ the model has
nonzero gaugino mass and negligible soft scalar mass.

For electroweak symmetry breaking,
the quartic terms of Higgs doublet
appear from $D$-term.
In the operator (\ref{bosop}), they are included in the term
$-({\cal D}^N\varphi)^*({\cal D}_N\varphi)$. 
Since there is no term such as Eq.(\ref{higgsino}),
$\mu$ 
term must be generated via
other mechanism.
In case with Chern-Simons terms,
it was proposed that combining $F$-components of 
compensator and radion as well as constant superpotentials
can generate $\mu$ term 
\cite{Hebecker:2008rk}.
In this mechanism, 
nontrivial radion potential is crucial.
It has also been shown that
a potential of radion with radius stabilization 
is generated in a model with
compensator, radion and constant 
superpotentials~\cite{Maru:2006id}\cite{Maru:2006ji}
and that this radius stabilization is sensitive to field 
content~\cite{Uekusa:2007sw}.  
Whether a radion potential relevant to 
generation of $\mu$ term arises in the present models 
is an open problem.

\section{Conclusion}
We have studied
gauge coupling corrections and Yukawa coupling constants
in 6D unified models based on the gauge group SU(6).
For a possibility of gauge-Higgs unification and inclusion of right-handed neutrino,
gauge group larger than the standard model gauge group has been motivated.
We have employed two extra dimensions to 
approach the doublet-triplet splitting,
no fermion mass relations for the first two generations and 
gaugino-mediated supersymmetry breaking (no supersymmetry flavor problem).
The source of supersymmetry breaking is spatially separated from 
fixed lines and fixed points where matter fields reside.
By orbifold, theory on fixed lines and fixed points
has restricted gauge symmetry.
Heavy fields decouple by boundary conditions.
The models are 6D and 4D anomaly free and
proton stability is achieved by R invariance
as a subgroup of SU(2)$_\textrm{\scriptsize R}$.

For energy dependence of 
gauge coupling constants, we have adopted 4D KK effective theory
by giving a 6D component action and the fixed-line and fixed-point actions.
The number of KK mode is regarded as energy-dependent.
At each scale $1/(2R_5)$, $1/R_5$, $3/(2R_5)$, $\cdots$, 
4D KK states with the corresponding masses become dynamical.
For the effective 4D theory, restricted gauge invariance has
the gauge group SU(3)$_\textrm{\scriptsize C}$$\times$
SU(2)$_\textrm{\scriptsize W}$$\times$U(1)$_Y$$\times$U(1)${}'$.
The gauge group U(1)$'$ is spontaneously broken by fixed-point couplings.
Up to the unification scale, zero mode contributes to renormalization
group evolution.
Since zero mode is right-handed neutrino and its superpartner
as well as the field content of MSSM,
successful SU(5) relations about the
gauge coupling unification, weak mixing angle and charge quantization
have been achieved.
Beyond the unification scale, gauge couplings are sensitive to the field content. 
In the model with 5D Higgs, 
the sum of the coefficients of $\beta$ function relevant to 
KK mode is
$b_{(1)}+b_{(2)}=4$
independently of gauge group.
This leads to strong coupling at high energies.
In the model with the Higgs superfields in the
6D gauge multiplet,
$b_{(1)}+b_{(2)}=-12$.
Asymptotic freedom is more strong compared to the 4D unified model.
These rapid energy dependence is likely to constrain the scale $1/R_6$.
In the two models, it may be chosen as $1/R_6<M_{Pl}$.
For $1/(2R_5)=M_X$, $1/R_6<M_{Pl}$ and $\alpha_X^{-1}=24.3$, 
realistic patterns of Yukawa couplings have been obtained
from fixed-point coupling in the model with 5D Higgs fields
and from the gauge interaction in the model with Higgs fields in
the 6D gauge multiplet.
For the values of paramters, small numbers required for the hierarchy of 
Yukawa coupling constants have been generated 
via volume suppression in the model with 5D Higgs fields
and exponential suppression dependent on 5D masses 
in the model with Higgs fields in the gauge multiplet.

We have fixed the smaller compactification radius $1/(2R_5)=M_X$.
From a viewpoint of LSP,
it may be important 
to examine the case $1/(2R_5)\neq M_X$
in a 6D setup where no fermion mass relations for the first two generations
and gaugino-mediated supersymmetry breaking still hold.
Gaugino masses as an input at high energies 
influence the whole values of supersymmetry breaking masses.
For a large $M_{1/2}$, spectrum avoiding the lightest stau as 
LSP was obtained
for a moderate $\tan\beta$ and 
for $1/(2R_5)\neq M_X$~\cite{Schmaltz:2000ei}.
In models with the two extra dimensions 
radiative correction
may also affect spectrum \cite{Azatov:2007fa}. 
There is still room to be examined about this point.

\vspace{4ex}

\subsubsection*{Acknowledgments}

I thank Masud Chaichian,
Nobuhito Maru and Norisuke Sakai
for useful discussions.
This work is supported by Bilateral exchange program between 
Japan Society for the Promotion of Science and the Academy of Finland.

\newpage

\begin{appendix}

\section{SU(6) group and 6D spinor}
\subsection{SU(6) generators and structure constants \label{lam}}

The generators of SU(6) transformations, $\lambda_a$ ($a=1,\cdots 35$)
are 
\begin{eqnarray*}
&&\!\!\!\!\!\!\!\!
 \lambda_1\!\!=\!\!\left(\!
  \begin{array}{cccccc}
   {}^\cdot_1\!&\! {}^1_\cdot\!&\! {}^\cdot_\cdot\!&\!
    {}^\cdot_\cdot\!&\! {}^\cdot_\cdot\!&\! {}^\cdot_\cdot\\ 
   {}^\cdot_\cdot\!&\! {}^\cdot_\cdot\!&\! {}^\cdot_\cdot\!&\!
    {}^\cdot_\cdot\!&\! {}^\cdot_\cdot\!&\! {}^\cdot_\cdot\\ 
   {}^\cdot_\cdot\!&\! {}^\cdot_\cdot\!&\! {}^\cdot_\cdot\!&\!
    {}^\cdot_\cdot\!&\! {}^\cdot_\cdot\!&\! {}^\cdot_\cdot\\ 
  \end{array}\!\right)\!\! ,~
 \lambda_2\!\!=\!\!\left(\!
  \begin{array}{cccccc}
   {}^\cdot_{i}\!&\! {}^{-i}_{~\cdot}\!&\! {}^\cdot_\cdot\!&\!
    {}^\cdot_\cdot\!&\! {}^\cdot_\cdot\!&\! {}^\cdot_\cdot\\ 
   {}^\cdot_\cdot\!&\! {}^\cdot_\cdot\!&\! {}^\cdot_\cdot\!&\!
    {}^\cdot_\cdot\!&\! {}^\cdot_\cdot\!&\! {}^\cdot_\cdot\\ 
   {}^\cdot_\cdot\!&\! {}^\cdot_\cdot\!&\! {}^\cdot_\cdot\!&\!
    {}^\cdot_\cdot\!&\! {}^\cdot_\cdot\!&\! {}^\cdot_\cdot\\ 
  \end{array}\!\right)\!\! ,~
 \lambda_3\!\!=\!\!\left(\!
  \begin{array}{cccccc}
   {}^1_\cdot\!&\! {}^{~\cdot}_{-1}\!&\! {}^\cdot_\cdot\!&\!
    {}^\cdot_\cdot\!&\! {}^\cdot_\cdot\!&\! {}^\cdot_\cdot\\ 
   {}^\cdot_\cdot\!&\! {}^\cdot_\cdot\!&\! {}^\cdot_\cdot\!&\!
    {}^\cdot_\cdot\!&\! {}^\cdot_\cdot\!&\! {}^\cdot_\cdot\\ 
   {}^\cdot_\cdot\!&\! {}^\cdot_\cdot\!&\! {}^\cdot_\cdot\!&\!
    {}^\cdot_\cdot\!&\! {}^\cdot_\cdot\!&\! {}^\cdot_\cdot\\ 
  \end{array}\!\right)\!\! ,~
 \lambda_4\!\!=\!\!\left(\!
  \begin{array}{cccccc}
   {}^\cdot_\cdot\!&\! {}^\cdot_\cdot\!&\! {}^1_\cdot\!&\!
    {}^\cdot_\cdot\!&\! {}^\cdot_\cdot\!&\! {}^\cdot_\cdot\\ 
   {}^1_\cdot\!&\! {}^\cdot_\cdot\!&\! {}^\cdot_\cdot\!&\!
    {}^\cdot_\cdot\!&\! {}^\cdot_\cdot\!&\! {}^\cdot_\cdot\\ 
   {}^\cdot_\cdot\!&\! {}^\cdot_\cdot\!&\! {}^\cdot_\cdot\!&\!
    {}^\cdot_\cdot\!&\! {}^\cdot_\cdot\!&\! {}^\cdot_\cdot\\ 
  \end{array}\!\right)\!\! ,~
\\ &&\!\!\!\!\!\!\!\!
 \lambda_5\!\!=\!\!\left(\!
  \begin{array}{cccccc}
   {}^\cdot_\cdot\!&\! {}^\cdot_\cdot\!&\! {}^{-i}_{~\cdot}\!&\!
    {}^\cdot_\cdot\!&\! {}^\cdot_\cdot\!&\! {}^\cdot_\cdot\\ 
   {}^i_\cdot\!&\! {}^\cdot_\cdot\!&\! {}^\cdot_\cdot\!&\!
    {}^\cdot_\cdot\!&\! {}^\cdot_\cdot\!&\! {}^\cdot_\cdot\\ 
   {}^\cdot_\cdot\!&\! {}^\cdot_\cdot\!&\! {}^\cdot_\cdot\!&\!
    {}^\cdot_\cdot\!&\! {}^\cdot_\cdot\!&\! {}^\cdot_\cdot\\ 
  \end{array}\!\right)\!\! ,~~
 \lambda_6\!\!=\!\!\left(\!
  \begin{array}{cccccc}
   {}^\cdot_\cdot\!&\! {}^\cdot_\cdot\!&\! {}^\cdot_1\!&\!
    {}^\cdot_\cdot\!&\! {}^\cdot_\cdot\!&\! {}^\cdot_\cdot\\ 
   {}^\cdot_\cdot\!&\! {}^1_\cdot\!&\! {}^\cdot_\cdot\!&\!
    {}^\cdot_\cdot\!&\! {}^\cdot_\cdot\!&\! {}^\cdot_\cdot\\ 
   {}^\cdot_\cdot\!&\! {}^\cdot_\cdot\!&\! {}^\cdot_\cdot\!&\!
    {}^\cdot_\cdot\!&\! {}^\cdot_\cdot\!&\! {}^\cdot_\cdot\\ 
  \end{array}\!\right)\!\! ,~~
 \lambda_7\!\!=\!\!\left(\!
  \begin{array}{cccccc}
   {}^\cdot_\cdot\!&\! {}^\cdot_\cdot\!&\! {}^{~\cdot}_{-i}\!&\!
    {}^\cdot_\cdot\!&\! {}^\cdot_\cdot\!&\! {}^\cdot_\cdot\\ 
   {}^\cdot_\cdot\!&\! {}^i_\cdot\!&\! {}^\cdot_\cdot\!&\!
    {}^\cdot_\cdot\!&\! {}^\cdot_\cdot\!&\! {}^\cdot_\cdot\\ 
   {}^\cdot_\cdot\!&\! {}^\cdot_\cdot\!&\! {}^\cdot_\cdot\!&\!
    {}^\cdot_\cdot\!&\! {}^\cdot_\cdot\!&\! {}^\cdot_\cdot\\ 
  \end{array}\!\right)\!\! ,~~
\\ &&\!\!\!\!\!\!\!\!
 \lambda_8\!\!=\!\!\textrm{${1\over \sqrt{3}}$}\left(\!
  \begin{array}{cccccc}
   {}^1_\cdot\!&\! {}^\cdot_1\!&\! {}^\cdot_\cdot\!&\!
    {}^\cdot_\cdot\!&\! {}^\cdot_\cdot\!&\! {}^\cdot_\cdot\\ 
   {}^\cdot_\cdot\!&\! {}^\cdot_\cdot\!&\! {}^{-2}_{~\cdot}\!&\!
    {}^\cdot_\cdot\!&\! {}^\cdot_\cdot\!&\! {}^\cdot_\cdot\\ 
   {}^\cdot_\cdot\!&\! {}^\cdot_\cdot\!&\! {}^\cdot_\cdot\!&\!
    {}^\cdot_\cdot\!&\! {}^\cdot_\cdot\!&\! {}^\cdot_\cdot\\ 
  \end{array}\!\right)\!\! ,~~
 \lambda_9\!\!=\!\!\left(\!
  \begin{array}{cccccc}
   {}^\cdot_\cdot\!&\! {}^\cdot_\cdot\!&\! {}^\cdot_\cdot\!&\!
    {}^1_\cdot\!&\! {}^\cdot_\cdot\!&\! {}^\cdot_\cdot\\ 
   {}^\cdot_1\!&\! {}^\cdot_\cdot\!&\! {}^\cdot_\cdot\!&\!
    {}^\cdot_\cdot\!&\! {}^\cdot_\cdot\!&\! {}^\cdot_\cdot\\ 
   {}^\cdot_\cdot\!&\! {}^\cdot_\cdot\!&\! {}^\cdot_\cdot\!&\!
    {}^\cdot_\cdot\!&\! {}^\cdot_\cdot\!&\! {}^\cdot_\cdot\\ 
  \end{array}\!\right)\!\! ,~~
 \lambda_{10}\!\!=\!\!\left(\!
  \begin{array}{cccccc}
   {}^\cdot_\cdot\!&\! {}^\cdot_\cdot\!&\! {}^\cdot_\cdot\!&\!
    {}^{-i}_{~\cdot}\!&\! {}^\cdot_\cdot\!&\! {}^\cdot_\cdot\\ 
   {}^\cdot_{i}\!&\! {}^\cdot_\cdot\!&\! {}^\cdot_\cdot\!&\!
    {}^\cdot_\cdot\!&\! {}^\cdot_\cdot\!&\! {}^\cdot_\cdot\\ 
   {}^\cdot_\cdot\!&\! {}^\cdot_\cdot\!&\! {}^\cdot_\cdot\!&\!
    {}^\cdot_\cdot\!&\! {}^\cdot_\cdot\!&\! {}^\cdot_\cdot\\ 
  \end{array}\!\right)\!\! ,~~
\\ &&\!\!\!\!\!\!\!\!
 \lambda_{11}\!\!=\!\!\left(\!
  \begin{array}{cccccc}
   {}^\cdot_\cdot\!&\! {}^\cdot_\cdot\!&\! {}^\cdot_\cdot\!&\!
    {}^\cdot_1\!&\! {}^\cdot_\cdot\!&\! {}^\cdot_\cdot\\ 
   {}^\cdot_\cdot\!&\! {}^\cdot_1\!&\! {}^\cdot_\cdot\!&\!
    {}^\cdot_\cdot\!&\! {}^\cdot_\cdot\!&\! {}^\cdot_\cdot\\ 
   {}^\cdot_\cdot\!&\! {}^\cdot_\cdot\!&\! {}^\cdot_\cdot\!&\!
    {}^\cdot_\cdot\!&\! {}^\cdot_\cdot\!&\! {}^\cdot_\cdot\\ 
  \end{array}\!\right)\!\! ,~
 \lambda_{12}\!\!=\!\!\left(\!
  \begin{array}{cccccc}
   {}^\cdot_\cdot\!&\! {}^\cdot_\cdot\!&\! {}^\cdot_\cdot\!&\!
    {}^{~\cdot}_{-i}\!&\! {}^\cdot_\cdot\!&\! {}^\cdot_\cdot\\ 
   {}^\cdot_\cdot\!&\! {}^\cdot_i\!&\! {}^\cdot_\cdot\!&\!
    {}^\cdot_\cdot\!&\! {}^\cdot_\cdot\!&\! {}^\cdot_\cdot\\ 
   {}^\cdot_\cdot\!&\! {}^\cdot_\cdot\!&\! {}^\cdot_\cdot\!&\!
    {}^\cdot_\cdot\!&\! {}^\cdot_\cdot\!&\! {}^\cdot_\cdot\\ 
  \end{array}\!\right)\!\! ,~
 \lambda_{13}\!\!=\!\!\left(\!
  \begin{array}{cccccc}
   {}^\cdot_\cdot\!&\! {}^\cdot_\cdot\!&\! {}^\cdot_\cdot\!&\!
    {}^\cdot_\cdot\!&\! {}^\cdot_\cdot\!&\! {}^\cdot_\cdot\\ 
   {}^\cdot_\cdot\!&\! {}^\cdot_\cdot\!&\! {}^\cdot_1\!&\!
    {}^1_\cdot\!&\! {}^\cdot_\cdot\!&\! {}^\cdot_\cdot\\ 
   {}^\cdot_\cdot\!&\! {}^\cdot_\cdot\!&\! {}^\cdot_\cdot\!&\!
    {}^\cdot_\cdot\!&\! {}^\cdot_\cdot\!&\! {}^\cdot_\cdot\\ 
  \end{array}\!\right)\!\! ,~
 \lambda_{14}\!\!=\!\!\left(\!
  \begin{array}{cccccc}
   {}^\cdot_\cdot\!&\! {}^\cdot_\cdot\!&\! {}^\cdot_\cdot\!&\!
    {}^\cdot_\cdot\!&\! {}^\cdot_\cdot\!&\! {}^\cdot_\cdot\\ 
   {}^\cdot_\cdot\!&\! {}^\cdot_\cdot\!&\! {}^\cdot_i \!&\!
    {}^{-i}_{~\cdot}\!&\! {}^\cdot_\cdot\!&\! {}^\cdot_\cdot\\ 
   {}^\cdot_\cdot\!&\! {}^\cdot_\cdot\!&\! {}^\cdot_\cdot\!&\!
    {}^\cdot_\cdot\!&\! {}^\cdot_\cdot\!&\! {}^\cdot_\cdot\\ 
  \end{array}\!\right)\!\! ,~
\\ &&\!\!\!\!\!\!\!\!
 \lambda_{15}\!\!=\!\!\textrm{${1\over \sqrt{6}}$}\left(\!
  \begin{array}{cccccc}
   {}^1_\cdot\!&\! {}^\cdot_1\!&\! {}^\cdot_\cdot\!&\!
    {}^\cdot_\cdot\!&\! {}^\cdot_\cdot\!&\! {}^\cdot_\cdot\\ 
   {}^\cdot_\cdot\!&\! {}^\cdot_\cdot\!&\! {}^1_\cdot\!&\!
    {}^{~\cdot}_{-3}\!&\! {}^\cdot_\cdot\!&\! {}^\cdot_\cdot\\ 
   {}^\cdot_\cdot\!&\! {}^\cdot_\cdot\!&\! {}^\cdot_\cdot\!&\!
    {}^\cdot_\cdot\!&\! {}^\cdot_\cdot\!&\! {}^\cdot_\cdot\\ 
  \end{array}\!\right)\!\! ,~~
 \lambda_{16}\!\!=\!\!\left(\!
  \begin{array}{cccccc}
   {}^\cdot_\cdot\!&\! {}^\cdot_\cdot\!&\! {}^\cdot_\cdot\!&\!
    {}^\cdot_\cdot\!&\! {}^1_\cdot\!&\! {}^\cdot_\cdot\\ 
   {}^\cdot_\cdot\!&\! {}^\cdot_\cdot\!&\! {}^\cdot_\cdot\!&\!
    {}^\cdot_\cdot\!&\! {}^\cdot_\cdot\!&\! {}^\cdot_\cdot\\ 
   {}^1_\cdot\!&\! {}^\cdot_\cdot\!&\! {}^\cdot_\cdot\!&\!
    {}^\cdot_\cdot\!&\! {}^\cdot_\cdot\!&\! {}^\cdot_\cdot\\ 
  \end{array}\!\right)\!\! ,~~
 \lambda_{17}\!\!=\!\!\left(\!
  \begin{array}{cccccc}
   {}^\cdot_\cdot\!&\! {}^\cdot_\cdot\!&\! {}^\cdot_\cdot\!&\!
    {}^\cdot_\cdot\!&\! {}^{-i}_{~\cdot}\!&\! {}^\cdot_\cdot\\ 
   {}^\cdot_\cdot\!&\! {}^\cdot_\cdot\!&\! {}^\cdot_\cdot\!&\!
    {}^\cdot_\cdot\!&\! {}^\cdot_\cdot\!&\! {}^\cdot_\cdot\\ 
   {}^i_\cdot\!&\! {}^\cdot_\cdot\!&\! {}^\cdot_\cdot\!&\!
    {}^\cdot_\cdot\!&\! {}^\cdot_\cdot\!&\! {}^\cdot_\cdot\\ 
  \end{array}\!\right)\!\! ,~~
\\ &&\!\!\!\!\!\!\!\!
 \lambda_{18}\!\!=\!\!\left(\!
  \begin{array}{cccccc}
   {}^\cdot_\cdot\!&\! {}^\cdot_\cdot\!&\! {}^\cdot_\cdot\!&\!
    {}^\cdot_\cdot\!&\! {}^\cdot_1\!&\! {}^\cdot_\cdot\\ 
   {}^\cdot_\cdot\!&\! {}^\cdot_\cdot\!&\! {}^\cdot_\cdot\!&\!
    {}^\cdot_\cdot\!&\! {}^\cdot_\cdot\!&\! {}^\cdot_\cdot\\ 
   {}^\cdot_\cdot\!&\! {}^1_\cdot\!&\! {}^\cdot_\cdot\!&\!
    {}^\cdot_\cdot\!&\! {}^\cdot_\cdot\!&\! {}^\cdot_\cdot\\ 
  \end{array}\!\right)\!\! ,~
 \lambda_{19}\!\!=\!\!\left(\!
  \begin{array}{cccccc}
   {}^\cdot_\cdot\!&\! {}^\cdot_\cdot\!&\! {}^\cdot_\cdot\!&\!
    {}^\cdot_\cdot\!&\! {}^{~\cdot}_{-i}\!&\! {}^\cdot_\cdot\\ 
   {}^\cdot_\cdot\!&\! {}^\cdot_\cdot\!&\! {}^\cdot_\cdot\!&\!
    {}^\cdot_\cdot\!&\! {}^\cdot_\cdot\!&\! {}^\cdot_\cdot\\ 
   {}^\cdot_\cdot\!&\! {}^i_\cdot\!&\! {}^\cdot_\cdot\!&\!
    {}^\cdot_\cdot\!&\! {}^\cdot_\cdot\!&\! {}^\cdot_\cdot\\ 
  \end{array}\!\right)\!\! ,~
 \lambda_{20}\!\!=\!\!\left(\!
  \begin{array}{cccccc}
   {}^\cdot_\cdot\!&\! {}^\cdot_\cdot\!&\! {}^\cdot_\cdot\!&\!
    {}^\cdot_\cdot\!&\! {}^\cdot_\cdot\!&\! {}^\cdot_\cdot\\ 
   {}^\cdot_\cdot\!&\! {}^\cdot_\cdot\!&\! {}^\cdot_\cdot\!&\!
    {}^\cdot_\cdot\!&\! {}^1_\cdot\!&\! {}^\cdot_\cdot\\ 
   {}^\cdot_\cdot\!&\! {}^\cdot_\cdot\!&\! {}^1_\cdot\!&\!
    {}^\cdot_\cdot\!&\! {}^\cdot_\cdot\!&\! {}^\cdot_\cdot\\ 
  \end{array}\!\right)\!\! ,~
 \lambda_{21}\!\!=\!\!\left(\!
  \begin{array}{cccccc}
   {}^\cdot_\cdot\!&\! {}^\cdot_\cdot\!&\! {}^\cdot_\cdot\!&\!
    {}^\cdot_\cdot\!&\! {}^\cdot_\cdot\!&\! {}^\cdot_\cdot\\ 
   {}^\cdot_\cdot\!&\! {}^\cdot_\cdot\!&\! {}^\cdot_\cdot\!&\!
    {}^\cdot_\cdot\!&\! {}^{-i}_{~\cdot}\!&\! {}^\cdot_\cdot\\ 
   {}^\cdot_\cdot\!&\! {}^\cdot_\cdot\!&\! {}^i_\cdot\!&\!
    {}^\cdot_\cdot\!&\! {}^\cdot_\cdot\!&\! {}^\cdot_\cdot\\ 
  \end{array}\!\right)\!\! ,~
\\ &&\!\!\!\!\!\!\!\!
 \lambda_{22}\!\!=\!\!\left(\!
  \begin{array}{cccccc}
   {}^\cdot_\cdot\!&\! {}^\cdot_\cdot\!&\! {}^\cdot_\cdot\!&\!
    {}^\cdot_\cdot\!&\! {}^\cdot_\cdot\!&\! {}^\cdot_\cdot\\ 
   {}^\cdot_\cdot\!&\! {}^\cdot_\cdot\!&\! {}^\cdot_\cdot\!&\!
    {}^\cdot_\cdot\!&\! {}^\cdot_1\!&\! {}^\cdot_\cdot\\ 
   {}^\cdot_\cdot\!&\! {}^\cdot_\cdot\!&\! {}^\cdot_\cdot\!&\!
    {}^1_\cdot\!&\! {}^\cdot_\cdot\!&\! {}^\cdot_\cdot\\ 
  \end{array}\!\right)\!\! ,~~
 \lambda_{23}\!\!=\!\!\left(\!
  \begin{array}{cccccc}
   {}^\cdot_\cdot\!&\! {}^\cdot_\cdot\!&\! {}^\cdot_\cdot\!&\!
    {}^\cdot_\cdot\!&\! {}^\cdot_\cdot\!&\! {}^\cdot_\cdot\\ 
   {}^\cdot_\cdot\!&\! {}^\cdot_\cdot\!&\! {}^\cdot_\cdot\!&\!
    {}^\cdot_\cdot\!&\! {}^{~\cdot}_{-i}\!&\! {}^\cdot_\cdot\\ 
   {}^\cdot_\cdot\!&\! {}^\cdot_\cdot\!&\! {}^\cdot_\cdot\!&\!
    {}^i_\cdot\!&\! {}^\cdot_\cdot\!&\! {}^\cdot_\cdot\\ 
  \end{array}\!\right)\!\! ,~~
 \lambda_{24}\!\!=\!\!\textrm{${1\over \sqrt{10}}$}\left(\!
  \begin{array}{cccccc}
   {}^1_\cdot\!&\! {}^\cdot_1\!&\! {}^\cdot_\cdot\!&\!
    {}^\cdot_\cdot\!&\! {}^\cdot_\cdot\!&\! {}^\cdot_\cdot\\ 
   {}^\cdot_\cdot\!&\! {}^\cdot_\cdot\!&\! {}^1_\cdot\!&\!
    {}^\cdot_1\!&\! {}^\cdot_\cdot\!&\! {}^\cdot_\cdot\\ 
   {}^\cdot_\cdot\!&\! {}^\cdot_\cdot\!&\! {}^\cdot_\cdot\!&\!
    {}^\cdot_\cdot\!&\! {}^{-4}_{~\cdot}\!&\! {}^\cdot_\cdot\\ 
  \end{array}\!\right)\!\! ,~~
\\ &&\!\!\!\!\!\!\!\!
 \lambda_{25}\!\!=\!\!\left(\!
  \begin{array}{cccccc}
   {}^\cdot_\cdot\!&\! {}^\cdot_\cdot\!&\! {}^\cdot_\cdot\!&\!
    {}^\cdot_\cdot\!&\! {}^\cdot_\cdot\!&\! {}^1_\cdot\\ 
   {}^\cdot_\cdot\!&\! {}^\cdot_\cdot\!&\! {}^\cdot_\cdot\!&\!
    {}^\cdot_\cdot\!&\! {}^\cdot_\cdot\!&\! {}^\cdot_\cdot\\ 
   {}^\cdot_1\!&\! {}^\cdot_\cdot\!&\! {}^\cdot_\cdot\!&\!
    {}^\cdot_\cdot\!&\! {}^\cdot_\cdot\!&\! {}^\cdot_\cdot\\ 
  \end{array}\!\right)\!\! ,~
 \lambda_{26}\!\!=\!\!\left(\!
  \begin{array}{cccccc}
   {}^\cdot_\cdot\!&\! {}^\cdot_\cdot\!&\! {}^\cdot_\cdot\!&\!
    {}^\cdot_\cdot\!&\! {}^\cdot_\cdot\!&\! {}^{-i}_{~\cdot}\\ 
   {}^\cdot_\cdot\!&\! {}^\cdot_\cdot\!&\! {}^\cdot_\cdot\!&\!
    {}^\cdot_\cdot\!&\! {}^\cdot_\cdot\!&\! {}^\cdot_\cdot\\ 
   {}^\cdot_i\!&\! {}^\cdot_\cdot\!&\! {}^\cdot_\cdot\!&\!
    {}^\cdot_\cdot\!&\! {}^\cdot_\cdot\!&\! {}^\cdot_\cdot\\ 
  \end{array}\!\right)\!\! ,~
 \lambda_{27}\!\!=\!\!\left(\!
  \begin{array}{cccccc}
   {}^\cdot_\cdot\!&\! {}^\cdot_\cdot\!&\! {}^\cdot_\cdot\!&\!
    {}^\cdot_\cdot\!&\! {}^\cdot_\cdot\!&\! {}^\cdot_1\\ 
   {}^\cdot_\cdot\!&\! {}^\cdot_\cdot\!&\! {}^\cdot_\cdot\!&\!
    {}^\cdot_\cdot\!&\! {}^\cdot_\cdot\!&\! {}^\cdot_\cdot\\ 
   {}^\cdot_\cdot\!&\! {}^\cdot_1\!&\! {}^\cdot_\cdot\!&\!
    {}^\cdot_\cdot\!&\! {}^\cdot_\cdot\!&\! {}^\cdot_\cdot\\ 
  \end{array}\!\right)\!\! ,~~
 \lambda_{28}\!\!=\!\!\left(\!
  \begin{array}{cccccc}
   {}^\cdot_\cdot\!&\! {}^\cdot_\cdot\!&\! {}^\cdot_\cdot\!&\!
    {}^\cdot_\cdot\!&\! {}^\cdot_\cdot\!&\! {}^{~\cdot}_{-i}\\ 
   {}^\cdot_\cdot\!&\! {}^\cdot_\cdot\!&\! {}^\cdot_\cdot\!&\!
    {}^\cdot_\cdot\!&\! {}^\cdot_\cdot\!&\! {}^\cdot_\cdot\\ 
   {}^\cdot_\cdot\!&\! {}^\cdot_i\!&\! {}^\cdot_\cdot\!&\!
    {}^\cdot_\cdot\!&\! {}^\cdot_\cdot\!&\! {}^\cdot_\cdot\\ 
  \end{array}\!\right)\!\! ,~~
\\ &&\!\!\!\!\!\!\!\!
 \lambda_{29}\!\!=\!\!\left(\!
  \begin{array}{cccccc}
   {}^\cdot_\cdot\!&\! {}^\cdot_\cdot\!&\! {}^\cdot_\cdot\!&\!
    {}^\cdot_\cdot\!&\! {}^\cdot_\cdot\!&\! {}^\cdot_\cdot\\ 
   {}^\cdot_\cdot\!&\! {}^\cdot_\cdot\!&\! {}^\cdot_\cdot\!&\!
    {}^\cdot_\cdot\!&\! {}^\cdot_\cdot\!&\! {}^1_\cdot\\ 
   {}^\cdot_\cdot\!&\! {}^\cdot_\cdot\!&\! {}^\cdot_1\!&\!
    {}^\cdot_\cdot\!&\! {}^\cdot_\cdot\!&\! {}^\cdot_\cdot\\ 
  \end{array}\!\right)\!\! ,~
 \lambda_{30}\!\!=\!\!\left(\!
  \begin{array}{cccccc}
   {}^\cdot_\cdot\!&\! {}^\cdot_\cdot\!&\! {}^\cdot_\cdot\!&\!
    {}^\cdot_\cdot\!&\! {}^\cdot_\cdot\!&\! {}^\cdot_\cdot\\ 
   {}^\cdot_\cdot\!&\! {}^\cdot_\cdot\!&\! {}^\cdot_\cdot\!&\!
    {}^\cdot_\cdot\!&\! {}^\cdot_\cdot\!&\! {}^{-i}_{~\cdot}\\ 
   {}^\cdot_\cdot\!&\! {}^\cdot_\cdot\!&\! {}^\cdot_i\!&\!
    {}^\cdot_\cdot\!&\! {}^\cdot_\cdot\!&\! {}^\cdot_\cdot\\ 
  \end{array}\!\right)\!\! ,~
 \lambda_{31}\!\!=\!\!\left(\!
  \begin{array}{cccccc}
   {}^\cdot_\cdot\!&\! {}^\cdot_\cdot\!&\! {}^\cdot_\cdot\!&\!
    {}^\cdot_\cdot\!&\! {}^\cdot_\cdot\!&\! {}^\cdot_\cdot\\ 
   {}^\cdot_\cdot\!&\! {}^\cdot_\cdot\!&\! {}^\cdot_\cdot\!&\!
    {}^\cdot_\cdot\!&\! {}^\cdot_\cdot\!&\! {}^\cdot_1\\ 
   {}^\cdot_\cdot\!&\! {}^\cdot_\cdot\!&\! {}^\cdot_\cdot\!&\!
    {}^\cdot_1\!&\! {}^\cdot_\cdot\!&\! {}^\cdot_\cdot\\ 
  \end{array}\!\right)\!\! ,~
 \lambda_{32}\!\!=\!\!\left(\!
  \begin{array}{cccccc}
   {}^\cdot_\cdot\!&\! {}^\cdot_\cdot\!&\! {}^\cdot_\cdot\!&\!
    {}^\cdot_\cdot\!&\! {}^\cdot_\cdot\!&\! {}^\cdot_\cdot\\ 
   {}^\cdot_\cdot\!&\! {}^\cdot_\cdot\!&\! {}^\cdot_\cdot\!&\!
    {}^\cdot_\cdot\!&\! {}^\cdot_\cdot\!&\! {}^{~\cdot}_{-i}\\ 
   {}^\cdot_\cdot\!&\! {}^\cdot_\cdot\!&\! {}^\cdot_\cdot\!&\!
    {}^\cdot_i\!&\! {}^\cdot_\cdot\!&\! {}^\cdot_\cdot\\ 
  \end{array}\!\right)\!\! ,~
\\ &&\!\!\!\!\!\!\!\!
 \lambda_{33}\!\!=\!\!\left(\!
  \begin{array}{cccccc}
   {}^\cdot_\cdot\!&\! {}^\cdot_\cdot\!&\! {}^\cdot_\cdot\!&\!
    {}^\cdot_\cdot\!&\! {}^\cdot_\cdot\!&\! {}^\cdot_\cdot\\ 
   {}^\cdot_\cdot\!&\! {}^\cdot_\cdot\!&\! {}^\cdot_\cdot\!&\!
    {}^\cdot_\cdot\!&\! {}^\cdot_\cdot\!&\! {}^\cdot_\cdot\\ 
   {}^\cdot_\cdot\!&\! {}^\cdot_\cdot\!&\! {}^\cdot_\cdot\!&\!
    {}^\cdot_\cdot\!&\! {}^\cdot_1\!&\! {}^1_\cdot\\ 
  \end{array}\!\right)\!\! ,~~
 \lambda_{34}\!\!=\!\!\left(\!
  \begin{array}{cccccc}
   {}^\cdot_\cdot\!&\! {}^\cdot_\cdot\!&\! {}^\cdot_\cdot\!&\!
    {}^\cdot_\cdot\!&\! {}^\cdot_\cdot\!&\! {}^\cdot_\cdot\\ 
   {}^\cdot_\cdot\!&\! {}^\cdot_\cdot\!&\! {}^\cdot_\cdot\!&\!
    {}^\cdot_\cdot\!&\! {}^\cdot_\cdot\!&\! {}^\cdot_\cdot\\ 
   {}^\cdot_\cdot\!&\! {}^\cdot_\cdot\!&\! {}^\cdot_\cdot\!&\!
    {}^\cdot_\cdot\!&\! {}^\cdot_i\!&\! {}^{-i}_{~\cdot}\\ 
  \end{array}\!\right)\!\! ,~~
 \lambda_{35}\!\!=\!\!\textrm{${1\over \sqrt{15}}$}\left(\!
  \begin{array}{cccccc}
   {}^1_\cdot\!&\! {}^\cdot_1\!&\! {}^\cdot_\cdot\!&\!
    {}^\cdot_\cdot\!&\! {}^\cdot_\cdot\!&\! {}^\cdot_\cdot\\ 
   {}^\cdot_\cdot\!&\! {}^\cdot_\cdot\!&\! {}^1_\cdot\!&\!
    {}^\cdot_1\!&\! {}^\cdot_\cdot\!&\! {}^\cdot_\cdot\\ 
   {}^\cdot_\cdot\!&\! {}^\cdot_\cdot\!&\! {}^\cdot_\cdot\!&\!
    {}^\cdot_\cdot\!&\! {}^1_\cdot\!&\! {}^{~\cdot}_{-5}\\ 
  \end{array}\!\right)\!\! ,~~
\end{eqnarray*}
where dots indicate 0.
The $\lambda_a$ obey the following commutation relationship:
\begin{eqnarray}
 \left[\lambda_a, \lambda_b\right]
  \equiv \lambda_a \lambda_b -\lambda_b \lambda_a
   =2i f_{abc}\lambda_c .
 \label{commu}
\end{eqnarray}
The $f_{abc}$ are odd under the permutation of any pair of indices.
The nonzero values are tabulated in Table~\ref{fabc}.
\begin{table}[h]
\caption{The $f_{abc}$. There are 125 independent nonzero structure constants. 
\label{fabc}}
\vspace{-7mm}
\begin{minipage}[t]{2.8cm}
\begin{eqnarray*}
 \begin{array}{lrl}
 a~b~c && f_{abc} \\ \hline 
 {}_1 ~{}_2 ~{}_3 && {}_1 \\
 {}_1 ~{}_4 ~{}_7 && {}_{1/2}\\
 {}_1 ~{}_5 ~{}_6 &{}_-\!\!\!\!\!& {}_{1/2}\\
 {}_1 ~{}_9 ~{}_{12} && {}_{1/2}\\
 {}_1 ~{}_{10}~{}_{11} &{}_-\!\!\!\!\!& {}_{1/2}\\
 {}_1~{}_{16}~{}_{19} &&{}_{1/2} \\
 {}_1~{}_{17}~{}_{18} &{}_-\!\!\!\!\!&{}_{1/2} \\
 {}_1~{}_{25}~{}_{28} &&{}_{1/2} \\
 {}_1~{}_{26}~{}_{27}&{}_-\!\!\!\!\!&{}_{1/2} \\
 {}_2~{}_4~{}_6&&{}_{1/2}\\
 {}_2~{}_5~{}_7&&{}_{1/2}\\
 {}_2~{}_9~{}_{11}&&{}_{1/2}\\
 {}_2~{}_{10}~{}_{12}&&{}_{1/2}\\
 {}_2~{}_{16}~{}_{18}&&{}_{1/2}\\
 {}_2~{}_{17}~{}_{19}&&{}_{1/2}\\
 {}_2~{}_{25}~{}_{27}&&{}_{1/2}\\
 {}_2~{}_{26}~{}_{28}&&{}_{1/2}\\
 {}_3~{}_4~{}_5&&{}_{1/2}\\
 {}_3~{}_6~{}_7&{}_-\!\!\!\!\!&{}_{1/2}\\
 {}_3~{}_9~{}_{10}&&{}_{1/2}\\
 {}_3~{}_{11}~{}_{12}&{}_-\!\!\!\!\!&{}_{1/2}\\
 {}_3~{}_{16}~{}_{17}&&{}_{1/2}\\
 {}_3~{}_{18}~{}_{19}&{}_-\!\!\!\!\!&{}_{1/2}\\
 {}_3~{}_{25}~{}_{26}&&{}_{1/2}\\
 {}_3~{}_{27}~{}_{28}&{}_-\!\!\!\!\!&{}_{1/2}\\
 \end{array}
\end{eqnarray*}
\end{minipage}
%
%
\begin{minipage}[t]{2.9cm}
\begin{eqnarray*}
 \begin{array}{|lrl}
 a~b~c && f_{abc} \\ \hline 
 {}_4 ~{}_5 ~{}_8 && {}_{\sqrt{3}/2} \\
 {}_4 ~{}_9 ~{}_{14} && {}_{1/2}\\
 {}_4 ~{}_{10} ~{}_{13} &{}_-\!\!\!\!\!& {}_{1/2}\\
 {}_4 ~{}_{16} ~{}_{21} && {}_{1/2}\\
 {}_4 ~{}_{17}~{}_{20} &{}_-\!\!\!\!\!& {}_{1/2}\\
 {}_4~{}_{25}~{}_{30} &&{}_{1/2} \\
 {}_4~{}_{26}~{}_{29} &{}_-\!\!\!\!\!&{}_{1/2} \\
 {}_5~{}_{9}~{}_{13} &&{}_{1/2} \\
 {}_5~{}_{10}~{}_{14}&&{}_{1/2} \\
 {}_5~{}_{16}~{}_{20}&&{}_{1/2}\\
 {}_5~{}_{17}~{}_{21}&&{}_{1/2}\\
 {}_5~{}_{25}~{}_{29}&&{}_{1/2}\\
 {}_5~{}_{26}~{}_{30}&&{}_{1/2}\\
 {}_6~{}_{7}~{}_{8}&&{}_{\sqrt{3}/2}\\
 {}_6~{}_{11}~{}_{14}&&{}_{1/2}\\
 {}_6~{}_{12}~{}_{13}&{}_-\!\!\!\!\!&{}_{1/2}\\
 {}_6~{}_{18}~{}_{21}&&{}_{1/2}\\
 {}_6~{}_{19}~{}_{20}&{}_-\!\!\!\!\!&{}_{1/2}\\
 {}_6~{}_{27}~{}_{30}&&{}_{1/2}\\
 {}_6~{}_{28}~{}_{29}&{}_-\!\!\!\!\!&{}_{1/2}\\
 {}_7~{}_{11}~{}_{13}&&{}_{1/2}\\
 {}_7~{}_{12}~{}_{14}&&{}_{1/2}\\
 {}_7~{}_{18}~{}_{20}&&{}_{1/2}\\
 {}_7~{}_{19}~{}_{21}&&{}_{1/2}\\
 {}_7~{}_{27}~{}_{29}&&{}_{1/2}\\
 \end{array}
\end{eqnarray*}
\end{minipage}
\begin{minipage}[t]{3cm}
\begin{eqnarray*}
 \begin{array}{|lrl}
 a~b~c && f_{abc} \\ \hline 
 {}_7 ~{}_{28} ~{}_{30} && {}_{1/2} \\
 {}_8 ~{}_9 ~{}_{10} && {}_{\sqrt{3}/6}\\
 {}_8 ~{}_{11} ~{}_{12} && {}_{\sqrt{3}/6}\\
 {}_8 ~{}_{13} ~{}_{14} &{}_-\!\!\!\!\!& {}_{\sqrt{3}/3}\\
 {}_8 ~{}_{16}~{}_{17} && {}_{\sqrt{3}/6}\\
 {}_8~{}_{18}~{}_{19} &&{}_{\sqrt{3}/6} \\
 {}_8~{}_{20}~{}_{21} &{}_-\!\!\!\!\!&{}_{\sqrt{3}/3} \\
 {}_8~{}_{25}~{}_{26} &&{}_{\sqrt{3}/6} \\
 {}_8~{}_{27}~{}_{28}&&{}_{\sqrt{3}/6} \\
 {}_8~{}_{29}~{}_{30}&{}_-\!\!\!\!\!&{}_{\sqrt{3}/3}\\
 {}_9~{}_{10}~{}_{15}&&{}_{\sqrt{6}/3}\\
 {}_9~{}_{16}~{}_{23}&&{}_{1/2}\\
 {}_9~{}_{17}~{}_{22}&{}_-\!\!\!\!\!&{}_{1/2}\\
 {}_9~{}_{25}~{}_{32}&&{}_{1/2}\\
 {}_9~{}_{26}~{}_{31}&{}_-\!\!\!\!\!&{}_{1/2}\\
 {}_{10}~{}_{16}~{}_{22}&&{}_{1/2}\\
 {}_{10}~{}_{17}~{}_{23}&&{}_{1/2}\\
 {}_{10}~{}_{25}~{}_{31}&&{}_{1/2}\\
 {}_{10}~{}_{26}~{}_{32}&&{}_{1/2}\\
 {}_{11}~{}_{12}~{}_{15}&&{}_{\sqrt{6}/3}\\
 {}_{11}~{}_{18}~{}_{23}&&{}_{1/2}\\
 {}_{11}~{}_{19}~{}_{22}&{}_-\!\!\!\!\!&{}_{1/2}\\
 {}_{11}~{}_{27}~{}_{32}&&{}_{1/2}\\
 {}_{11}~{}_{28}~{}_{31}&{}_-\!\!\!\!\!&{}_{1/2}\\
 {}_{12}~{}_{18}~{}_{22}&&{}_{1/2}\\
 \end{array}
\end{eqnarray*}
\end{minipage}
\begin{minipage}[t]{3.1cm}
\begin{eqnarray*}
 \begin{array}{|lrl}
 a~b~c && f_{abc} \\ \hline 
 {}_{12} ~{}_{19} ~{}_{23} && {}_{1/2} \\
 {}_{12} ~{}_{27} ~{}_{31} && {}_{1/2}\\
 {}_{12} ~{}_{28} ~{}_{32} && {}_{1/2}\\
 {}_{13} ~{}_{14} ~{}_{15} && {}_{\sqrt{6}/3}\\
 {}_{13} ~{}_{20}~{}_{23} && {}_{1/2}\\
 {}_{13}~{}_{21}~{}_{22} &{}_-\!\!\!\!\!&{}_{1/2} \\
 {}_{13}~{}_{29}~{}_{32} &&{}_{1/2} \\
 {}_{13}~{}_{30}~{}_{31} &{}_-\!\!\!\!\!&{}_{1/2} \\
 {}_{14}~{}_{20}~{}_{22}&&{}_{1/2} \\
 {}_{14}~{}_{21}~{}_{23}&&{}_{1/2}\\
 {}_{14}~{}_{29}~{}_{31}&&{}_{1/2}\\
 {}_{14}~{}_{30}~{}_{32}&&{}_{1/2}\\
 {}_{15}~{}_{16}~{}_{17}&&{}_{\sqrt{6}/12}\\
 {}_{15}~{}_{18}~{}_{19}&&{}_{\sqrt{6}/12}\\
 {}_{15}~{}_{20}~{}_{21}&&{}_{\sqrt{6}/12}\\
 {}_{15}~{}_{22}~{}_{23}&{}_-\!\!\!\!\!&{}_{\sqrt{6}/4}\\
 {}_{15}~{}_{25}~{}_{26}&&{}_{\sqrt{6}/12}\\
 {}_{15}~{}_{27}~{}_{28}&&{}_{\sqrt{6}/12}\\ 
 {}_{15}~{}_{29}~{}_{30}&&{}_{\sqrt{6}/12}\\
 {}_{15}~{}_{31}~{}_{32}&{}_-\!\!\!\!\!&{}_{\sqrt{6}/4}\\
 {}_{16}~{}_{17}~{}_{24}&&{}_{\sqrt{10}/4}\\
 {}_{16}~{}_{25}~{}_{34}&&{}_{1/2}\\
 {}_{16}~{}_{26}~{}_{33}&{}_-\!\!\!\!\!&{}_{1/2}\\
 {}_{17}~{}_{25}~{}_{33}&&{}_{1/2}\\
 {}_{17}~{}_{26}~{}_{34}&&{}_{1/2}\\
 \end{array}
\end{eqnarray*}
\end{minipage}
\begin{minipage}[t]{2.8cm}
\begin{eqnarray*}
 \begin{array}{|lrl}
 a~b~c && f_{abc} \\ \hline 
 {}_{18} ~{}_{19} ~{}_{24} && {}_{\sqrt{10}/4} \\
 {}_{18} ~{}_{27} ~{}_{34} && {}_{1/2}\\
 {}_{18} ~{}_{28} ~{}_{33} &{}_-\!\!\!\!\!& {}_{1/2}\\
 {}_{19} ~{}_{27} ~{}_{33} && {}_{1/2}\\
 {}_{19} ~{}_{28}~{}_{34} && {}_{1/2}\\
 {}_{20}~{}_{21}~{}_{24} &&{}_{\sqrt{10}/4} \\
 {}_{20}~{}_{29}~{}_{34} &&{}_{1/2} \\
 {}_{20}~{}_{30}~{}_{33} &{}_-\!\!\!\!\!&{}_{1/2} \\
 {}_{21}~{}_{29}~{}_{33}&&{}_{1/2} \\
 {}_{21}~{}_{30}~{}_{34}&&{}_{1/2}\\
 {}_{22}~{}_{23}~{}_{24}&&{}_{\sqrt{10}/4}\\
 {}_{22}~{}_{31}~{}_{34}&&{}_{1/2}\\
 {}_{22}~{}_{32}~{}_{33}&{}_-\!\!\!\!\!&{}_{1/2}\\
 {}_{23}~{}_{31}~{}_{33}&&{}_{1/2}\\
 {}_{23}~{}_{32}~{}_{34}&&{}_{1/2}\\
 {}_{24}~{}_{25}~{}_{26}&&{}_{\sqrt{10}/20}\\
 {}_{24}~{}_{27}~{}_{28}&&{}_{\sqrt{10}/20}\\
 {}_{24}~{}_{29}~{}_{30}&&{}_{\sqrt{10}/20}\\
 {}_{24}~{}_{31}~{}_{32}&&{}_{\sqrt{10}/20}\\
 {}_{24}~{}_{33}~{}_{34}&{}_-\!\!\!\!\!&{}_{\sqrt{10}/5}\\
 {}_{25}~{}_{26}~{}_{35}&&{}_{\sqrt{15}/5}\\
 {}_{27}~{}_{28}~{}_{35}&&{}_{\sqrt{15}/5}\\
 {}_{29}~{}_{30}~{}_{35}&&{}_{\sqrt{15}/5}\\
 {}_{31}~{}_{32}~{}_{35}&&{}_{\sqrt{15}/5}\\
 {}_{33}~{}_{34}~{}_{35}&&{}_{\sqrt{15}/5}\\
 \end{array}
\end{eqnarray*}
\end{minipage}
\end{table}

Among the generators of SU(6) transformations,  the 15 generators 
\begin{eqnarray*}
\lambda_2,~ \lambda_5,~ \lambda_7,~ \lambda_{10},~ \lambda_{12},~ 
\lambda_{14},~ \lambda_{17},~ \lambda_{19},~\lambda_{21},~\lambda_{23},~
\lambda_{26},~\lambda_{28},~\lambda_{30},~\lambda_{32},~\lambda_{34}
\end{eqnarray*}
form a subalgebra. 
The 15 generators listed above are written also
as generators with antisymmetric two indices, 
${\cal T}_{ab}$ $(a=1,\cdots, 6)$.
The generators are given in a matrix form as
\begin{eqnarray*}
 {\cal T}\equiv\left(\begin{array}{cccccc}
 0&\lambda_2 &\lambda_5 &\lambda_{10} &\lambda_{17}  &\lambda_{26} \\
 -\lambda_2 &0& \lambda_{7}&\lambda_{12}&\lambda_{19}&\lambda_{28} \\
 -\lambda_5&-\lambda_{7}&0&\lambda_{14}&\lambda_{21}&\lambda_{30}\\
 -\lambda_{10}&-\lambda_{12}&-\lambda_{14}&0&\lambda_{23}&\lambda_{32}\\
 -\lambda_{17}&-\lambda_{19}&-\lambda_{21}&-\lambda_{23}&0&\lambda_{34}\\
 -\lambda_{26}&-\lambda_{28}&-\lambda_{30}&-\lambda_{32}&-\lambda_{34}&0\\
	      \end{array}\right) .
\end{eqnarray*}
The ${\cal T}_{ab}$ obey the following commutation relationship
of SO(6) transformations:
\begin{eqnarray*}
 \left[{\cal T}_{ab},{\cal T}_{cd}\right]
 =-i(\delta_{bc}{\cal T}_{ad}-\delta_{ac}{\cal T}_{bd}
     -\delta_{bd}{\cal T}_{ac}+\delta_{ad}{\cal T}_{bc}).
\end{eqnarray*}

We define $\lambda_{\widehat{a}}$ ($\widehat{a}
=\widehat{1},\cdots, \widehat{15}$)
\begin{eqnarray*} &&
 \left(\begin{array}{c}
  \lambda_{\widehat{1}}\\ \lambda_{\widehat{10}}
       \end{array}\right)
 \equiv
 \left(\begin{array}{cc}
   1& 1 \\
  -1&1 \\
       \end{array}\right)
 \left(\begin{array}{c}
  \lambda_{12}\\ \lambda_{17}
       \end{array}\right) ,~~
\left(\begin{array}{c}
  \lambda_{\widehat{2}}\\ \lambda_{\widehat{11}}
       \end{array}\right)
 \equiv
 \left(\begin{array}{cc}
   1& 1 \\
  -1&1 \\
       \end{array}\right)
 \left(\begin{array}{c}
  \lambda_{2}\\ \lambda_{23}
       \end{array}\right) ,
\\ && ~~~~~~~~~~~~~~
\left(\begin{array}{c}
  \lambda_{\widehat{3}}\\ \lambda_{\widehat{8}}
 \\ \lambda_{\widehat{9}}
       \end{array}\right)
 \equiv
 \left(\begin{array}{ccc}
   1& -1 &0 \\
  \textrm{${1\over \sqrt{3}}$}&\textrm{${1\over \sqrt{3}}$} &
  \textrm{$-{2\over \sqrt{3}}$}\\
  \textrm{$\sqrt{2\over 3}$}&\textrm{$\sqrt{2\over 3}$}
 &\textrm{$\sqrt{2\over 3}$} \\
       \end{array}\right)
 \left(\begin{array}{c}
  \lambda_{10}\\ \lambda_{19} \\ \lambda_{30}
       \end{array}\right) ,
\\ &&
\left(\begin{array}{c}
  \lambda_{\widehat{4}}\\ \lambda_{\widehat{12}}
       \end{array}\right)
 \equiv
 \left(\begin{array}{cc}
   1& 1 \\
  -1&1 \\
       \end{array}\right)
 \left(\begin{array}{c}
  \lambda_{14}\\ \lambda_{26}
       \end{array}\right) ,~~
\left(\begin{array}{c}
  \lambda_{\widehat{5}}\\ \lambda_{\widehat{13}}
       \end{array}\right)
 \equiv
 \left(\begin{array}{cc}
   1& 1 \\
  -1&1 \\
       \end{array}\right)
 \left(\begin{array}{c}
  \lambda_{5}\\ \lambda_{32}
       \end{array}\right) , 
\\ &&
\left(\begin{array}{c}
  \lambda_{\widehat{6}}\\ \lambda_{\widehat{14}}
       \end{array}\right)
 \equiv
 \left(\begin{array}{cc}
   1& 1 \\
  -1&1 \\
       \end{array}\right)
 \left(\begin{array}{c}
  \lambda_{21}\\ \lambda_{28}
       \end{array}\right) ,~~
\left(\begin{array}{c}
  \lambda_{\widehat{7}}\\ \lambda_{\widehat{15}}
       \end{array}\right)
 \equiv
 \left(\begin{array}{cc}
   1& 1 \\
  -1&1 \\
       \end{array}\right)
 \left(\begin{array}{c}
  \lambda_{7}\\ \lambda_{34}
       \end{array}\right) .
\end{eqnarray*}
Then Eq.(\ref{commu}) becomes
\begin{eqnarray*}&&
  \left[\lambda_a, \lambda_b\right]
   =2i f_{abc}\lambda_c +2i f_{ab\widehat{c}}\lambda_{\widehat{c}},
\\ && 
\left[\lambda_a, \lambda_{\widehat{b}}\right]
   =2i f_{a\widehat{b}c}\lambda_c 
   +2i f_{a\widehat{b}\widehat{c}}\lambda_{\widehat{c}} ,
\\ &&
\left[\lambda_{\widehat{a}}, \lambda_{\widehat{b}}\right]
   =2i f_{\widehat{a}\widehat{b}c}\lambda_c 
  +2i f_{\widehat{a}\widehat{b}\widehat{c}}\lambda_{\widehat{c}} ,
\end{eqnarray*}
where $a$ run except for $\widehat{a}$.
From Table~\ref{fabc}, it is immediately seen that
\begin{eqnarray*}
f_{abc}=0  \quad\textrm{and}\quad
f_{a\widehat{b}\widehat{c}}=0 .
\end{eqnarray*}
The nonzero $f_{\widehat{a}bc}$ are tabulated in Table~\ref{fwabc}.
\begin{table}[htbp]
\caption{The $f_{\widehat{a}bc}$ with 
$\widehat{a}=\widehat{1},\widehat{2},\widehat{3},\widehat{4},
\widehat{5},\widehat{10},\widehat{11},\widehat{12},\widehat{13}$. 
There are 123 independent nonzero structure constants. 
\label{fwabc}}
\vspace{-7mm}
\begin{minipage}[t]{4.8cm}
\begin{eqnarray*}
 \begin{array}{lrl}
 \widehat{a}~b~c && f_{\widehat{a}bc}\\ \hline 
 {}_{\widehat{1}}{}_{(\widehat{10})}~{}_1 ~{}_9 &\!\!&{}_{1/2}
 {}_{(-1/2)}\\
 {}_{\widehat{1}}{}_{(\widehat{10})}~{}_1 ~{}_{18} && {}_{1/2}{}_{(1/2)}\\
 {}_{\widehat{1}}{}_{(\widehat{10})}~{}_3 ~{}_{11} &{}_-\!\!\!\!\!& {}_{1/2}{}_{(1/2)}\\
 {}_{\widehat{1}}{}_{(\widehat{10})}~{}_3 ~{}_{16} && {}_{1/2}{}_{(1/2)}\\
 {}_{\widehat{1}}{}_{(\widehat{10})}~{}_4 ~{}_{20} && {}_{1/2}{}_{(1/2)}\\
 {}_{\widehat{1}}{}_{(\widehat{10})}~{}_{6}~{}_{13} && {}_{1/2}{}_{(-1/2)}\\
 {}_{\widehat{1}}{}_{(\widehat{10})}~{}_{8}~{}_{11} &&{}_{\sqrt{3}/6}{}_{(-\sqrt{3}/6)} \\
 {}_{\widehat{1}}{}_{(\widehat{10})}~{}_{8}~{}_{16} &&{}_{\sqrt{3}/6}{}_{(\sqrt{3}/6)} \\
 {}_{\widehat{1}}{}_{(\widehat{10})}~{}_{9}~{}_{22} &&{}_{1/2}{}_{(1/2)} \\
 {}_{\widehat{1}}{}_{(\widehat{10})}~{}_{11}~{}_{15}&{}_-\!\!\!\!\!&{}_{\sqrt{6}/3}{}_{(\sqrt{6}/3)} \\
 {}_{\widehat{1}}{}_{(\widehat{10})}~{}_{15}~{}_{16}&&{}_{\sqrt{6}/12}{}_{(\sqrt{6}/12)}\\
 {}_{\widehat{1}}{}_{(\widehat{10})}~{}_{16}~{}_{24}&{}_-\!\!\!\!\!&{}_{\sqrt{10}/4}{}_{(-\sqrt{10}/4)}\\
 {}_{\widehat{1}}{}_{(\widehat{10})}~{}_{18}~{}_{22}&&{}_{1/2}{}_{(-1/2)}\\
 {}_{\widehat{1}}{}_{(\widehat{10})}~{}_{25}~{}_{33}&&{}_{1/2}{}_{(1/2)}\\
 {}_{\widehat{1}}{}_{(\widehat{10})}~{}_{27}~{}_{31}&&{}_{1/2}{}_{(-1/2)}\\
 {}_{\widehat{2}}{}_{(\widehat{11})}~{}_{1}~{}_{3}&&{}_{-1}{}_{(1)}\\
 {}_{\widehat{2}}{}_{(\widehat{11})}~{}_{4}~{}_{6}&&{}_{1/2}{}_{(-1/2)}\\
 {}_{\widehat{2}}{}_{(\widehat{11})}~{}_{9}~{}_{11}&&{}_{1/2}{}_{(-1/2)}\\
 {}_{\widehat{2}}{}_{(\widehat{11})}~{}_9~{}_{16}&&{}_{1/2}{}_{(1/2)}\\
 {}_{\widehat{2}}{}_{(\widehat{11})}~{}_{11}~{}_{18}&&{}_{1/2}{}_{(1/2)}\\
 {}_{\widehat{2}}{}_{(\widehat{11})}~{}_{13}~{}_{20}&&{}_{1/2}{}_{(1/2)}\\
 {}_{\widehat{2}}{}_{(\widehat{11})}~{}_{15}~{}_{22}&{}_-\!\!\!\!\!&{}_{\sqrt{6}/4}{}_{(-\sqrt{6}/4)}\\
 {}_{\widehat{2}}{}_{(\widehat{11})}~{}_{16}~{}_{18}&&{}_{1/2}{}_{(-1/2)}\\
\end{array}
\end{eqnarray*}
\end{minipage}
\begin{minipage}[t]{5cm}
\begin{eqnarray*}
 \begin{array}{|lrl}
 \widehat{a}~b~c && f_{\widehat{a}bc} \\ \hline 
 {}_{\widehat{2}}{}_{(\widehat{11})}~{}_{22}~{}_{24}&{}_-\!\!\!\!\!&{}_{\sqrt{10}/4}{}_{(-\sqrt{10}/4)}\\
 {}_{\widehat{2}}{}_{(\widehat{11})}~{}_{25}~{}_{27}&&{}_{1/2}{}_{(-1/2)}\\
 {}_{\widehat{2}}{}_{(\widehat{11})}~{}_{31}~{}_{33}&&{}_{1/2}{}_{(1/2)}\\
 {}_{\widehat{3}} ~{}_{1} ~{}_{11} && {}_{1/2} \\
 {}_{\widehat{3}}~{}_1 ~{}_{16} &{}_-\!\!\!\!\!& {}_{1/2}\\
 {}_{\widehat{3}}~{}_{3} ~{}_{9} && {}_{1/2}\\
 {}_{\widehat{3}}~{}_{3} ~{}_{18} && {}_{1/2}\\
 {}_{\widehat{3}}~{}_{4}~{}_{13} && {}_{1/2}\\
 {}_{\widehat{3}}~{}_{6}~{}_{20} &{}_-\!\!\!\!\!&{}_{1/2}\\
 {}_{\widehat{3}}~{}_{8}~{}_{9} &&{}_{\sqrt{3}/6}\\
 {}_{\widehat{3}}~{}_{8}~{}_{18} &{}_-\!\!\!\!\!&{}_{\sqrt{3}/6}\\
 {}_{\widehat{3}}~{}_{9}~{}_{15}&{}_-\!\!\!\!\!&{}_{{\sqrt{6}}/3} \\
 {}_{\widehat{3}}~{}_{11}~{}_{22}&{}_-\!\!\!\!\!&{}_{1/2}\\
 {}_{\widehat{3}}~{}_{15}~{}_{18}&{}_-\!\!\!\!\!&{}_{\sqrt{6}/12}\\
 {}_{\widehat{3}}~{}_{16}~{}_{22}&&{}_{1/2}\\
 {}_{\widehat{3}}~{}_{18}~{}_{24}&&{}_{\sqrt{10}/4}\\
 {}_{\widehat{3}}~{}_{25}~{}_{31}&&{}_{1/2}\\
 {}_{\widehat{3}}~{}_{27}~{}_{33}&{}_-\!\!\!\!\!&{}_{1/2}\\
 {}_{\widehat{4}}{}_{(\widehat{12})}~{}_{1}~{}_{27}&&{}_{1/2}{}_{(1/2)}\\
 {}_{\widehat{4}}{}_{(\widehat{12})}~{}_{3}~{}_{25}&&{}_{1/2}{}_{(1/2)}\\
 {}_{\widehat{4}}{}_{(\widehat{12})}~{}_{4}~{}_{9}&&{}_{1/2}{}_{(-1/2)}\\
 {}_{\widehat{4}}{}_{(\widehat{12})}~{}_{4}~{}_{29}&&{}_{1/2}{}_{(1/2)}\\
 {}_{\widehat{4}}{}_{(\widehat{12})}~{}_{6}~{}_{11}&&{}_{1/2}{}_{(-1/2)}\\
 {}_{\widehat{4}}{}_{(\widehat{12})}~{}_{8}~{}_{13}&{}_-\!\!\!\!\!&{}_{\sqrt{3}/3}{}_{(\sqrt{3}/3)}\\
\end{array}
\end{eqnarray*}
\end{minipage}
\begin{minipage}[t]{5cm}
\begin{eqnarray*}
 \begin{array}{|lrl}
 \widehat{a}~b~c && f_{\widehat{a}bc} \\ \hline 
 {}_{\widehat{4}}{}_{(\widehat{12})}~{}_{8}~{}_{25}&&{}_{\sqrt{3}/6}{}_{(\sqrt{3}/6)}\\
 {}_{\widehat{4}}{}_{(\widehat{12})}~{}_{9}~{}_{31}&&{}_{1/2}{}_{(1/2)}\\
 {}_{\widehat{4}}{}_{(\widehat{12})}~{}_{13}~{}_{15}&{}_-\!\!\!\!\!&{}_{\sqrt{6}/3}{}_{(\sqrt{6}/3)}\\
 {}_{\widehat{4}}{}_{(\widehat{12})} ~{}_{15} ~{}_{25} && {}_{\sqrt{6}/12}{}_{(\sqrt{6}/12)}\\
 {}_{\widehat{4}}{}_{(\widehat{12})} ~{}_{16} ~{}_{33} && {}_{1/2}{}_{(1/2)}\\
 {}_{\widehat{4}}{}_{(\widehat{12})}~{}_{20}~{}_{22}&&{}_{1/2}{}_{(-1/2)}\\
 {}_{\widehat{4}}{}_{(\widehat{12})} ~{}_{24} ~{}_{25} && {}_{\sqrt{10}/20}{}_{(\sqrt{10}/20)}\\
 {}_{\widehat{4}}{}_{(\widehat{12})} ~{}_{25}~{}_{35} &{}_-\!\!\!\!\!&{}_{\sqrt{15}/5}{}_{(-\sqrt{15}/5)}\\
 {}_{\widehat{4}}{}_{(\widehat{12})} ~{}_{29} ~{}_{31} && {}_{1/2}
  {}_{(-1/2)}\\
 {}_{\widehat{5}}{}_{(\widehat{13})}~{}_{1}~{}_{6} &&{}_{1/2} {}_{(-1/2)}\\
 {}_{\widehat{5}}{}_{(\widehat{13})}~{}_{3}~{}_{4} &&{}_{1/2} {}_{(-1/2)}\\
 {}_{\widehat{5}}{}_{(\widehat{13})}~{}_{4}~{}_{8} &{}_-\!\!\!\!\!&{}_{\sqrt{3}/2} {}_{(\sqrt{3}/2)}\\
 {}_{\widehat{5}}{}_{(\widehat{13})}~{}_{9}~{}_{13}&&{}_{1/2} {}_{(-1/2)}\\
 {}_{\widehat{5}}{}_{(\widehat{13})}~{}_{9}~{}_{25}&&{}_{1/2}{}_{(1/2)}\\
 {}_{\widehat{5}}{}_{(\widehat{13})}~{}_{11}~{}_{27}&&{}_{1/2}{}_{(1/2)}\\
 {}_{\widehat{5}}{}_{(\widehat{13})}~{}_{13}~{}_{29}&&{}_{1/2}{}_{(1/2)}\\
 {}_{\widehat{5}}{}_{(\widehat{13})}~{}_{15}~{}_{31}&{}_-\!\!\!\!\!&{}_{\sqrt{6}/4}{}_{(-\sqrt{6}/4)}\\
 {}_{\widehat{5}}{}_{(\widehat{13})}~{}_{16}~{}_{20}&&{}_{1/2}{}_{(-1/2)}\\
 {}_{\widehat{5}}{}_{(\widehat{13})}~{}_{22}~{}_{33}&&{}_{1/2}{}_{(1/2)}\\
 {}_{\widehat{5}}{}_{(\widehat{13})}~{}_{24}~{}_{31}&&{}_{\sqrt{10}/20}{}_{(\sqrt{10}/20)}\\
 {}_{\widehat{5}}{}_{(\widehat{13})}~{}_{25}~{}_{29}&&{}_{1/2}{}_{(-1/2)}\\
 {}_{\widehat{5}}{}_{(\widehat{13})}~{}_{31}~{}_{35}&{}_-\!\!\!\!\!&{}_{\sqrt{15}/5}{}_{(-\sqrt{15}/5)}\\
\end{array}
\end{eqnarray*}
\end{minipage}
\end{table}
\begin{table}[htbp]
\caption{The $f_{\widehat{a}bc}$ with
$\widehat{a}=\widehat{6},\widehat{7},\widehat{8},\widehat{9},
\widehat{14},\widehat{15}$. 
There are 102 independent nonzero structure constants. 
\label{fwabc2}}
\vspace{-7mm}
\begin{minipage}[t]{5cm}
\begin{eqnarray*}
 \begin{array}{|lrl}
 \widehat{a}~b~c && f_{\widehat{a}bc} \\ \hline
 {}_{\widehat{6}}{}_{(\widehat{14})}~{}_{1}~{}_{25}&&{}_{1/2}{}_{(1/2)}\\
 {}_{\widehat{6}}{}_{(\widehat{14})}~{}_{3}~{}_{27}&{}_-\!\!\!\!\!&{}_{1/2}{}_{(-1/2)}\\
 {}_{\widehat{6}}{}_{(\widehat{14})}~{}_{4}~{}_{16}&&{}_{1/2}{}_{(-1/2)}\\
 {}_{\widehat{6}}{}_{(\widehat{14})}~{}_{6}~{}_{18}&&{}_{1/2}{}_{(-1/2)}\\
 {}_{\widehat{6}}{}_{(\widehat{14})}~{}_{6}~{}_{29}&&{}_{1/2}{}_{(1/2)}\\
 {}_{\widehat{6}}{}_{(\widehat{14})}~{}_{8}~{}_{20}&{}_-\!\!\!\!\!&{}_{\sqrt{3}/3}{}_{(\sqrt{3}/3)}\\
 {}_{\widehat{6}}{}_{(\widehat{14})}~{}_{8}~{}_{27}&&{}_{\sqrt{3}/6}{}_{(\sqrt{3}/6)}\\
 {}_{\widehat{6}}{}_{(\widehat{14})} ~{}_{11} ~{}_{31} && {}_{1/2} {}_{(1/2)}\\
 {}_{\widehat{6}}{}_{(\widehat{14})} ~{}_{13} ~{}_{22} &{}_-\!\!\!\!\!& {}_{1/2}{}_{(1/2)}\\
 {}_{\widehat{6}}{}_{(\widehat{14})} ~{}_{15} ~{}_{20} && {}_{\sqrt{6}/12}{}_{(-\sqrt{6}/12)}\\
 {}_{\widehat{6}}{}_{(\widehat{14})} ~{}_{15} ~{}_{27} && {}_{\sqrt{6}/12}{}_{(\sqrt{6}/12)}\\
 {}_{\widehat{6}}{}_{(\widehat{14})} ~{}_{18}~{}_{33} && {}_{1/2}{}_{(1/2)}\\
 {}_{\widehat{6}}{}_{(\widehat{14})}~{}_{20}~{}_{24}&{}_-\!\!\!\!\!&{}_{\sqrt{10}/4} {}_{(\sqrt{10}/4)}\\
 {}_{\widehat{6}}{}_{(\widehat{14})}~{}_{24}~{}_{27} &&{}_{\sqrt{10}/20} {}_{(\sqrt{10}/20)}\\
 {}_{\widehat{6}}{}_{(\widehat{14})}~{}_{27}~{}_{35}&{}_-\!\!\!\!\!&{}_{\sqrt{15}/5} {}_{(-\sqrt{15}/5)}\\
 {}_{\widehat{6}}{}_{(\widehat{14})}~{}_{29}~{}_{33} &&{}_{1/2}
  {}_{(-1/2)}\\
 {}_{\widehat{7}}{}_{(\widehat{15})}~{}_{1}~{}_{4}&&{}_{1/2}{}_{(-1/2)}\\
 \end{array}
\end{eqnarray*}
\end{minipage}
\begin{minipage}[t]{5.1cm}
\begin{eqnarray*}
 \begin{array}{|lrl}
 \widehat{a}~b~c && f_{\widehat{a}bc} \\ \hline 
 {}_{\widehat{7}}{}_{(\widehat{15})}~{}_{3}~{}_{6}&{}_-\!\!\!\!\!&{}_{1/2}{}_{(1/2)}\\
 {}_{\widehat{7}}{}_{(\widehat{15})}~{}_{6}~{}_{8}&{}_-\!\!\!\!\!&{}_{\sqrt{3}/2}{}_{(\sqrt{3}/2)}\\
 {}_{\widehat{7}}{}_{(\widehat{15})}~{}_{11}~{}_{13}&&{}_{1/2}{}_{(-1/2)}\\
 {}_{\widehat{7}}{}_{(\widehat{15})}~{}_{16}~{}_{25}&&{}_{1/2}{}_{(1/2)}\\
 {}_{\widehat{7}}{}_{(\widehat{15})}~{}_{18}~{}_{20}&&{}_{1/2}{}_{(-1/2)}\\
 {}_{\widehat{7}}{}_{(\widehat{15})}~{}_{18}~{}_{27}&&{}_{1/2}{}_{(1/2)}\\
 {}_{\widehat{7}}{}_{(\widehat{15})}~{}_{20}~{}_{29}&&{}_{1/2}{}_{(1/2)}\\
 {}_{\widehat{7}}{}_{(\widehat{15})}~{}_{22}~{}_{31}&&{}_{1/2}{}_{(1/2)}\\ 
 {}_{\widehat{7}}{}_{(\widehat{15})}~{}_{24}~{}_{33}&{}_-\!\!\!\!\!&{}_{\sqrt{10}/5}{}_{(-\sqrt{10}/5)}\\
 {}_{\widehat{7}}{}_{(\widehat{15})}~{}_{27}~{}_{29}&&{}_{1/2}{}_{(-1/2)}\\
 {}_{\widehat{7}}{}_{(\widehat{15})}~{}_{33}~{}_{35}&{}_-\!\!\!\!\!&{}_{\sqrt{15}/5}{}_{(-\sqrt{15}/5)}\\
 {}_{\widehat{8}}{}_{(\widehat{9})} ~{}_{1} ~{}_{11} && {}_{\sqrt{3}/6}{}_{(\sqrt{6}/6)} \\
 {}_{\widehat{8}}{}_{(\widehat{9})}~{}_1 ~{}_{16} && {}_{\sqrt{3}/6}{}_{(\sqrt{6}/6)}\\
 {}_{\widehat{8}}{}_{(\widehat{9})}~{}_{3} ~{}_{9} && {}_{\sqrt{3}/6}{}_{(\sqrt{6}/6)}\\
 {}_{\widehat{8}}{}_{(\widehat{9})}~{}_{3} ~{}_{18} &{}_-\!\!\!\!\!& {}_{\sqrt{3}/6}{}_{(-\sqrt{6}/6)}\\
 {}_{\widehat{8}}{}_{(\widehat{9})}~{}_{4}~{}_{13} && {}_{\sqrt{3}/6}{}_{(\sqrt{6}/6)}\\
 {}_{\widehat{8}}{}_{(\widehat{9})}~{}_{4}~{}_{25}&{}_-\!\!\!\!\!&{}_{\sqrt{3}/3}{}_{(\sqrt{6}/6)}\\
\end{array}
\end{eqnarray*}
\end{minipage}
\begin{minipage}[t]{4.5cm}
\begin{eqnarray*}
 \begin{array}{|lrl}
 \widehat{a}~b~c && f_{\widehat{a}bc} \\ \hline 
 {}_{\widehat{8}}{}_{(\widehat{9})}~{}_{6}~{}_{20}
  &&{}_{\sqrt{3}/6}{}_{(\sqrt{6}/6)} \\
 {}_{\widehat{8}}{}_{(\widehat{9})}~{}_{6}~{}_{27}&{}_-\!\!\!\!\!&{}_{\sqrt{3}/3}{}_{(\sqrt{6}/6)}\\
 {}_{\widehat{8}}{}_{(\widehat{9})}~{}_{8}~{}_{9} &&{}_{1/6}{}_{(\sqrt{2}/6)} \\
 {}_{\widehat{8}}{}_{(\widehat{9})}~{}_{8}~{}_{18} &&{}_{1/6}{}_{(\sqrt{2}/6)} \\
 {}_{\widehat{8}}{}_{(\widehat{9})}~{}_{8}~{}_{29}&&{}_{2/3}{}_{(-\sqrt{2}/3)}\\
 {}_{\widehat{8}}{}_{(\widehat{9})}~{}_{9}~{}_{15}&{}_-\!\!\!\!\!&{}_{\sqrt{2}/3}{}_{(-2/3)} \\
 {}_{\widehat{8}}{}_{(\widehat{9})}~{}_{11}~{}_{22}&&{}_{\sqrt{3}/6}{}_{(\sqrt{6}/6)}\\
 {}_{\widehat{8}}{}_{(\widehat{9})}~{}_{13}~{}_{31}&{}_-\!\!\!\!\!&{}_{\sqrt{3}/3}{}_{(\sqrt{6}/6)}\\
 {}_{\widehat{8}}{}_{(\widehat{9})}~{}_{15}~{}_{18}&&{}_{\sqrt{2}/12}{}_{(1/6)}\\
 {}_{\widehat{8}}{}_{(\widehat{9})}~{}_{15} ~{}_{29} &{}_-\!\!\!\!\!&{}_{\sqrt{2}/6} {}_{(1/6)}\\
 {}_{\widehat{8}}{}_{(\widehat{9})}~{}_{16}~{}_{22}&&{}_{\sqrt{3}/6}{}_{(\sqrt{6}/6)}\\
 {}_{\widehat{8}}{}_{(\widehat{9})}~{}_{18}~{}_{24}&{}_-\!\!\!\!\!&{}_{\sqrt{30}/12}{}_{(-\sqrt{60}/12)}\\
 {}_{\widehat{8}}{}_{(\widehat{9})} ~{}_{20} ~{}_{33} &{}_-\!\!\!\!\!& {}_{\sqrt{3}/3}{}_{(\sqrt{6}/6)}\\
 {}_{\widehat{8}}{}_{(\widehat{9})} ~{}_{24} ~{}_{29} &{}_-\!\!\!\!\!&  {}_{\sqrt{30}/30}{}_{(\sqrt{60}/60)}\\
 {}_{\widehat{8}}{}_{(\widehat{9})}~{}_{25}~{}_{31}&&{}_{\sqrt{3}/6}{}_{(\sqrt{6}/6)}\\
 {}_{\widehat{8}}{}_{(\widehat{9})}~{}_{27}~{}_{33}&&{}_{\sqrt{3}/6}{}_{(\sqrt{6}/6)}\\
 {}_{\widehat{8}}{}_{(\widehat{9})} ~{}_{29} ~{}_{35} &&  {}_{2\sqrt{5}/5}{}_{(-\sqrt{10}/5)}\\
 \end{array}
\end{eqnarray*}
\end{minipage}
\end{table}
The nonzero $f_{\widehat{a}\widehat{b}\widehat{c}}$ are 
tabulated in Table~\ref{fabcwww}.
\begin{table}[htbp]
\caption{The $f_{\widehat{a}\widehat{b}\widehat{c}}$. 
There are 29 independent nonzero structure constants. 
\label{fabcwww}}
\vspace{-7mm}
\begin{minipage}[t]{2.8cm}
\begin{eqnarray*}
 \begin{array}{lrl}
 \widehat{a}~\widehat{b}~\widehat{c} && f_{\widehat{a}\widehat{b}\widehat{c}} \\ \hline 
 {}_{\widehat{1}} ~{}_{\widehat{2}} ~{}_{\widehat{3}} &&{}_{1}\\
 {}_{\widehat{1}} ~{}_{\widehat{4}} ~{}_{\widehat{7}} &&{}_{1/2}\\
 {}_{\widehat{1}} ~{}_{\widehat{5}} ~{}_{\widehat{6}} &{}_-\!\!\!\!\!&{}_{1/2}\\
 {}_{\widehat{1}} ~{}_{\widehat{12}} ~{}_{\widehat{15}} &&{}_{1/2}\\
 {}_{\widehat{1}} ~{}_{\widehat{13}} ~{}_{\widehat{14}} &{}_-\!\!\!\!\!&{}_{1/2}\\
 {}_{\widehat{2}} ~{}_{\widehat{4}} ~{}_{\widehat{6}} &&{}_{1/2}\\
 \end{array}
\end{eqnarray*}
\end{minipage}
\begin{minipage}[t]{2.8cm}
\begin{eqnarray*}
 \begin{array}{|lrl}
 \widehat{a}~\widehat{b}~\widehat{c} && f_{\widehat{a}\widehat{b}\widehat{c}} \\ \hline 
 {}_{\widehat{2}} ~{}_{\widehat{5}} ~{}_{\widehat{7}} &&{}_{1/2}\\
 {}_{\widehat{2}} ~{}_{\widehat{12}} ~{}_{\widehat{14}} &&{}_{1/2}\\
 {}_{\widehat{2}} ~{}_{\widehat{13}} ~{}_{\widehat{15}} &&{}_{1/2}\\
 {}_{\widehat{3}} ~{}_{\widehat{4}} ~{}_{\widehat{5}} &&{}_{1/2}\\
 {}_{\widehat{3}} ~{}_{\widehat{6}} ~{}_{\widehat{7}} &{}_-\!\!\!\!\!&{}_{1/2}\\
 {}_{\widehat{3}} ~{}_{\widehat{12}} ~{}_{\widehat{13}} &&{}_{1/2}\\
 \end{array}
\end{eqnarray*}
\end{minipage}
\begin{minipage}[t]{2.8cm}
\begin{eqnarray*}
 \begin{array}{|lrl}
 \widehat{a}~\widehat{b}~\widehat{c} && f_{\widehat{a}\widehat{b}\widehat{c}} \\ \hline 
 {}_{\widehat{3}} ~{}_{\widehat{14}} ~{}_{\widehat{15}} &{}_-\!\!\!\!\!&{}_{1/2}\\
 {}_{\widehat{4}} ~{}_{\widehat{5}} ~{}_{\widehat{8}} &&{}_{\sqrt{3}/2}\\
 {}_{\widehat{4}} ~{}_{\widehat{10}} ~{}_{\widehat{15}} &{}_-\!\!\!\!\!&{}_{1/2}\\
 {}_{\widehat{4}} ~{}_{\widehat{11}} ~{}_{\widehat{14}} &&{}_{1/2}\\
 {}_{\widehat{5}} ~{}_{\widehat{10}} ~{}_{\widehat{14}} &{}_-\!\!\!\!\!&{}_{1/2}\\
 {}_{\widehat{5}} ~{}_{\widehat{11}} ~{}_{\widehat{15}} &{}_-\!\!\!\!\!&{}_{1/2}\\
 \end{array}
\end{eqnarray*}
\end{minipage}
\begin{minipage}[t]{2.8cm}
\begin{eqnarray*}
 \begin{array}{|lrl}
 \widehat{a}~\widehat{b}~\widehat{c} && f_{\widehat{a}\widehat{b}\widehat{c}} \\ \hline 
 {}_{\widehat{6}} ~{}_{\widehat{7}} ~{}_{\widehat{8}} &&{}_{\sqrt{3}/2}\\
 {}_{\widehat{6}} ~{}_{\widehat{10}} ~{}_{\widehat{13}} &&{}_{1/2}\\
 {}_{\widehat{6}} ~{}_{\widehat{11}} ~{}_{\widehat{12}} &{}_-\!\!\!\!\!&{}_{1/2}\\
 {}_{\widehat{7}} ~{}_{\widehat{10}} ~{}_{\widehat{12}} &&{}_{1/2}\\
 {}_{\widehat{7}} ~{}_{\widehat{11}} ~{}_{\widehat{13}} &&{}_{1/2}\\
 {}_{\widehat{8}} ~{}_{\widehat{10}} ~{}_{\widehat{11}} &&{}_{\sqrt{3}/3}\\
 \end{array}
\end{eqnarray*}
\end{minipage}
\begin{minipage}[t]{2.8cm}
\begin{eqnarray*}
 \begin{array}{|lrl}
 \widehat{a}~\widehat{b}~\widehat{c} && f_{\widehat{a}\widehat{b}\widehat{c}} \\ \hline 
 {}_{\widehat{8}} ~{}_{\widehat{12}} ~{}_{\widehat{13}} &{}_-\!\!\!\!\!&{}_{\sqrt{3}/6}\\
 {}_{\widehat{8}} ~{}_{\widehat{14}} ~{}_{\widehat{15}} &{}_-\!\!\!\!\!&{}_{\sqrt{3}/6}\\
 {}_{\widehat{9}} ~{}_{\widehat{10}} ~{}_{\widehat{11}} &&{}_{\sqrt{6}/3}\\
 {}_{\widehat{9}} ~{}_{\widehat{12}} ~{}_{\widehat{13}} &&{}_{\sqrt{6}/3}\\
 {}_{\widehat{9}} ~{}_{\widehat{14}} ~{}_{\widehat{15}}  &&{}_{\sqrt{6}/3}\\
 &&{}_{}\\
 \end{array}
\end{eqnarray*}
\end{minipage}
\end{table}
From Tables~\ref{fwabc} and \ref{fabcwww}, it is seen that
$\lambda_{\widehat{1}}, \cdots,\lambda_{\widehat{8}}$ 
form the algebra of SU(3) transformations
and that $\lambda_{\widehat{9}}$ which commutes 
$\lambda_{\widehat{1}},\cdots,\lambda_{\widehat{9}}$ form
the algebra of U(1) transformations.

\newpage

\subsection{Spinor notation \label{spinor}}

In 5D and 6D, the spinor representation is
pseudo real \cite{Strathdee:1986jr}
\begin{eqnarray}
 \Gamma_0 {\cal C} \Gamma_0^* {\cal C}^* =
 -{\cal C}{\cal C}^\dag ,
 \label{pser}
\end{eqnarray}
where ${\cal C}$ denotes 5D and 6D charge conjugation matrices
collectively and
$\Gamma_0$ denotes 0-component of
5D and 6D gamma matrices
collectively.
The gamma matrices and charge conjugation 
matrix to satisfy the pseudo-real condition (\ref{pser})
are listed in the following.

The 4D gamma matrices are
\begin{eqnarray}
 \gamma^m
  &\!\!\!=\!\!\!&\left(
  \begin{array}{cc}
   0 & \sigma^m \\
  \bar{\sigma}^m & 0
  \end{array}
 \right) ,~~~~~~
   \sigma^m=({\bf 1}_2,\vec{\sigma}) ,~~~
   \bar{\sigma}^m=({\bf 1}_2,-\vec{\sigma}) , 
\end{eqnarray}
where 
$\lbrace\gamma^m,\gamma^n\rbrace
   =-2\eta^{mn}$ and
$\eta^{mn}=\textrm{diag}(-1,1,1,1)$.
The 4D chirality matrix is
\begin{eqnarray}
  -i\gamma^5=\left(
   \begin{array}{cc}
    -{\bf 1}_2& \\
      & {\bf 1}_2 \\
   \end{array}
   \right) .
\end{eqnarray}
The 4D charge conjugation matrix is
\begin{eqnarray}
  C=\gamma^0\gamma^2
   =\left(
 \begin{array}{cccc}
  & i& & \\
 -i & & & \\
& & & -i\\
& & i& \\
 \end{array}\right) ,~~~
 C^T =-C ,~~~ 
  C^{-1}=C .
\end{eqnarray}
where $\gamma^m{}^T =-C^{-1}\gamma^m C$.
4D Dirac fermions and the conjugate are
\begin{eqnarray}
     \psi=\left(
  \begin{array}{c}
   \psi_L\\
   \psi_R\\
  \end{array}
     \right) , ~~~
  \bar{\psi}=\psi^\dag\gamma^0 .
\end{eqnarray}
 
\vspace{4ex}

The 5D gamma matrices are 
$\gamma^m$ and $\gamma^5$.
The 5D charge conjugation matrix is
\begin{eqnarray}
 C_5=\left(\begin{array}{cc}
      i\sigma^2& \\
      &i\sigma^2 \\
	   \end{array}\right)
    =\left(\begin{array}{cccc}
      &1&& \\
      -1&&& \\
      &&&1 \\
     &&-1& \\
	   \end{array}\right) .
\end{eqnarray}
The matrix has the property,
\begin{eqnarray}
 C_5^{-1}=-C_5 ,~~~ C_5^T=-C_5, ~~~ 
   C_5^\dag C_5={\bf 1}_4 . 
\end{eqnarray}
\begin{eqnarray}
  (\gamma^m)^T=C_5^{-1}\gamma^m C_5 ,\quad~
 (\gamma^5)^T=C_5^{-1}\gamma^5 C_5 .
\end{eqnarray}

The 5D minimal spinor is written as
two symplectic Majorana spinors or 
a Dirac spinor.
5D symplectic Majorana spinors are
\begin{eqnarray}
  \psi^1=\left(
  \begin{array}{c}
   (\psi_L)_\alpha \\
    (\bar{\psi}_R)^{\dot{\alpha}} \\
  \end{array}\right) ,~~~~
  \psi^2=\left(
  \begin{array}{c}
   (\psi_R)_\alpha \\
    -(\bar{\psi}_L)^{\dot{\alpha}} \\
  \end{array}\right) .
 \label{5dsp}
\end{eqnarray}
where $\psi_L$ and $\psi_R$ are 4D two-component Weyl spinors.
These satisfy the symplectic Majorana condition
\begin{eqnarray}
 \psi^i=\epsilon^{ij}C_5\bar{\psi}_j^T ,~~~~
 \epsilon^{12}=-\epsilon^{21}=1.
\end{eqnarray}
A 5D Dirac spinor is written in terms of 4D Weyl spinor as
\begin{eqnarray}
     \psi=\left(
  \begin{array}{c}
   \psi_L\\
   \psi_R\\
  \end{array}
     \right) .
\end{eqnarray}

\vspace{4ex}

The 6D gamma matrices are
\begin{eqnarray}
 \Gamma^M=(\Gamma^m,\Gamma^5,\Gamma^6) .
\end{eqnarray}
\begin{eqnarray}
 \Gamma^m
  &\!\!\!=\!\!\!&\gamma^m \otimes {\bf 1}_2
  =\left(
  \begin{array}{cc}
   \gamma^m& \\
   &\gamma^m \\
  \end{array}\right) ,
\\
 \Gamma^5
  &\!\!\!=\!\!\!&\gamma^5 \otimes \sigma^1
  =\left(
  \begin{array}{cc}
   &\gamma^5 \\
   \gamma^5& \\
  \end{array}\right) ,
\\
 \Gamma^6
  &\!\!\!=\!\!\!&\gamma^5 \otimes \sigma^2
  =\left(
  \begin{array}{cc}
   &-i\gamma^5 \\
   i\gamma^5 &\\
  \end{array}\right) ,
\end{eqnarray}
where 
$\lbrace\Gamma^M,\Gamma^N\rbrace
   =-2\eta^{MN}$ and
$\eta^{MN}=\textrm{diag}(-1,1,1,1,1,1)$.
The 6D chirality matrix is
\begin{eqnarray}
 -i\Gamma^7
  &\!\!\!=\!\!\!&-i\gamma^5 \otimes \sigma^3
  =\left(
  \begin{array}{ccc}
   -{\bf 1}_2&  & \\
     &{\bf 1}_4&   \\
     & &-{\bf 1}_2\\
  \end{array}\right) .
\end{eqnarray}
The 6D charge conjugation matrix is
\begin{eqnarray}
 C_6=\gamma^5 C_5 \otimes \sigma^2
  =\left(
\begin{array}{cc}
 & -i\gamma^5 C_5 \\
 i\gamma^5 C_5 & \\
\end{array}\right) .
\end{eqnarray}
\begin{eqnarray}
 C_6^{-1}=C_6, ~~~
 C_6^T=C_6,~~~
 C_6^\dag C_6={\bf 1}_8 .
\end{eqnarray}
\begin{eqnarray}
 (\Gamma^M)^T=-C_6\Gamma^M C_6 .
\end{eqnarray}
The 6D symplectic Majorana spinors are written as
\begin{eqnarray}
 \Psi^1=\left(
 \begin{array}{r}
  \psi_\alpha\\
 \bar{\eta}^{\dot{\alpha}} \\
  -u_\beta \\
  -\bar{v}^{\dot{\beta}}
 \end{array}\right) ,~~~
\Psi^2=\left(
 \begin{array}{r}
  v_\alpha\\
  \bar{u}^{\dot{\alpha}} \\
  \eta_\beta \\
  \bar{\psi}^{\dot{\beta}}
 \end{array}\right) ,~~~
\bar{\Psi}_1=\left(
 \begin{array}{r}
  \eta^\alpha\\
  \bar{\psi}_{\dot{\alpha}} \\
  -v^\beta \\
  -\bar{u}_{\dot{\beta}}
 \end{array}\right)^T ,~~~
\bar{\Psi}_2=\left(
 \begin{array}{r}
  u^\alpha\\
  \bar{v}_{\dot{\alpha}} \\
  \psi^\beta \\
  \bar{\eta}_{\dot{\beta}}
 \end{array}\right)^T ,
\end{eqnarray}
where $\psi_\alpha, \eta_\alpha, u_\alpha, v_\alpha$ are
four 4D two-component left-handed Weyl spinors.
These satisfy the symplectic Majorana condition 
given in (\ref{6dsymp}).

\section{Formula for superfields and Lagrangian}

\subsection{Some formula in superfield formalism \label{ap:sf}}

The spinor superfield is given by
\begin{eqnarray}
 W_\alpha(y,\theta,x^i) 
   &\!\!\!=\!\!\!& -\textrm{${1\over 4}$}\bar{D}\bar{D}e^{-2V} 
   D_\alpha e^{2V} ,
\nonumber
\\
  &\!\!\!=\!\!\!& 2\bigg[ -i\lambda_\alpha(y,x^i)
   +\theta_\beta\left(
    \delta_\alpha^\beta
 D(y,x^i)-{1\over 2}i
(\sigma^n\bar{\sigma}^m)_\alpha^{\cdot \beta}
  F_{nm}(y,x^i)\right)
\nonumber
\\
 &&+\theta^2 \sigma_{\alpha\dot{\alpha}}^n
   ({\partial\over \partial y^n}\bar{\lambda}^{\dot{\alpha}}
    +iA_n \bar{\lambda}^{\dot{\alpha}}
    -i\bar{\lambda}^{\dot{\alpha}}A_n)(y,x^i)\bigg] ,
\nonumber
\end{eqnarray}
where 
 $F_{nm}(y,x^i)=(\partial_n A_m - \partial_m A_n
   +iA_n A_m -iA_m A_n)(y,x^i)$.

On the fixed line $x^6=0$,
the left-handed chiral superfield
$\Phi_l(y,x^5)$ and its complex conjugate
$\Phi_l^c(y,x^5)$ are expanded with
component fields as
\begin{eqnarray}
 \Phi_l(y,x^5)&\!\!\!=\!\!\!&\phi_l(y,x^5) +\sqrt{2}\theta\chi_l(y,x^5) +\theta\theta F_l(y,x^5) ,
\\
 \Phi_l^c(y,x^5)&\!\!\!=\!\!\!&\phi_l^c(y,x^5)+\sqrt{2}\theta\chi_l^c(y,x^5)
  +\theta\theta F_l^c(y,x^5) .
\end{eqnarray}
The action $S_5$ (\ref{actions5}) is
written in terms of component fields as
\begin{eqnarray}
 S_5&\!\!\!\!=\!\!\!\!&\int d^6x ~\delta (x^6)
\nonumber
\\
 &\times\!\!\!\!&\bigg[
  -({\cal D}_n \phi_l)^*({\cal D}^n \phi_l)
   +F_l^* F_l-i\bar{\chi}_l\bar{\sigma}^n{\cal D}_n\chi_l
   +\sqrt{2}i(\phi_l^*\lambda\chi_l-\bar{\chi}_l\bar{\lambda}\phi_l)
    +\phi_l^*D\phi_l
\nonumber
\\
&&
  -({\cal D}_n \phi_l^c)({\cal D}^n \phi_l^c)^*
   +F_l^c F_l^{c*}
   -i\chi_l^c\sigma^n{\cal D}_n\bar{\chi}_l^c
  +\sqrt{2}i(\phi_l^c\bar{\lambda}\bar{\chi}_l^c
   -\chi_l^c\lambda \phi_l^{c*})
  -\phi_l^c D \phi_l^{c*}
\nonumber
\\
 &&+\bigg(
F_l^c \nabla_5 \phi_l-\chi_l^c\lambda_5 \phi_l
  -\chi_l^c \nabla_5 \chi_l +\phi_l^c F_5 \phi_l
  -\phi_l^c\lambda_5 \chi_l +\phi_l^c \nabla_5 F_l
  +\textrm{H.c.}
 \bigg)
 \bigg] ,
 \label{s5after}
\end{eqnarray}
where
\begin{eqnarray}
 {\cal D}_n \phi_l&\!\!=\!\!&\partial_n \phi_l+i A_n \phi_l ,\qquad~
 {\cal D}_n \phi_l^c =\partial_n \phi_l^c-i \phi_l^c A_n ,
\\
 {\cal D}_n\chi_l&\!\!=\!\!&\partial_n\chi_l+iA_n \chi_l, \qquad~
 {\cal D}_n\chi_l^c=\partial_n\chi_l^c -i\chi_l^c A_n ,
\\
 \nabla_5 \phi_l&\!\!=\!\!& (\partial_5+\sqrt{2}
\sigma_5)\phi_l ,\quad
\nabla_5 \chi_l= (\partial_5+\sqrt{2}
\sigma_5)\chi_l ,\quad
\nabla_5 F_l= (\partial_5+\sqrt{2}
\sigma_5)F_l .
\end{eqnarray}
For $\phi_l^c$, the operator $\nabla_5$ is defined as
\begin{eqnarray}
 \nabla_5 \phi_l^c =\partial_5 \phi_l^c -\sqrt{2}
\phi_l^c\sigma_5  ,
\end{eqnarray}
and 
\begin{eqnarray}
 (\nabla_5 \phi_l)^*=\partial_5 \phi_l^*
+\sqrt{2}\phi_l^*\sigma_5^* ,\qquad~ 
 (\nabla_5 \phi_l^c)^*=\partial_5 \phi_l^{c*} 
-\sqrt{2}\sigma_5^* \phi_l^{c*} . 
\end{eqnarray}
Fields confined on $x^6=\pi R_6$,
$x^5=0,\pi R_5$ and their actions 
are obtained in a parallel way.

For a generic 4D field
\begin{eqnarray}
 \Phi_p&\!\!=\!\!&\phi_p(y) +\sqrt{2}\theta\chi_p(y) 
 +\theta\theta F_p(y) ,
\end{eqnarray}
the action $S_4$ (\ref{actions4}) becomes
\begin{eqnarray}
&\!\!\!\!& S_4=\int d^6 x ~ \delta(x^5)\delta(x^6)
\nonumber
\\
 &\!\!\!\!\!\times\!\!\!&\bigg[
 -({\cal D}_n \phi_p)^*({\cal D}^n \phi_p)
  +F_p^* F_p -i\bar{\chi}_p\bar{\sigma}^n {\cal D}_n\chi_p
 +\sqrt{2}i(\phi_p^*\lambda\chi_p-\bar{\chi}_p\bar{\lambda}\phi_p)
 +\phi_p^* D\phi_p
\bigg] .
\end{eqnarray}
with
\begin{eqnarray}
 {\cal D}_n \phi_p&\!\!\!=\!\!\!&\partial_n \phi_p+i A_n \phi_p ,\qquad~ 
{\cal D}_n \chi_p=\partial_n \chi_p+i A_n \chi_p .
\end{eqnarray}
Here superpotential and supersymmetry-breaking terms
are omitted.

\vspace{4ex}

From $S_6$, $S_5$ and $S_4$, the auxiliary field part of the 
action is 
\begin{eqnarray}
 &&S_{\textrm{\scriptsize aux}}
\nonumber
\\
&\!\!=\!\!\!\!& 
\int d^6x \bigg\lbrace
  {1\over kg^2}\textrm{Tr}
  \bigg[
   {1\over 2}D^2+
  \bigg(\phi(\nabla_5 F_6-\nabla_6 F_5)
  +F(\partial_5\sigma_6-\partial_6\sigma_5
   +\sqrt{2}\left[\sigma_5,\sigma_6\right]) 
   +\textrm{H.c.}
 \bigg)
\nonumber
\\
  &&
   +{1\over 2}\sum_{j=5}^6
   (-\sqrt{2}
(\partial_j D)\sigma_j -\sqrt{2}
\sigma_j^* \partial_j D
  +2F_j^*F_j +2\sigma_j^* D\sigma_j 
-2\sigma_j^*\sigma_j D)
   +F^*F+\phi^*D\phi -\phi^*\phi D
\bigg]
\nonumber
\\
 &&+\delta(x^6)
   \bigg[
   F_l^* F_l +\phi_l^* D\phi_l+F_l^c F_l^{c*} -\phi_l^c D\phi_l^{c*}
   +\bigg(
  F_l^c\nabla_5 \phi_l 
+\sqrt{2}\phi_l^c F_5 \phi_l +\phi_l^c \nabla_5 F_l 
   +\textrm{H.c.}\bigg)
\bigg]
\nonumber
\\
 &&+\delta(x^5)\delta(x^6)
   \left[F_p^* F_p+ \phi_p^* D\phi_p\right]
 \bigg\rbrace   .
\end{eqnarray}
The equations of motion for auxiliary fields are 
\begin{eqnarray}
 &&
  \textrm{${1\over g^2}$}D^a 
  +\textrm{${\sqrt{2}\over 2g^2}$}\sum_{j=5}^6 
\left(
  \partial_j\sigma_j^a
  +\partial_j\sigma_j^a{}^*
 +\sqrt{2}it^{bac}\sigma_j^{*b}\sigma_j^c\right)
  +\textrm{${1\over g^2}$}it^{bac}\phi^{*b}\phi^c
\nonumber
\\
  &&\qquad
   +\delta(x^6)
(\phi_l^*T^a \phi_l-\phi_l^c T^a \phi_l^{c*}) 
 +\delta(x^5)\delta(x^6)
   \phi_p^* T^a \phi_p =0 ,
   \label{ap:deq}
\\
 &&
 \partial_5\sigma_6^a -\partial_6\sigma_5^a
  +\sqrt{2}it^{abc}\sigma_5^b\sigma_6^c 
  +F^{*a}=0 ,
\\
 && 
   \partial_6\phi^a
   -\sqrt{2}\phi^b \sigma_6^c it^{abc}
   +F_5^{*a}
  +\sqrt{2}g^2\delta(x^6)\phi_l^c T^a \phi_l
   =0 ,
  \label{ap:f5eq}
\\
 && 
   -\partial_5\phi^a
   +\sqrt{2}\phi^b \sigma_5^c it^{abc}
   +F_6^{*a}
   =0 ,
  \label{ap:f6eq}
\\
 &&
  F_l^* -\partial_5 \phi_l^c +\sqrt{2}\phi_l^c \sigma_5=0,
   \label{ap:fheq}
\\
 &&
    F_l^{c*}+\partial_5 \phi_l +\sqrt{2}\sigma_5 \phi_l=0 ,
 \label{ap:fheqc}
\\
 &&
  F_p=0.
\end{eqnarray}
Here partial integrals are used in obtaining $\partial_6 \phi^a$
in Eq.(\ref{ap:f5eq}), $\partial_5 \phi^a$
in Eq.(\ref{ap:f6eq}) and
$\partial_5 \phi_l^c$ in Eq.(\ref{ap:fheq}).

\subsection{Bulk field action}

The operator of
bosonic part of 6D bulk action (\ref{componentl6})
is 
\begin{eqnarray}
 {\cal O}_B&\!\!\!=\!\!\!&
 -{1\over 4}F_{nm}F^{nm}
+{1\over 2}
 \bigg(-\partial_j A^n \partial_j A_n
   -\sqrt{2}i(\partial_j A^n)
     ({\cal D}_n \sigma_j)
   +\sqrt{2}i({\cal D}^n \sigma_j)^*
    (\partial_j A_n)
\nonumber
\\
 &&  -2({\cal D}^n\sigma_j)^*
  ({\cal D}_n\sigma_j)
  -2({\cal D}_n\phi)^*({\cal D}_n\phi)    \bigg) 
\nonumber
\\
 &&-{1\over 2}
  \bigg(
  {\sqrt{2}\over 2}\left(
   \partial_j\sigma_j
    +\partial_j\sigma_j^*
   +\sqrt{2}\left[\sigma_j^*,\sigma_j\right]\right)
  +\left[\phi^*,\phi\right]
\bigg)^2
\nonumber
\\
  &&-
  \left(\partial_5\sigma_6-\partial_6\sigma_5
   +\sqrt{2}\left[\sigma_5,\sigma_6\right]\right)
  \left(\partial_5\sigma_6^*-\partial_6\sigma_5^*
   -\sqrt{2}\left[\sigma_5^*,\sigma_6^*\right]\right)
\nonumber
\\
  &&-
   \left(\partial_j\phi+\sqrt{2}\left[\sigma_j,\phi\right]\right)
   \left(\partial_j\phi^*-\sqrt{2}\left[\sigma_j^*,\phi^*\right]\right) .
\nonumber
\\
 &\!\!\!\cong\!\!\!&
  -{1\over 4}F_{nm}F^{nm}
  +\sum_{j=5}^{6}\bigg(
  -{1\over 2}\partial_jA^n\partial_j A_n
  +(\partial_j A^n)(\partial_n A_j
    +i\left[A_n,A_j\right])
\nonumber
\\
 &&-{1\over 2}(\partial^n A_j -i\left[A_j, A^n\right])
   (\partial_n A_j +i\left[A_n,A_j\right])
  \bigg)
 -{1\over 2}
   \left(\partial_5 A_6-\partial_6 A_5
    +i\left[A_5,A_6\right]\right)^2
\nonumber
\\
 &&-\sum_{s=1}^2
    (\partial^n\pi_s+i\left[A^n,\pi_s\right])
    (\partial_n\pi_s+i\left[A_n,\pi_s\right])
   -
   ({\cal D}^n \phi)^*({\cal D}_n\phi)
\nonumber
\\
 &&-\sum_{j=5}^{6}\sum_{s=1}^2
   (\partial_j\pi_s+i\left[A_j,\pi_s\right])^2
-\sum_{j=5}^{6}(\partial_j\phi
  +i\left[A_j,\phi\right])
  (\partial_j\phi^*-i\left[\phi^*,A_j\right])
\nonumber
\\
 &&-{1\over 2}
   (\left[\phi^*,\phi\right])^2
  -\sum_{s=1}^2\left[\pi_s,\phi\right]
   \left[\phi^*,\pi_s\right] .
  \label{lagbos}
\end{eqnarray}
Here $\cong$ indicates that both hand sides are equal to each other
when the trace $\textrm{Tr}$ is taken into account.
The bosonic operator (\ref{lagbos}) is written in terms of bulk 
bosons as in Eq.(\ref{bosop}).

The operator of fermionic part of 6D bulk 
action (\ref{componentl6}) is
\begin{eqnarray}
 {\cal O}_F&\!\!\!\!=\!\!\!\!&
   -i\lambda \sigma^n{\cal D}_n\bar{\lambda}
  +\left(
    -\chi(\nabla_5\lambda_6
   -\nabla_6\lambda_5)
   -\sqrt{2}\phi\left[\lambda_5,\lambda_6\right]
  +\textrm{H.c.}\right)
\nonumber
\\
 &&+\sum_{j=5}^6
  \left(-i\bar{\lambda}_j\bar{\sigma}^n
    {\cal D}_n\lambda_j
  -i(\mathfrak{D}_j\lambda)\lambda_j
   +i\bar{\lambda}_j
(\mathfrak{D}_j\bar{\lambda})\right)
\nonumber
\\
 &&
   -i\bar{\chi}\bar{\sigma^n}{\cal D}_n\chi
   +\left(-\sqrt{2}i\bar{\chi}\left[\bar{\lambda},\phi\right]
   +\textrm{H.c.}\right) .
 \label{lagfer}
\end{eqnarray}
Here
\begin{eqnarray}
 \mathfrak{D}_j\lambda=\partial_j\lambda
   -\sqrt{2}\sigma_j^*\lambda+\sqrt{2}\lambda\sigma_j^* , 
\quad~
 \mathfrak{D}_j\bar{\lambda}=
   \partial_j\bar{\lambda}
  -\sqrt{2}\bar{\lambda}\sigma_j
+\sqrt{2}\sigma_j\bar{\lambda} .
\end{eqnarray}
The operator (\ref{lagfer}) becomes Eq.(\ref{ferop}).

\section{Mode and parity of fields}

\subsection{Mode expansion \label{xpan}}

A generic 6D massless field 
$\phi_B(x,x^5,x^6)$ is mode-expanded with possible boundary conditions
at $x^5=0,\pi R_5$, $x^6=0,\pi R_6$ as follow:
\begin{eqnarray}
 &&\phi_B(x,x^5,x^6)=
\nonumber
\\
&&
  \phi_{00}^{++++}(x)
  +\sum_{n=1}^\infty
   \sqrt{2}\phi_{n0}^{++++}(x)
   \cos\left({nx^5\over R_5}\right)
\nonumber
\\
 &&+\sum_{n=1}^\infty
   \sqrt{2}
   \phi_{0n}^{++++}(x)\cos\left(nx^6\over R_6\right)
 +\sum_{n=1}^\infty \sum_{m=1}^\infty
     2\phi_{nm}^{++++}(x)
   \cos\left({nx^5\over R_5}\right)
   \cos\left({mx^6\over R_6}\right) 
\nonumber
\\
 && \qquad\qquad\qquad\qquad\qquad\qquad \qquad\qquad\qquad\qquad\qquad\qquad\qquad
 \textrm{for}~ (++++) ,
\\
&&
  \sum_{n=1}^\infty \sqrt{2}
   \phi_{n0}^{+-++}(x)\cos\left((n-\textrm{${1\over 2}$})x^5\over R_5\right)
 +\sum_{n=1}^\infty \sum_{m=1}^\infty
     2\phi_{nm}^{+-++}(x)
   \cos\left({(n-\textrm{${1\over 2}$})x^5\over R_5}\right)
   \cos\left({mx^6\over R_6}\right) 
\nonumber
\\
 && \qquad\qquad\qquad\qquad\qquad\qquad \qquad\qquad\qquad\qquad\qquad\qquad\qquad
 \textrm{for}~ (+-++) ,
  \label{modepmpp}
\\
&& 
 \sum_{n=1}^\infty \sum_{m=1}^\infty
     2\phi_{nm}^{+-++}(x)
   \cos\left({(n-\textrm{${1\over 2}$})x^5\over R_5}\right)
   \cos\left({(m-\textrm{${1\over 2}$})x^6\over R_6}\right) 
\nonumber
\\
 && \qquad\qquad\qquad\qquad\qquad\qquad \qquad\qquad\qquad\qquad\qquad\qquad\qquad
 \textrm{for}~ (+-+-) ,
\\
&&
  \sum_{n=1}^\infty
   \sqrt{2}
   \phi_{0n}^{+++-}(x)
 \cos\left((n-\textrm{${1\over 2}$})x^6\over R_6\right)
+\sum_{n=1}^\infty \sum_{m=1}^\infty
     2\phi_{nm}^{+++-}(x)
   \cos\left({nx^5\over R_5}\right)
   \cos\left({(m-\textrm{${1\over 2}$})x^6\over R_6}\right) 
\nonumber
\\
 && \qquad\qquad\qquad\qquad\qquad\qquad \qquad\qquad\qquad\qquad\qquad\qquad\qquad
 \textrm{for}~ (+++-) ,
\\
&&
  \sum_{n=1}^\infty
   \sqrt{2}
   \phi_{n0}^{--++}(x)\sin\left(nx^5\over R_5\right)
 +\sum_{n=1}^\infty \sum_{m=1}^\infty
     2\phi_{nm}^{--++}(x)
   \sin\left({nx^5\over R_5}\right)
   \cos\left({mx^6\over R_6}\right) 
\nonumber
\\
 && \qquad\qquad\qquad\qquad\qquad\qquad \qquad\qquad\qquad\qquad\qquad\qquad\qquad
 \textrm{for}~ (--++) ,
  \label{modemmpp}
\\
&&
  \sum_{n=1}^\infty
   \sqrt{2}
   \phi_{n0}^{-+++}(x)\sin\left((n-\textrm{${1\over 2}$})
 x^5\over R_5\right)
 +\sum_{n=1}^\infty \sum_{m=1}^\infty
     2\phi_{nm}^{-+++}(x)
   \sin\left({(n-\textrm{${1\over 2}$})x^5\over R_5}\right)
   \cos\left({mx^6\over R_6}\right) 
\nonumber
\\
 && \qquad\qquad\qquad\qquad\qquad\qquad \qquad\qquad\qquad\qquad\qquad\qquad\qquad
 \textrm{for}~ (-+++) ,
  \label{modemppp}
\\
&&
  \sum_{n=1}^\infty \sum_{m=1}^\infty
     2\phi_{nm}^{-++-}(x)
   \sin\left({(n-\textrm{${1\over 2}$})x^5\over R_5}\right)
   \cos\left({(m-\textrm{${1\over 2}$})x^6\over R_6}\right) 
\nonumber
\\
 && \qquad\qquad\qquad\qquad\qquad\qquad \qquad\qquad\qquad\qquad\qquad\qquad\qquad
 \textrm{for}~ (-++-) ,
\\
&&
  \sum_{n=1}^\infty \sum_{m=1}^\infty
     2\phi_{nm}^{--+-}(x)
   \sin\left({nx^5\over R_5}\right)
   \cos\left({(m-\textrm{${1\over 2}$})x^6\over R_6}\right) 
\nonumber
\\
 && \qquad\qquad\qquad\qquad\qquad\qquad \qquad\qquad\qquad\qquad\qquad\qquad\qquad
 \textrm{for}~ (--+-) ,
\\
&&
  \sum_{n=1}^\infty
   \sqrt{2}
   \phi_{0n}^{++--}(x)\sin\left(nx^6\over R_6\right)
 +\sum_{n=1}^\infty \sum_{m=1}^\infty
     2\phi_{nm}^{++--}(x)
   \cos\left({nx^5\over R_5}\right)
   \sin\left({mx^6\over R_6}\right) 
\nonumber
\\
 && \qquad\qquad\qquad\qquad\qquad\qquad \qquad\qquad\qquad\qquad\qquad\qquad\qquad
 \textrm{for}~ (++--) ,
\\
&&
  \sum_{n=1}^\infty \sum_{m=1}^\infty
     2\phi_{nm}^{+---}(x)
   \cos\left({(n-\textrm{${1\over 2}$})x^5\over R_5}\right)
   \sin\left({mx^6\over R_6}\right) 
\nonumber
\\
 && \qquad\qquad\qquad\qquad\qquad\qquad \qquad\qquad\qquad\qquad\qquad\qquad\qquad
 \textrm{for}~ (+---) ,
\\
&&
  \sum_{n=1}^\infty \sum_{m=1}^\infty
     2\phi_{nm}^{+--+}(x)
   \cos\left({(n-\textrm{${1\over 2}$})x^5\over R_5}\right)
   \sin\left({(m-\textrm{${1\over 2}$})x^6\over R_6}\right) 
\nonumber
\\
 && \qquad\qquad\qquad\qquad\qquad\qquad \qquad\qquad\qquad\qquad\qquad\qquad\qquad
 \textrm{for}~ (+--+) ,
\\
&&
  \sum_{n=1}^\infty
   \sqrt{2}
   \phi_{0n}^{++-+}(x)\sin\left((n-\textrm{${1\over 2}$})x^6\over R_6\right)
 +\sum_{n=1}^\infty \sum_{m=1}^\infty
     2\phi_{nm}^{++-+}(x)
   \cos\left({nx^5\over R_5}\right)
   \sin\left({(m-\textrm{${1\over 2}$})x^6\over R_6}\right) 
\nonumber
\\
 && \qquad\qquad\qquad\qquad\qquad\qquad \qquad\qquad\qquad\qquad\qquad\qquad\qquad
 \textrm{for}~ (++-+) ,
\\
&&
  \sum_{n=1}^\infty \sum_{m=1}^\infty
     2\phi_{nm}^{----}(x)
   \sin\left({nx^5\over R_5}\right)
   \sin\left({mx^6\over R_6}\right) 
\nonumber
\\
 && \qquad\qquad\qquad\qquad\qquad\qquad \qquad\qquad\qquad\qquad\qquad\qquad\qquad
 \textrm{for}~ (----) ,
\\
&&
  \sum_{n=1}^\infty \sum_{m=1}^\infty
     2\phi_{nm}^{-+--}(x)
   \sin\left({(n-\textrm{${1\over 2}$})x^5\over R_5}\right)
   \sin\left({mx^6\over R_6}\right) 
\nonumber
\\
 && \qquad\qquad\qquad\qquad\qquad\qquad \qquad\qquad\qquad\qquad\qquad\qquad\qquad
 \textrm{for}~ (-+--) ,
\\
&&
  \sum_{n=1}^\infty \sum_{m=1}^\infty
     2\phi_{nm}^{-+-+}(x)
   \sin\left({(n-\textrm{${1\over 2}$})x^5\over R_5}\right)
   \sin\left({(m-\textrm{${1\over 2}$})x^6\over R_6}\right) 
\nonumber
\\
 && \qquad\qquad\qquad\qquad\qquad\qquad \qquad\qquad\qquad\qquad\qquad\qquad\qquad
 \textrm{for}~ (-+-+) ,
\\
&&
  \sum_{n=1}^\infty \sum_{m=1}^\infty
     2\phi_{nm}^{---+}(x)
   \sin\left({nx^5\over R_5}\right)
   \sin\left({(m-\textrm{${1\over 2}$})x^6\over R_6}\right) 
\nonumber
\\
 && \qquad\qquad\qquad\qquad\qquad\qquad\qquad\qquad\qquad\qquad \qquad\qquad\qquad
 \textrm{for}~ (---+) .
\end{eqnarray}

\subsection{Parity of fields\label{ap:par}}

\subsubsection{Model with Higgs as 5D multiplets}

For the parity matrices (\ref{parity1})
\begin{eqnarray}
 P_1=P_3={\bf 1}_6 ,\quad~ 
 P_2=\left(
  \begin{array}{cc}
   -{\bf 1}_{2}& \\
         &{\bf 1}_{4} \\
  \end{array}\right) ,\quad~
 P_4=\left(
  \begin{array}{ccccc|c}
    {\bf 1}_{5} & \\
     &-1 \\ 
  \end{array}\right) ,
\end{eqnarray}
the parities of fields are obtained in the following:
for $V(A_m,\lambda)$ and $\Sigma_5(\sigma_5,\lambda_5)$, 
the parities are given in (\ref{vparity5dh}) and (\ref{5parity5dh}).

For $\Sigma_6(\sigma_6,\lambda_6) $,
\begin{eqnarray}
 \left(\begin{array}{cc|ccc|c}
   _{(++--)} & _{(++--)} & _{(+---)} & _{(+---)} & _{(+---)} & _{(+--+)} \\
   _{(++--)} & _{(++--)} & _{(+---)} & _{(+---)} & _{(+---)} & _{(+--+)} \\ \hline
   _{(+---)} & _{(+---)} & _{(++--)} & _{(++--)} & _{(++--)} & _{(++-+)} \\
   _{(+---)} & _{(+---)} & _{(++--)} & _{(++--)} & _{(++--)} & _{(++-+)} \\
   _{(+---)} & _{(+---)} & _{(++--)} & _{(++--)} & _{(++--)} & _{(++-+)} \\ \hline
   _{(+--+)} & _{(+--+)} & _{(++-+)} & _{(++-+)} & _{(++-+)} & _{(++--)} \\
 \end{array}\right) .
\end{eqnarray}

For $\Phi(\phi,\chi) $,
\begin{eqnarray}
 \left(\begin{array}{cc|ccc|c}
   _{(----)} & _{(----)} & _{(-+--)} & _{(-+--)} & _{(-+--)} & _{(-+-+)} \\
   _{(----)} & _{(----)} & _{(-+--)} & _{(-+--)} & _{(-+--)} & _{(-+-+)} \\ \hline
   _{(-+--)} & _{(-+--)} & _{(----)} & _{(----)} & _{(----)} & _{(---+)} \\
   _{(-+--)} & _{(-+--)} & _{(----)} & _{(----)} & _{(----)} & _{(---+)} \\
   _{(-+--)} & _{(-+--)} & _{(----)} & _{(----)} & _{(----)} & _{(---+)} \\ \hline
   _{(-+-+)} & _{(-+-+)} & _{(---+)} & _{(---+)} & _{(---+)} & _{(----)} \\
 \end{array}\right) .
\end{eqnarray}

\vspace{4ex}

For another choice of the parity (\ref{anotherp})
\begin{eqnarray}
 P_1=P_3={\bf 1}_6,\quad~
 P_2=\left(
  \begin{array}{cc}
   {\bf 1}_{3}& \\
         &-{\bf 1}_{3} \\
  \end{array}\right) ,\quad~
 P_4=\left(
  \begin{array}{ccccc|c}
    {\bf 1}_{5} & \\
     &-1 \\ 
  \end{array}\right) ,
\end{eqnarray}
the orbifold parity for $V$ is
\begin{eqnarray}
 \left(\begin{array}{ccc|cc|c}
   _{(++++)} & _{(++++)} & _{(++++)} & _{(+-++)} & _{(+-++)} & _{(+-+-)} \\
   _{(++++)} & _{(++++)} & _{(++++)} & _{(+-++)} & _{(+-++)} & _{(+-+-)} \\ 
   _{(++++)} & _{(++++)} & _{(++++)} & _{(+-++)} & _{(+-++)} & _{(+-+-)} \\ \hline
   _{(+-++)} & _{(+-++)} & _{(+-++)} & _{(++++)} & _{(++++)} & _{(+++-)} \\
   _{(+-++)} & _{(+-++)} & _{(+-++)} & _{(++++)} & _{(++++)} & _{(+++-)} \\ \hline
   _{(+-+-)} & _{(+-+-)} & _{(+-+-)} & _{(+++-)} & _{(+++-)} & _{(++++)} \\
 \end{array}\right) .
\end{eqnarray}
For $\Sigma_5$,
\begin{eqnarray}
 \left(\begin{array}{ccc|cc|c}
   _{(--++)} & _{(--++)} & _{(--++)} & _{(-+++)} & _{(-+++)} & _{(-++-)} \\
   _{(--++)} & _{(--++)} & _{(--++)} & _{(-+++)} & _{(-+++)} & _{(-++-)} \\ 
   _{(--++)} & _{(--++)} & _{(--++)} & _{(-+++)} & _{(-+++)} & _{(-++-)} \\ \hline
   _{(-+++)} & _{(-+++)} & _{(-+++)} & _{(--++)} & _{(--++)} & _{(--+-)} \\
   _{(-+++)} & _{(-+++)} & _{(-+++)} & _{(--++)} & _{(--++)} & _{(--+-)} \\ \hline
   _{(-++-)} & _{(-++-)} & _{(-++-)} & _{(--+-)} & _{(--+-)} & _{(--++)} \\
 \end{array}\right) .
\end{eqnarray}
For $\Sigma_6$, 
\begin{eqnarray}
 \left(\begin{array}{ccc|cc|c}
   _{(++--)} & _{(++--)} & _{(++--)} & _{(+---)} & _{(+---)} & _{(+--+)} \\
   _{(++--)} & _{(++--)} & _{(++--)} & _{(+---)} & _{(+---)} & _{(+--+)} \\ 
   _{(++--)} & _{(++--)} & _{(++--)} & _{(+---)} & _{(+---)} & _{(+--+)} \\ \hline
   _{(+---)} & _{(+---)} & _{(+---)} & _{(++--)} & _{(++--)} & _{(++-+)} \\
   _{(+---)} & _{(+---)} & _{(+---)} & _{(++--)} & _{(++--)} & _{(++-+)} \\ \hline
   _{(+--+)} & _{(+--+)} & _{(+--+)} & _{(++-+)} & _{(++-+)} & _{(++--)} \\
 \end{array}\right) .
\end{eqnarray}
For $\Phi$, 
\begin{eqnarray}
 \left(\begin{array}{ccc|cc|c}
   _{(----)} & _{(----)} & _{(----)} & _{(-+--)} & _{(-+--)} & _{(-+-+)} \\
   _{(----)} & _{(----)} & _{(----)} & _{(-+--)} & _{(-+--)} & _{(-+-+)} \\ 
   _{(----)} & _{(----)} & _{(----)} & _{(-+--)} & _{(-+--)} & _{(-+-+)} \\ \hline
   _{(-+--)} & _{(-+--)} & _{(-+--)} & _{(----)} & _{(----)} & _{(---+)} \\
   _{(-+--)} & _{(-+--)} & _{(-+--)} & _{(----)} & _{(----)} & _{(---+)} \\ \hline
   _{(-+-+)} & _{(-+-+)} & _{(-+-+)} & _{(---+)} & _{(---+)} & _{(----)} \\
 \end{array}\right) .
\end{eqnarray}

\subsubsection{Model with Higgs in gauge multiplet}

For the parity matrices (\ref{parity2})
\begin{eqnarray}
 P_1=\left(\begin{array}{cc}
    {\bf 1}_5& \\
    & -1 \\
   \end{array}\right) ,\quad~
 P_2=\left(\begin{array}{cc}
    {\bf 1}_2 & \\
    & -{\bf 1}_4 \\
    \end{array}\right) ,\quad~
 P_3=P_4={\bf 1}_6,
\end{eqnarray}
$V(A_m,\lambda)$ and 
$\Sigma_5(\sigma_5,\lambda_5)$ have the boundary conditions 
given in Eqs.(\ref{vparitygh}) and (\ref{5paritygh}). \\
For $\Sigma_6(\sigma_6,\lambda_6) $,
\begin{eqnarray}
  \left(\begin{array}{cc|ccc|c}
  _{(++--)} &_{(++--)}& _{(+---)}& _{(+---)}& _{(+---)}& _{(----)} \\
  _{(++--)} &_{(++--)}& _{(+---)}& _{(+---)}& _{(+---)}& _{(----)} \\ \hline
  _{(+---)} &_{(+---)}& _{(++--)}& _{(++--)}& _{(++--)}& _{(-+--)} \\
  _{(+---)} &_{(+---)}& _{(++--)}& _{(++--)}& _{(++--)}& _{(-+--)} \\
  _{(+---)} &_{(+---)}& _{(++--)}& _{(++--)}& _{(++--)}& _{(-+--)} \\ \hline
  _{(----)} &_{(----)}& _{(-+--)}& _{(-+--)}& _{(-+--)}& _{(++--)} \\
\end{array}\right) .
\end{eqnarray}
For $\Phi(\phi,\chi) $,
\begin{eqnarray}
  \left(\begin{array}{cc|ccc|c}
  _{(----)} &_{(----)}& _{(-+--)}& _{(-+--)}& _{(-+--)}& _{(++--)} \\
  _{(----)} &_{(----)}& _{(-+--)}& _{(-+--)}& _{(-+--)}& _{(++--)} \\ \hline
  _{(-+--)} &_{(-+--)}& _{(----)}& _{(----)}& _{(----)}& _{(+---)} \\
  _{(-+--)} &_{(-+--)}& _{(----)}& _{(----)}& _{(----)}& _{(+---)} \\
  _{(-+--)} &_{(-+--)}& _{(----)}& _{(----)}& _{(----)}& _{(+---)} \\ \hline
  _{(++--)} &_{(++--)}& _{(+---)}& _{(+---)}& _{(+---)}& _{(----)} \\
\end{array}\right) . 
\end{eqnarray}

\end{appendix}

\newpage

\vspace*{10mm}


\end{document}